\documentclass[fleqn,usenatbib]{mnras}

\usepackage{newtxtext,newtxmath}

\usepackage[T1]{fontenc}
\usepackage{ae,aecompl}

\usepackage{graphicx}	
\usepackage{amsmath}	
\usepackage{amssymb}	
\usepackage{bm}
\usepackage{tabularx}

\title[WASP-12b Spitzer Phase Curve Observations]{Mass Loss from the Exoplanet WASP-12b Inferred from \textit{Spitzer} Phase Curves}

\author[T.~J.~Bell et al.]{%
Taylor J.~Bell,$^{1,2}$\thanks{E-mail: taylor.bell@mail.mcgill.ca (TJB)}
Michael Zhang,$^{3}$
Patricio E.~Cubillos,$^{4}$
Lisa Dang,$^{1,2}$
\newauthor Luca Fossati,$^{4}$
Kamen O.~Todorov,$^{5}$
Nicolas B.~Cowan,$^{1,2,6}$
Drake Deming,$^{7}$
\newauthor
Robert T.~Zellem,$^{8}$
Kevin B.~Stevenson,$^{9}$
Ian J.~M.~Crossfield,$^{10}$
\newauthor
Ian Dobbs-Dixon,$^{11}$
Jonathan J.~Fortney,$^{12}$
Heather A.~Knutson,$^{13}$
\newauthor
and Michael R.~Line$^{14}$
\\
$^{1}$Department of Physics, McGill University, 3600 rue University, Montr\'eal, QC H3A 2T8, Canada\\
$^{2}$McGill Space Institute; Institute for Research on Exoplanets; Centre for Research in Astrophysics of Quebec\\
$^{3}$Department of Astronomy, California Institute of Technology, 1216 E California Blvd, Pasadena, CA 91125, USA\\
$^{4}$Space Research Institute, Austrian Academy of Sciences, Schmiedlstrasse 6, A-8042 Graz, Austria\\
$^{5}$Anton Pannekoek Institute for Astronomy, University of Amsterdam, Science Park 904, 1090 GE Amsterdam, The Netherlands\\
$^{6}$Department of Earth \& Planetary Sciences, McGill University, 3450 rue University, Montr\'eal, QC H3A 0E8, Canada\\
$^{7}$Department of Astronomy, University of Maryland, College Park, MD 20742, USA\\
$^{8}$Jet Propulsion Laboratory, California Institute of Technology, 4800 Oak Grove Drive, Pasadena, CA 91109, USA\\
$^{9}$Space Telescope Science Institute, Baltimore, MD 21218, USA\\
$^{10}$Department of Physics, Massachusetts Institute of Technology, Cambridge, MA, USA\\
$^{11}$Department of Physics, NYU Abu Dhabi, P.O. Box 129188, Abu Dhabi, UAE\\
$^{12}$Other Worlds Laboratory, Department of Astronomy and Astrophysics, University of California, Santa Cruz, California 95064, USA\\
$^{13}$Division of Geological and Planetary Sciences, California Institute of Technology, Pasadena, CA 91125, USA\\
$^{14}$School of Earth \& Space Exploration, Arizona State University, Tempe AZ 85287, USA
}

\date{Accepted 2019 July 17. Received 2019 July 16; in original form 2019 June 6}

\pubyear{2019}

\begin{document}
\label{firstpage}
\pagerange{\pageref{firstpage}--\pageref{lastpage}}
\maketitle

\begin{abstract}
The exoplanet WASP-12b is the prototype for the emerging class of ultra-hot, Jupiter-mass exoplanets. Past models have predicted---and near ultra-violet observations have shown---that this planet is losing mass. We present an analysis of two sets of 3.6~$\mu$m and 4.5~$\mu$m \textit{Spitzer} phase curve observations of the system which show clear evidence of infrared radiation from gas stripped from the planet, and the gas appears to be flowing directly toward or away from the host star. This accretion signature is only seen at 4.5~$\mu$m, not at 3.6~$\mu$m, which is indicative either of CO emission at the longer wavelength or blackbody emission from cool, $\lesssim$600~K gas. It is unclear why WASP-12b is the only ultra-hot Jupiter to exhibit this mass loss signature, but perhaps WASP-12b's orbit is decaying as some have claimed, while the orbits of other exoplanets may be more stable; alternatively, the high energy irradiation from WASP-12A may be stronger than the other host stars. We also find evidence for phase offset variability at the level of $6.4\sigma$ ($46.2^{\circ}$) at 3.6~$\mu$m.
\end{abstract}




\section{Introduction}

The exoplanet WASP-12b \citep{hebb2009} is one of the hottest planets known to date and, as a result of its exceedingly tight orbit and inflated radius \citep[\mbox{$a/R_* = 3.039$}, \mbox{$R_{p} = 1.900~R_{J}$};][]{collins2017}, it is one of the best-studied exoplanets. WASP-12b is also the archetype of an emerging class of exoplanets called ultra-hot Jupiters (UHJs). Planets in this regime are so strongly irradiated by their host star that many of the molecules (e.g., H$_2$ \& H$_2$O) in their dayside atmospheres thermally dissociate \citep{bell2017,bell2018,arcangeli2018,kreidberg2018b,lothringer2018,mansfield2018,parmentier2018} and may recombine nearer the nightside \citep{bell2018,komacek2018,parmentier2018}. UHJs also bear some similarities to cataclysmic variable star (CV) systems and may undergo significant tidal distortion and mass loss, depending on the specifics of the star-planet system \citep[e.g.,][]{bisikalo2013b,burton2014}.

While tidal distortion is expected for WASP-12b, a 2010 \textit{Spitzer} Infrared Array Camera (IRAC) thermal phase curve observation of WASP-12b at 4.5~$\mu$m demonstrated second order sinusoidal variations (with two maxima per planetary orbit) that were far greater than predicted \citep{cowan2012}. The substellar axis would have to be $1.8$ times as long as the dawn--dusk and polar axes if the observed variations were entirely due to the tidally distorted shape of the planet. Additionally, no evidence of these second order sinusoidal variations was found in a \textit{Spitzer}/IRAC 3.6~$\mu$m phase curve also taken in 2010 \citep{cowan2012}.

In this paper, we combine a new set of phase curves taken in 2013 with a reanalysis of the data from 2010 to determine the source of the unusually strong second order sinusoidal variations at 4.5~$\mu$m reported by \cite{cowan2012}. The observations are described in Section \ref{sec:data}. Our three astrophysical models are described in Section \ref{sec:astroModel}, and our three independent reduction and decorrelation methods are described in Section \ref{sec:decorrelation}. Results and their physical implications are presented in Section \ref{sec:results}, and Section \ref{sec:discussion} contains the discussion and conclusion.

\section{Observations}\label{sec:data}
We combine two sets of two-channel (3.6~$\mu$m and 4.5~$\mu$m) \textit{Spitzer}/IRAC observations taken in 2010 (PID 70060, PI Machalek) and 2013 (PID 90186, PI Todorov), all during the Post-Cryogenic \textit{Spitzer} Mission. In all four phase curves, the system was observed nearly-continuously for $\sim$33 hours (breaking only once or twice to repoint the telescope), beginning shortly before one secondary eclipse and ending shortly after the subsequent secondary eclipse. The reduced and detrended observations are shown in Figure \ref{fig:obs}.

For both data sets, the sub-array mode was used with 2~s exposures which produced data cubes of 64 images with $32\times32$ pixel ($39^{\prime\prime}\times39^{\prime\prime}$) dimensions. The 2010 observations were divided into 2 Astronomical Observation Requests (AORs) with a total of 902 data cubes (57\,728 exposures), while the 2013 observations were divided into 3 AORs with a total of 909 data cubes (58\,176 exposures). The 2010 full-phase observations were published by \citet{cowan2012}, the eclipse timings from the 2013 observations were published by \citet{patra2017}, and some derived parameters from all four phase curves were published as part of a broad comparison between different planets \citep{zhang2018}.

Past observations of WASP-12 show a nearby M-dwarf binary system WASP-12B,C 1\rlap.$^{\prime\prime}$06 away from WASP-12A \citep{bergfors2011,crossfield2012,bechter2014}. As this binary system lies too close to WASP-12A to be resolved by \textit{Spitzer}, we correct for blended light after analyzing the light curves, following past work \citep{stevenson2014b}; see Appendix A for more details.

\section{Light Curve Analysis}

\subsection{Astrophysical Models}\label{sec:astroModel}
We model the observations as
\begin{equation*}
	F_{\rm model}(t) = A(t) \times \tilde{D}(t)
\end{equation*}
where $\tilde{D}(t)$ is the normalized detector model; see Section \ref{sec:decorrelation} for details on the specific models used which consist of both parametric (2D polynomials and pixel level decorrelation) and non-parametric models (bilinear interpolated subpixel sensitivity mapping). The astrophysical model is
\begin{equation*}
	A(t) = F_*(t) + F_p(t),
\end{equation*}
where $F_*$ is the flux from the host star (assumed to be constant except during transits) and $F_p$ is the planetary flux. Transits and eclipses are modelled using \texttt{batman} \citep{kreidberg2015b}, assuming a quadratic limb-darkening model for the host star and a uniform disk for the planet. The planetary flux is modelled as
\begin{equation*}
	F_p(t) = F_{\rm day}\Phi\big(\psi(t)\big),
\end{equation*}
where $F_{\rm day}$ is the instantaneous eclipse depth at phase 0.5 (assumed to be constant over each $\sim$33 hour phase curve), $\Phi$ describes the phase variations, and the orbital phase with respect to eclipse is $\psi(t)=2\pi(t-t_e)/P$, where $t_e$ is the time of eclipse and $P$ is the planet's orbital period.

We consider three different models for the astrophysical phase variations in the lightcurve. The simplest astrophysical model we consider is a first order sinusoid
\begin{equation*}
	\Phi_{\rm 1}(\psi) = 1 + C_1\bigg(\cos(\psi)-1\bigg) + D_1\sin(\psi),
\end{equation*}
and we also consider a second order sinusoid
\begin{equation*}
	\Phi_{\rm 2}(\psi) = \Phi_{\rm 1}(\psi) + C_2\bigg(\cos(2\psi)-1\bigg) + D_2\sin(2\psi),
\end{equation*}
where $C_1$, $D_1$, $C_2$, and $D_2$ are all constants. If the previously reported double peaked phase curve \citep{cowan2012} is astrophysical in nature, one potential interpretation is that some/all of the power in the second order sinusoidal variations is from tidal distortion of the planet. We model this scenario with
\begin{equation*}
	\Phi_{\rm 1,ellipsoid}(\psi) = S(\psi)\Phi_{\rm 1}(\psi),
\end{equation*}
where $S(\psi)$ describes the projected area of an ellipsoid as it rotates. Rather than model a triaxial ellipsoid, we constrain the polar and dawn--dusk axes to share the same length since rotational deformation is expected to be negligible compared to tidal deformation \citep{leconte2011}. To find the deviations in the projected area of this biaxial ellipsoid, we adapt an equation from past work \citep{leconteCorr2011},
\begin{align*}
	S(\psi) = \bigg[&\sin^2(i)\Bigg(\bigg(\frac{R_{p,2}}{R_{p}}\bigg)^2\sin^2(\psi) + \cos^2(\psi)\Bigg)  \\
			 	    &+ \bigg(\frac{R_{p,2}}{R_{p}}\bigg)^2\cos^2(i)\bigg]^{1/2},
\end{align*}
where $i$ is the orbital inclination, $R_{p}$ is the planetary radius along the polar and dawn-dusk axes (the two axes observed during transit and eclipse if $i=90^{\circ}$), and $R_{p,2}$ is the planetary radius along the line connecting the planet and star (the sub-stellar axis).

\subsection{Decorrelation Procedures}\label{sec:decorrelation}
To ensure our results are robust and independent of the methods used, we perform three independent reductions and analyses following previously employed methods \citep{zhang2018,dang2018,cubillos2014} which are summarized below. Each analysis considers all three phase variation models. The model priors for each analysis are described below and summarized in Table \ref{tab:priors} for convenience. Within each analysis pipeline, models are selected based on the Bayesian Information Criterion (BIC). We cannot choose our fiducial models between our three analyses using the BIC as there are significant differences between the number of data used in each analysis because of different $\sigma$-clipping and binning. Instead, we choose to discriminate between the three analyses by selecting the model with the largest log-likelihood per datum, $\ln(L)/N_{\rm data}$; we therefore adopt the preferred models from M.~Zhang's analyses as our fiducial models. The fiducial reductions of the four data sets are presented in Figure \ref{fig:obs} and Table \ref{tab:fiducial} (see also the Appendix and Supplementary Information).

\subsubsection{Fiducial Reduction and Decorrelation Procedure}
For reasons described below, M.~Zhang's analyses were selected as our fiducial analyses and follow their previous work \citep{zhang2018}. In this analysis, we perform aperture photometry with a radius of 2.7 pixels on the Spitzer BCD files to get the raw flux for all frames. The background is calculated by excluding all pixels within a radius of 12 pixels from the star, rejecting outliers using sigma clipping, and then calculating the biweight location of the remaining pixels. We then bin the background-subtracted raw fluxes with a bin size of 64, discard the first 0.05 days of data, and perform fitting with \texttt{emcee} \citep{foremanMackey2013_emcee}. The fitting uses 250 walkers that walk for 20\,000 burn-in steps and 20\,000 post-burn-in steps. Our instrumental model uses first order PLD for all data except the 2010 3.6~$\mu$m data, in which case we find that second order PLD minimizes BIC. Aside from PLD, the instrumental model also includes a linear slope with respect to time. We fit for the following parameters, all with uniform priors: transit time, eclipse time, $R_p/R_*$, eclipse depth (assumed to be constant over each ${\sim33}$ hour phase curve), sinusoidal phase variation amplitudes ($C_1$ and $D_1$ for the first order sinusoid, and $C_2$ and $D_2$ if running a second order sinusoidal model), photometric error, slope in flux with time, and PLD coefficients. We fixed $P$, $a/R_*$, and $i$ to the highly precise values from the literature \citep{collins2017} as they are poorly constrained by our observations. As limb darkening is not that important in the \textit{Spitzer} bands, we adopt the closest model from a grid of 1D stellar models \citep{sing2010}.

Our fiducial analyses find that the photon noise limits are 652~ppm and 637~ppm for the 2010 and 2013 3.6~$\mu$m observations, respectively, and the limits for the 2010 and 2013 4.5~$\mu$m observations are 866 ppm and 860 ppm, respectively. The differences between these two is likely due to the star falling on parts of the detector with slightly different sensitivities, as well as varying aperture sizes. The fitted photometric standard deviation from our fiducial analyses are 950~ppm and 976~ppm for the 2010 and 2013 3.6~$\mu$m observations (1.46 and 1.53 times greater than the photon noise limit). For the 2010 and 2013 4.5~$\mu$m observations, the fitted photometric standard deviations are 1134~ppm and 1158~ppm (1.31 and 1.35 times greater than the photon noise limit). Figures showing the normalized raw, decorrelated, and residual fluxes from all four phase curves analyzed with M.~Zhang's pipeline can be found in the Appendix (Figures \ref{fig:zhang1} and \ref{fig:zhang2}).

\subsubsection{T.~Bell's Reduction and Decorrelation Procedure}
Reduction and decorrelation of these data follow \citet{dang2018} and are summarized here. We convert the pixel intensity from \mbox{MJy/str} to electron counts and mask bad pixels, i.e., $4\sigma$ outliers with respect to the median of that pixel in the datacube as well as any NaN pixels. We discard all frames with a bad pixel within the aperture used for photometry. We also discard every first frame from each data cube from the 2010 observations and every first and second frame from each data cube for the 2013 observations because these frames consistently show the presence of significant outliers compared to other frames within the same data cube. The effect of this sigma clipping is minimal, given that model fitting is performed on the median binned values from each data cube. There is another star (other than WASP-12A,B,C) that falls on the detector but lies outside the considered photometric apertures (${\sim}10^{\prime\prime}$ away); we place a $3\times3$ pixel mask around this star to ensure that it does not bias the background subtraction.

We then perform aperture photometry on each individual frame, with an aperture at the fixed pixel-location (15,15), and centroids were found using a flux-weighted mean algorithm and later used for decorrelation. Apertures ranging from 2 to 5 pixels in radius were considered as well as two different aperture edges: hard (the pixel's flux is included if the centre of the pixel lies within the aperture) and soft (each pixel is weighed by the exact fraction of its area included within the aperture). While some flux will be lost by smaller apertures, a smaller aperture better allows us to remove intra-pixel sensitivity variations, which are the dominant source of noise in our data. We select the aperture radius and edge which resulted in the lowest RMS after a copy of the raw data were smoothed by a boxcar filter of width 5 data cubes (${\sim}11$ minutes which is approximately half the ingress/egress duration) to remove features such as transits, eclipses, and phase variations. Tests run with apertures centred on the flux-weighted mean derived centroids showed that the RMS was $>100$~ppm higher than the fixed position apertures. For the 2010 data, we selected a hard-edged 2.5 pixel radius aperture for the 4.5~$\mu$m data and an soft-edged 4.3 pixel radius aperture for the 3.6~$\mu$m data; the previous analysis of these data \citep{cowan2012} used IDL's approximation on a soft-edged 2.5 pixel radius aperture for both wavelengths. For the 2013 data, we selected a hard-edged 3.2 pixel radius aperture for the 4.5~$\mu$m data and an soft-edged 2.9 pixel radius aperture for the 3.6~$\mu$m data. Before decorrelating and analyzing the data, we first bin the flux and centroid measurements from all 64 frames within a data cube using a median to reduce noise and decrease computation time. On average, each of our models take ${\sim}0.5$ hour to fit to the binned data, and computation time grows linearly with the number of data points, so running each of the different models on unbinned data is not feasible.

T.~Bell's analyses used various systematic models as implemented in the open-source Spitzer Phase Curve Analysis \citep[\texttt{SPCA};][]{dang2018} pipeline\footnote{\url{https://github.com/lisadang27/SPCA}}. In particular, we used two-dimensional polynomials of order 2 through 5 and BiLinear Interpolated Subpixel Sensitivity (BLISS) mapping. The two-dimensional polynomials \citep{charbonneau2008} assume the sensitivity of the detector can be described by an $n$th-order 2D polynomial in the measured centroid. BLISS mapping \citep{stevenson2012a,ingalls2016,schwartz2017a} is a non-parametric method to account for the intra-pixel sensitivity variations which requires accurate centroid measurements; when fitting BLISS models we adopt an 8$\times$8 grid of knots. For the 2013 observations at 3.6~$\mu$m, we also needed to add a slope in time to remove residual red noise.

Models were fit using the Markov Chain Ensemble Sampler \texttt{emcee} \citep{foremanMackey2013_emcee}. The orbital parameters of WASP-12b in the literature \citep{collins2017} have smaller errors than we can achieve with our photometry. Additionally, numerous searches for eccentricity have found that WASP-12b's orbit is best described by a circular orbit \citep{campo2011,croll2011,bailey2019}, so we set the orbital eccentricity to zero. Several orbital parameters are poorly constrained by a single phase curve observation compared to the literature values, so we adopted the following Gaussian priors to marginalize over the uncertainties in the literature values: \mbox{$t_0 = 56176.16825800\pm0.00007765$} (BMJD), \mbox{$a/R_*=3.039\pm0.034$}, \mbox{$i=83.37^{\circ}\pm0.68^{\circ}$} \citep{collins2017}. The orbital period is known to within 12~ms, so we simply fixed it at $1.09142030$~days \citep{collins2017}. The parameters that were always fitted were $t_0$, $R_p/R_*$, $a/R_*$, $i$, $F_{\rm day}/F_{*}$, two quadratic limb darkening parameters \citep{kipping2013} $q_1$ and $q_2$, and the first order sinusoidal amplitudes $C_1$ and $D_1$. In some models, we also fitted $R_{p,2}/R_*$ or $C_2$ and $D_2$. A number of detector parameters were also fitted, with the exact number depending on the detector model used.

For T.~Bell's apertures, the photon noise limits are 578~ppm and 566~ppm for the 2010 and 2013 3.6~$\mu$m observations, respectively, and the limits for the 2010 and 2013 4.5~$\mu$m observations are 795~ppm and 791~ppm, respectively. The fitted photometric standard deviation from T.~Bell's analysis are 1493~ppm and 1246~ppm for the 2010 and 2013 3.6~$\mu$m observations (2.58 and 2.20 times greater than the photon noise limit). For the 2010 and 2013 4.5~$\mu$m observations, the fitted photometric standard deviations are both 1440~ppm (1.81 and 1.82 times greater than the photon noise limit). The fitted data and red noise tests from these analyses can be found in the Supplementary Information (Figures A7 and A8).

\subsubsection{P.~Cubillos' Reduction and Decorrelation Procedure}
The models run by P.~Cubillos use the Photometry for Orbits, Eclipses, and Transits (\texttt{POET}) pipeline \citep{stevenson2010,stevenson2012a,stevenson2012b,campo2011,nymeyer2011, cubillos2013, cubillos2014}. The \texttt{POET} pipeline starts by flagging bad pixels from the Spitzer BCD files using the permanent bad pixel masks and performing a sigma-rejection routine. Next, it estimates the target center position either fitting a two-dimensional Gaussian function or calculating the least asymmetry \citep{lust2014}. Then it obtains raw light curves by applying a circular interpolated aperture photometry, testing several aperture radii between 2.0 and 4.0 pixels.

To determine the optimal centroiding method and photometry aperture, \texttt{POET} minimizes the standard deviation of the residuals, and minimizes time-correlated noise at timescales equal and larger than the transit duration (estimated through the time-averaging method). Least asymmetry centroiding outperformed Gaussian centring for all datasets, except the 2013 4.5~$\mu$m observation.  The optimal apertures were 2.5 and 3.0 pixels (2010) and 4.0 and 2.0 pixels (2013) for the 3.6 and 4.5~$\mu$m observations, respectively.  In any case, all relevant astrophysical parameters vary within their uncertainties as we vary the centroiding and photometry.

\texttt{POET} models the unbinned light curves, simultaneously fitting the astrophysical phase curve and the telescope systematics. The systematics model consists of the non-parametric BLISS intrapixel model, for which we set the map's bin size equal to the RMS of the frame-to-frame target position (0.01~pixels), and require at least 8 points per bin.  For the 2013 4.5~$\mu$m observation, we also apply a linear time-dependent ramp with the slope as a free parameter.

The astrophysical model consists of transit and eclipse models \citep{mandel2002}, combined with the sinusoidal and ellipsoid models described in the Methods section. The transit free fitting parameters are the epoch, ratio between the planetary and stellar radii, cosine of inclination, semi-major axis to stellar radius ratio, stellar flux, and quadratic limb-darkening coefficients.  The eclipse free fitting parameters are the midpoint, duration, depth, and ingress duration (setting the egress duration equal to the ingress duration).  We adopt uniform priors for all parameters, except for $\cos(i)$ and $a/R_{\rm *}$, which have Gaussian priors, and kept the orbital period fixed \citep[same values as in T.~Bell's reduction and decorrelation procedure;][]{collins2017}.

\texttt{POET} incorporates the MC3 statistical package \citep{cubillos2017} to find the best-fitting parameter values (using Levenberg-Marquardt optimization) and uncertainties \citep[using a differential-evolution Markov Chain Monte Carlo algorithm;][]{braak2008}, requiring the Gelman-Rubin statistic \citep{gelman1992} to be within 1\% of unity for each free parameter for convergence. \texttt{POET} uses Bayesian hypothesis testing to select the model best supported by the data, selecting the lowest BIC model. The \texttt{POET} results support the independent results of the other pipelines. Both 4.5 $\mu$m observations strongly favour the second order sinusoidal model, while both 3.6 $\mu$m observations strongly favour the first order sinusoidal model.

For P.~Cubillos' apertures, the photon noise limits are 4886~ppm and 6664~ppm for the 2010 and 2013 3.6~$\mu$m observations, respectively, and the limits for the 2010 and 2013 4.5~$\mu$m observations are 8273~ppm and 7988~ppm, respectively. The fitted photometric standard deviation from P.~Cubillos' analysis are 6915~ppm and 7360~ppm for the 2010 and 2013 3.6~$\mu$m observations (1.41 and 1.10 times greater than the photon noise limit). For the 2010 and 2013 4.5~$\mu$m observations, the fitted photometric standard deviations are 9130~ppm and 8658~ppm (1.10 and 1.08 times greater than the photon noise limit). The fitted data and red noise tests from these analyses can be found in the Supplementary Information (Figures A9 and A10).

\begin{figure*}
	\centering
	\includegraphics[width=\linewidth]{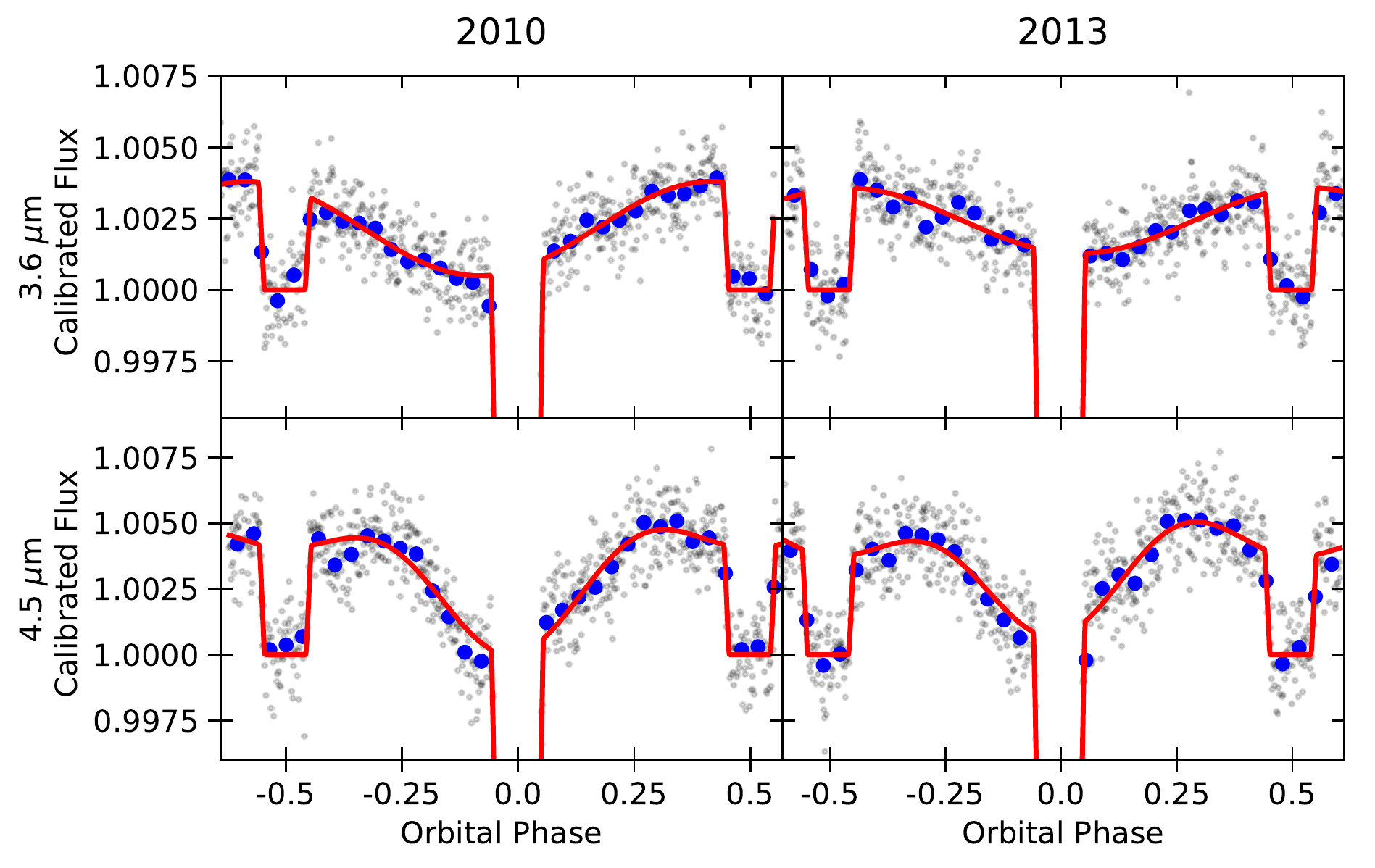}
	\caption{Fiducial analyses of 3.6~$\mu$m (top) and 4.5~$\mu$m (bottom) \textit{Spitzer}/IRAC phase curve observations of WASP-12b taken in 2010 (left) and 2013 (right). Both 3.6~$\mu$m phase curves show one maximum per planetary orbit, while both 4.5~$\mu$m phase curves exhibit two maxima per planetary orbit. The detector systematics have been removed from the data, and our fiducial astrophysical models for each data set are overplotted in red. Grey data points show binned values from each \textit{Spitzer} data cube (64 frames), and the blue points show more coarsely binned values (1664 frames).}\label{fig:obs}
\end{figure*}

\begin{table*}
\begin{center}
\caption{Key fiducial light curve parameters}\label{tab:fiducial}
\begin{tabular}{cccccc}
\hline %
                &                         & 1st Order                & 2nd Order                &                                 \\ 
                &                         & Phase Offset$^{\dagger}$ & Phase Offset$^{\dagger}$ & $F_{\rm day}/F_*$ $^{\ddagger}$ \\ 
Data Set        & $R_p/R_*$ $^{\ddagger}$ & (degrees)                & (degrees)                & (ppm)                           \\ \hline 
2010, 3.6~$\mu$m & $0.11642\pm0.00063$     & $-32.6\pm6.2$            & ---                      & $3870\pm130$                    \\
2013, 3.6~$\mu$m & $0.11327\pm0.00068$     & $13.6\pm3.8$             & ---                      & $3840\pm120$                    \\ \hline
2010, 4.5~$\mu$m & $0.10656\pm0.00085$     & $-9.5\pm2.3$             & $94.7\pm1.6$             & $4360\pm140$                    \\
2013, 4.5~$\mu$m & $0.1049\pm0.0010$       & $-19.1\pm3.9$            & $93.2\pm1.9$             & $3920\pm150$                    \\ \hline 
\end{tabular}
\end{center}
{$^{\dagger}$ These phase offsets are measured in degrees \textit{after} eclipse and are derived quantities.\\}
{$^{\ddagger}$ These quantities have been corrected for dilution from WASP-12BC (see Appendix A).}
\end{table*}

\section{Results}\label{sec:results}

\subsection{Comparison Between Pipelines and Epochs}
All three independent analyses confirm the presence of strong and persistent second order sinusoidal variations at 4.5~$\mu$m and the non-detection of these variations at 3.6~$\mu$m. The fitted phase curves parameters for the preferred models from all three independent pipelines are summarized in Figure \ref{fig:obs} and Table \ref{tab:s1}. See the Supplementary Information for tabulated values for all considered models. The astrophysical parameters at 3.6~$\mu$m and 4.5~$\mu$m are mostly consistent between all three analyses, with the preferred models from the three analyses generally differing by $<2\sigma$. In the few cases where one model differs from the others by more than $2\sigma$, the other two models are consistent with each other at a level of $<1\sigma$. Also, there is low-frequency noise in the 2010 4.5~$\mu$m residuals between the first eclipse and the end of the transit that is seen by all three analysis pipelines; the source of these variations is not understood.

From 2010 to 2013, the three pipelines show that all 4.5~$\mu$m phase curve and planetary parameters remain constant within $<2\sigma$. Most of the phase curve and planetary parameters at 3.6~$\mu$m also remain constant between the two observing epochs, with the main exception being the phase offset calculated using the first order sinusoidal terms. M.~Zhang's, P.~Cubillos', and T.~Bell's pipelines find that it changes by $6.4\sigma$ ($46.2^{\circ}$), $7.7\sigma$ ($46.6^{\circ}$), and $3.1\sigma$ ($28.1^{\circ}$), respectively. All three pipelines also agree that the sign of the hotspot offset changes between the two observing epochs, with the offset being ``eastward'' (before eclipse) in 2010 and ``westward'' (after eclipse) in 2013. It is interesting to note, however, that over this same time span both the first and second order sinusoidal phase offsets from the 4.5~$\mu$m observations do not change. No other parameter is found by all three analyses to vary by more than $3\sigma$ between the two observing epochs. Finally, if the first order sinusoidal phase variations are entirely attributable to WASP-12b's temperature map, our 2013 observations at 3.6~$\mu$m exhibit a $13\rlap.^{\circ}6\pm3\rlap.^{\circ}8$ westward hotspot offset. This may be a demonstration that eastward hotspot offsets are less ubiquitous than previously believed, with westward hotspot offsets reported for planets with irradiation temperatures spanning 2200--3700~K \citep{dang2018,zhang2018,wong2016}.

\subsection{Physical Sources}
The previously favoured explanation for the double peaked phase curve reported for WASP-12b by \citet{cowan2012} was detector systematics, but this hypothesis is now strongly disfavoured. To date, 23 papers have been published with new \textit{Spitzer} phase curves of 18 different exoplanets \citep{harrington2006, knutson2007, cowan2007, knutson2009hd149026b, knutson2009hd189733b, laughlin2009, crossfield2010upsilonAndb, cowan2012, knutson2012, crossfield2012hd209MIPS, lewis2013, maxted2013, zellem2014, wong2015, deWit2016, wong2016, krick2016,  demory2016, wong2016, stevenson2017, deWit2017, zhang2018, dang2018, kreidberg2018b}. Of these numerous observations, WASP-12 is the only system which has shown strong a double peaked phase curve not once, but twice. The observing strategy also differed between these two sets of WASP-12b phase curves, with the number and timing of AORs changing and the addition of PCRS Peak-Up before the 2013 observations. The consistency between the two sets of phase curves suggests that the observations probe an astrophysical source which does not vary significantly over a $\sim$3 year timescale. \citet{cowan2012} suggested that tidal distortion and/or mass loss might be able to explain the \textit{Spitzer} observations, if this signal was indeed astrophysical in nature. We explore these and other potential sources of emission below.

\begin{figure*}
	\centering
	\includegraphics[width=0.35\linewidth]{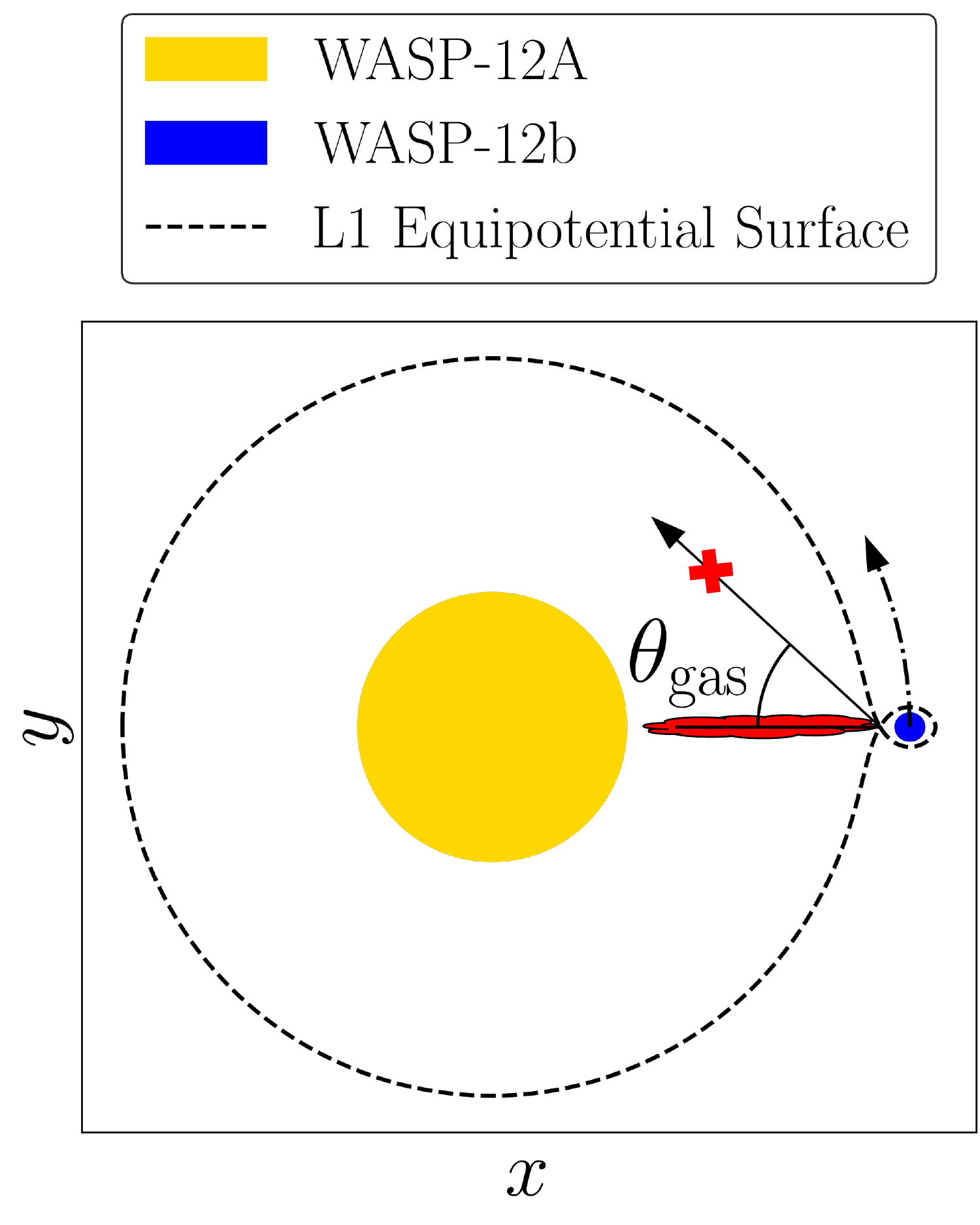}\hspace{2cm}%
	\includegraphics[width=0.4\linewidth]{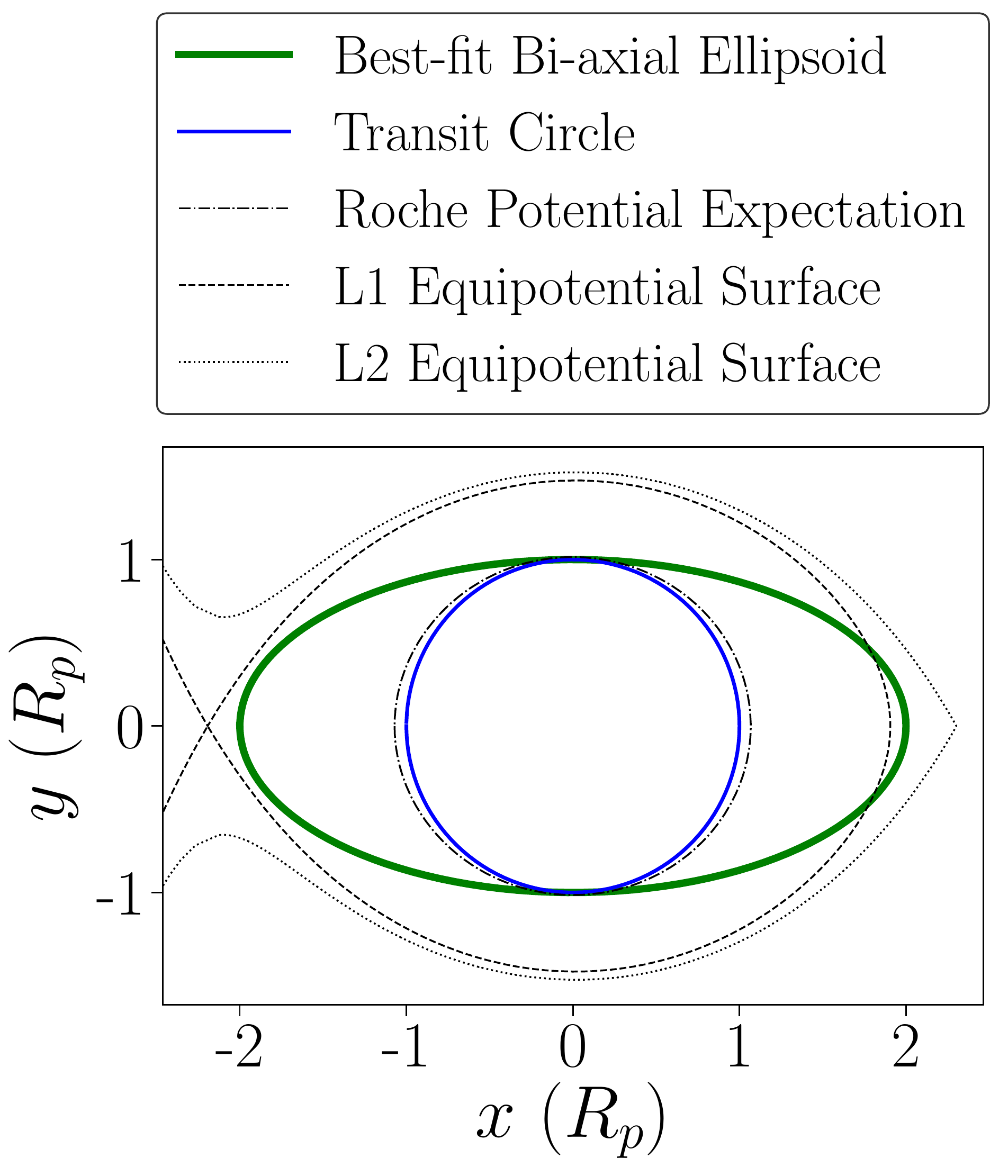}
	\caption{Bird's-eye views of the WASP-12 system to scale. \textit{Left:} While the planet is appreciably filling its Roche lobe and is expected to be tidally distorted, the star should not. Over-plotted is a depiction of the direction that gas would flow after passing through the L1 Lagrange point ($\theta_{\rm gas}$) previously predicted to be 53.4$^{\circ}$ \citep{lai2010}. If our observations probe the gas stream, we can firmly reject this ballistic trajectory hypotesis as the gas appears to be aligned along the star--planet axis (indicated by the red elongated patch of gas). The direction of the planet's orbit is shown with a dash-dotted arrow. 
	\textit{Right:} Best-fit bi-axial ellipsoid model fit to the 2013 phase curve observation at 4.5~$\mu$m, placed in the context of the planet's Roche lobe. This shape varies drastically from that of a Roche lobe and instead suggests that our observations are probing something other than the planet's tidally distorted shape.
	Also shown is a circle with the area seen at transit, the equipotential surface which would give that transit area, and the L1 and L2 equipotential surfaces. The $x$ and $y$ axes lie within the orbital plane; during transit the $x$-axis is parallel to our line of sight.}\label{fig:overview}
\end{figure*}

\subsubsection{Tidal Distortion}
One potential cause of second order sinusoidal variations is tidal deformation of the host star, as is seen at optical wavelengths for HAT-P-7 \citep{welsh2010} and WASP-18 \citep{shporer2019}. However, stellar distortion is expected to be negligible for WASP-12 \citep{leconte2011}. We verified this by numerically solving for the equipotential stellar/planetary surfaces using the dimensionless Roche potential (see Appendix C). However, since the star contributes significantly more flux than the planet, we ran simple simulations of both the star and planet including the effects of gravity darkening to assess their expected amplitudes of ellipsoidal variations. We find that the stellar ellipsoidal variations are approximately the same amplitude at 3.6~$\mu$m and 4.5~$\mu$m and the amplitude of the stellar variations are far smaller than the observed amplitudes; we therefore conclude that tidal bulges on the host star cannot be the source of the strong second order variations observed at 4.5~$\mu$m. Our predicted ellipsoidal variations for WASP-12b are consistent with our limits on second order sinusoidal variations at 3.6~$\mu$m, but significantly under-predict the observed amplitude at 4.5~$\mu$m (also see the implied dimensions of the best-fit ellipsoidal variation model shown in the right panel of Figure \ref{fig:overview}). If we interpret the second order sinusoidal variations at 4.5~$\mu$m as planetary ellipsoidal variations, this would require the 4.5~$\mu$m photosphere to be significantly higher up than 3.6~$\mu$m as the layers nearer the Roche lobe are more distorted. However, this increased radius at 4.5~$\mu$m is inconsistent with the smaller transit depth at 4.5~$\mu$m compared to 3.6~$\mu$m. We therefore conclude that tidal distortion of the planet is also not the source of the strong second order variations observed at 4.5~$\mu$m.

\subsubsection{Stellar Variability and Inhomogeneities}
Stellar variability is also unlikely to be the cause of these observations given the comparable phase of the second order variations in the two data sets. For reference, the WASP-12BC dilution correction term is ${\sim}400$~ppm while the observed amplitude of the second order sinusoidal variations is ${\sim}2000$~ppm. Additionally, variability in WASP-12A is only predicted to modulate the planetary signal at a level of $\sim$1~ppm, and variability in WASP-12B,C should also only contribute at the level of ${\sim}1$~ppm \citep[while these M-dwarfs should be ${\sim}10\times$ more variable, they contribute ${\sim}10\times$ less flux;][]{zellem2017}. We therefore rule out standard stellar variability as the source of the strong second order sinusoidal variations seen at 4.5~$\mu$m. If the second order variations were produced by unusually strong inhomogeneities on the host star, both the sub-planet longitude and the anti-planet longitude would need to be darker than intermediate longitudes --- this would imply star--planet interactions. However, these inhomogeneities would also need to be much more pronounced at 4.5~$\mu$m which would not be expected for the ${\sim}6000$~K star.

\subsubsection{Mass Loss}
There is significant observational evidence from near ultra-violet (NUV) transit observations that WASP-12b is undergoing mass loss and that there is a bow shock in the system \citep{fossati2010b,haswell2012,nichols2015} (see the Supplementary Information). A potential explanation for the unusual 4.5~$\mu$m phase curve is that there is gas being stripped from the planet which emits more strongly within the 4.5~$\mu$m bandpass than the 3.6~$\mu$m bandpass. The observations favour a stream of dense gas stripped from the planet flowing directly toward/away from the host star or some other elongated patch of hot gas whose long axis is parallel to the star--planet axis, such as an accretion hot spot. Double-peaked phase curves have been seen for dwarf novae CVs, such as WZ Sge \citep{skidmore1997}, however this feature was seen through the ultra-violet to infrared; for CVs these variations have been attributed to tidal distortion or an optically thick hot spot in an otherwise optically thin accretion disk \citep[e.g.,][]{skidmore1997}

The source of the 4.5~$\mu$m variations in the WASP-12 system must lie near the star--planet axis since there is no significant detection of an occultation of the source when the planet is not in transit or eclipse. Additionally, our \textit{Spitzer} observations demonstrate that the planetary radius appears ${\sim}8$\% ($11\sigma$) smaller at 4.5~$\mu$m than at 3.6~$\mu$m which is in disagreement with model predictions \citep{burrows2007,burrows2008,cowan2012}; this rules out the transit of a large exosphere that is opaque at 4.5~$\mu$m as this would make the planetary radii at the two wavelengths even more discrepant.

As shown in Table \ref{tab:fiducial}, the fitted second order offsets at 4.5~$\mu$m are consistent with being oriented along the star--planet axis ($90^{\circ}$). However, previously published 3D magnetohydrodynamic (MHD) numerical simulations of hypothetical exoplanet systems mostly produced gas flows that significantly lead the star--planet axis \citep{matsakos2015}. Indeed, gas streaming from the planet's L1 Lagrange point on a ballistic trajectory should flow $\theta_{\rm gas}=53$\rlap{$^{\circ}$}$.\,4$ \textit{ahead} of the star--planet axis \citep{lai2010} as angular momentum is conserved (see Figure \ref{fig:overview} for a schematic depiction). Assuming our observations probe the gas stream, this prediction is $27\sigma$ discrepant with our offset of $4$\rlap{$^{\circ}$}.\,$0\pm2$\rlap{$^{\circ}$}$.\,1$ \textit{behind} the star--planet axis found by averaging the offsets from the two fitted second order sinusoids at 4.5~$\mu$m. This discrepancy could potentially be explained if the 4.5~$\mu$m emitting area is much closer to the planet and is still aligned along the star--planet axis and then becomes more diffuse and flows ahead of the planet as it continues to fall toward the host star.

Alternatively, stellar effects could channel the infalling stream directly toward the star, but this may be inconsistent with past NUV transit observations \citep{fossati2010b,haswell2012,nichols2015}. One previously published 3D MHD model \citep{matsakos2015} did exhibit a stream of gas directly along the star--planet axis (their name for this model was `FvrB'). This model has high stellar ultra-violet (UV) flux, a low escape speed from the planet, the planet near to its host star, and a strong planetary magnetic field. In this model, the planet is experiencing Roche lobe overflow with a planetary wind that is weak compared to the stellar wind, producing an approximately linear stream of gas along the star--planet axis as well as a lower density tail trailing behind the planet \citep{matsakos2015}. The non-detection of the gas trailing behind the planet could be explained if the gas has a lower density and/or has a lower temperature. As the dense gas stream in the `FvrB' model is aligned along the star--planet axis, it may not contribute significantly to the transit depth and may remain consistent with the smaller apparent radius at 4.5~$\mu$m compared to 3.6~$\mu$m. Radiative transfer simulations based on the `FvrB' mass-loss model \citep{matsakos2015} would allow for this hypothesis to be tested.

\begin{figure*}
	\centering
	\includegraphics[width=0.75\linewidth]{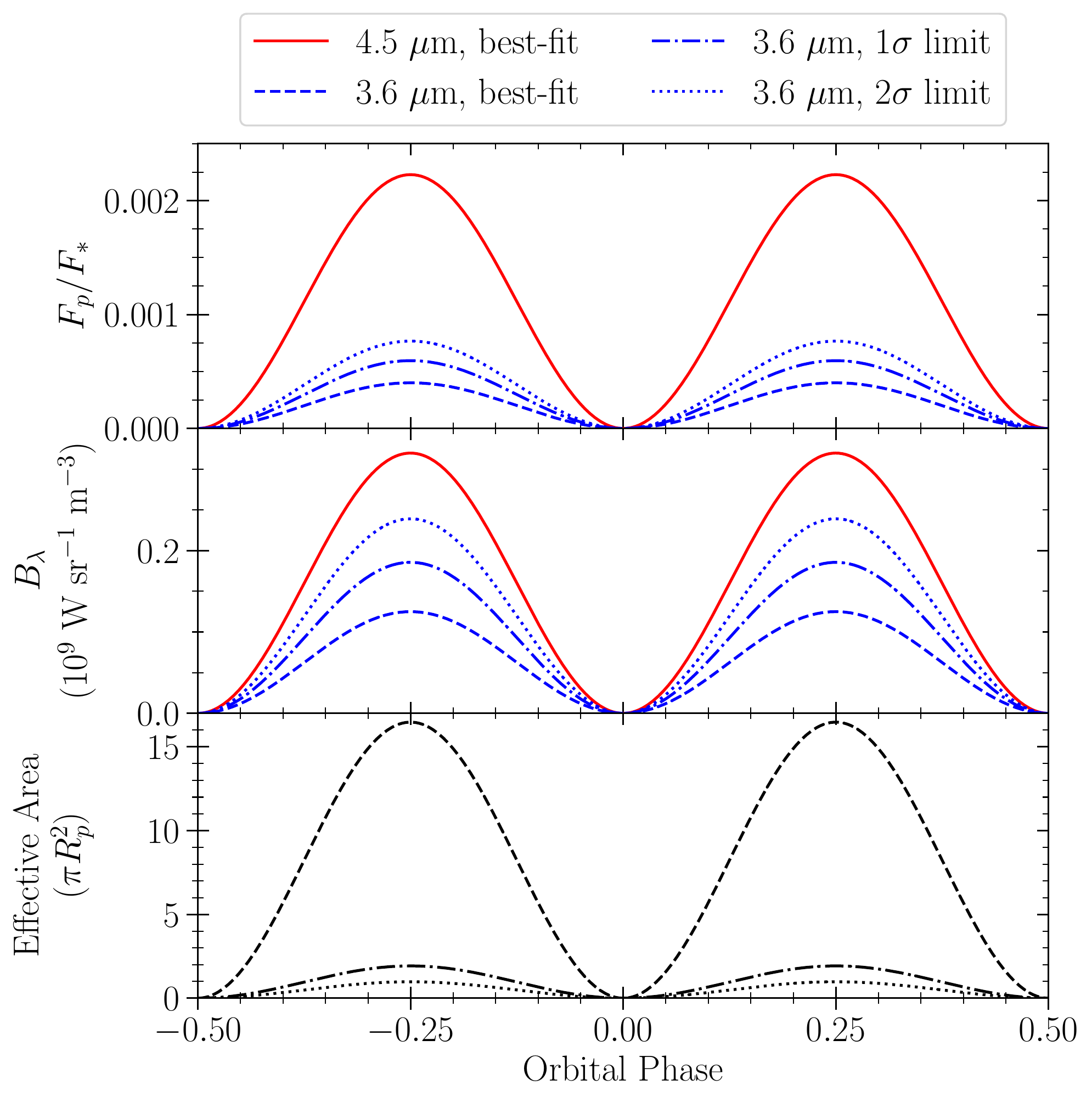}
	\caption{Limits on the emitting area required to explain the strong detection of second order sinusoidal variations at 4.5~$\mu$m but not at 3.6~$\mu$m. \textit{Top}: the emitting-body to star flux ratio for the second-order sinusoidal component of the 4.5~$\mu$m and 3.6~$\mu$m data (temporarily assuming a constant radius of $R_p$). \textit{Middle}: the emitting-body's blackbody flux assuming both wavelengths probe the same area. \textit{Bottom}: the effective emitting area of the emitting blackbody required to explain the observations. The inferred gas temperatures for the $0\sigma$, $1\sigma$, and $2\sigma$ limits are $420~K$, $549$~K, and $619$~K, respectively.}\label{fig:blackbodies}
\end{figure*}

\subsection{Radiation Mechanisms}

\subsubsection{Blackbody Emission}

The discrepant second order sinusoidal amplitudes at 3.6~$\mu$m and 4.5~$\mu$m can be explained by one of two emission mechanisms. First, blackbody emission could allow for greater flux at 4.5~$\mu$m than at 3.6~$\mu$m if the gas is sufficiently cool that the 3.6~$\mu$m bandpass lies on the Wien side of the blackbody curve; this scenario would allow us to place an upper limit on the temperature and spatial extent of the emitting gas, which we pursue below.

Using the host star's effective temperature of $6300\pm150$~K \citep{hebb2009}, we assume the host star emits as a blackbody and convert the second order sinusoidal curves from units of $F_{\rm day}/F_*$ to $B_{\lambda}$ as shown in the middle panel of Figure \ref{fig:blackbodies}. We adopt the fiducial 4.5~$\mu$m parameters from 2010, but set the phase offset to 90$^{\circ}$ since there is no evidence for an appreciable offset from the star--planet axis. We then take the best-fit and the $1\sigma$ and $2\sigma$ upper limits on the amplitude of the 3.6~$\mu$m second order sinusoidal variations from M.~Zhang's analysis using the second order astrophysical model. We assume that none of the flux seen during planetary transit/eclipse is from emission by the gas. By assuming the emitting area is the same at 3.6~$\mu$m and 4.5~$\mu$m, we can use the relative amounts of flux at these two wavelengths to determine the blackbody temperature of the gas.

Given the assumption that our observations are explained by blackbody emission, we can then place a $2\sigma$ upper limit of $619$~K on the gas temperature. For reference, a temperature of $816$~K would provide equal flux in both bandpasses. Attributing any of the ``nightside'' flux to emission from the gas only lowers this limit further. Also, as WASP-12b's skin temperature \citep{goody1972} is $0.5^{0.25}\,T_{\rm b, day} \approx 2500$~K, this gas cannot be the upper layers of the planet's atmosphere.

By taking the ratio between the flux emitted by the gas and that emitted by the star, we can determine the effective emitting area required to produce the phase curve observations. As shown in the bottom panel of Figure \ref{fig:blackbodies}, less emission at 3.6~$\mu$m requires lower temperature gas and therefore a larger emitting area. We can therefore place a $2\sigma$ lower-limit on the effective emitting area of the gas of $0.98$ times the planet's transiting area when seen at planetary quadrature, given the assumption that our observations are explained by blackbody emission. Attributing any of the nightside flux to emission from the gas slightly increases this limit and allows for a non-zero emitting area during planetary transit and eclipse.

\subsubsection{CO Emission}

An alternative explanation for the increased flux at 4.5~$\mu$m is emission by CO which has its strong $\Delta V=1$ band around 4.5~$\mu$m (see Figure \ref{fig:coIntensities} in the Appendix for the CO line intensities); CO emission has previously been predicted for gas lost from WASP-12b \citep{li2010,deming2011}. The CO molecule should be dissociated in the planetary upper atmosphere due to the strong UV and X-ray flux from the host star which also drives most of the observed atmospheric escape seen at NUV wavelengths \citep{fossati2010b,haswell2012,nichols2015}; the dissociation energy of CO corresponds to a wavelength of roughly 110~nm. However, the atomic carbon and oxygen from the upper layers of the planet's atmosphere could recombine in a gas stream where the density gets higher because of stellar wind confinement and the ``shadow effect'' from the material in the stream closer to the star. Given a gas temperature profile \citep{salz2016} and our calculations of the thermal dissociation fraction of CO using the Saha equation \citep{bell2018}, we find that any CO emission must either be produced within ${\sim}0.1~R_p$ of the planet's surface or beyond $2.5~R_p$. In the case of a bow shock supported by mass loss from the planet, gas temperatures are predicted to reach $10^3$--$10^4$~K \citep{turner2016} which should allow for stable CO, provided there is sufficient UV shielding from gas nearer to the star. Simulations of the behaviour of CO in these environments are required to determine the feasibility of this molecule recombining once in a stream and emitting sufficiently strongly to explain our observations.

\subsection{A Note on Eclipse Depths}
\begin{figure*}
	\centering
	\includegraphics[width=0.48\linewidth]{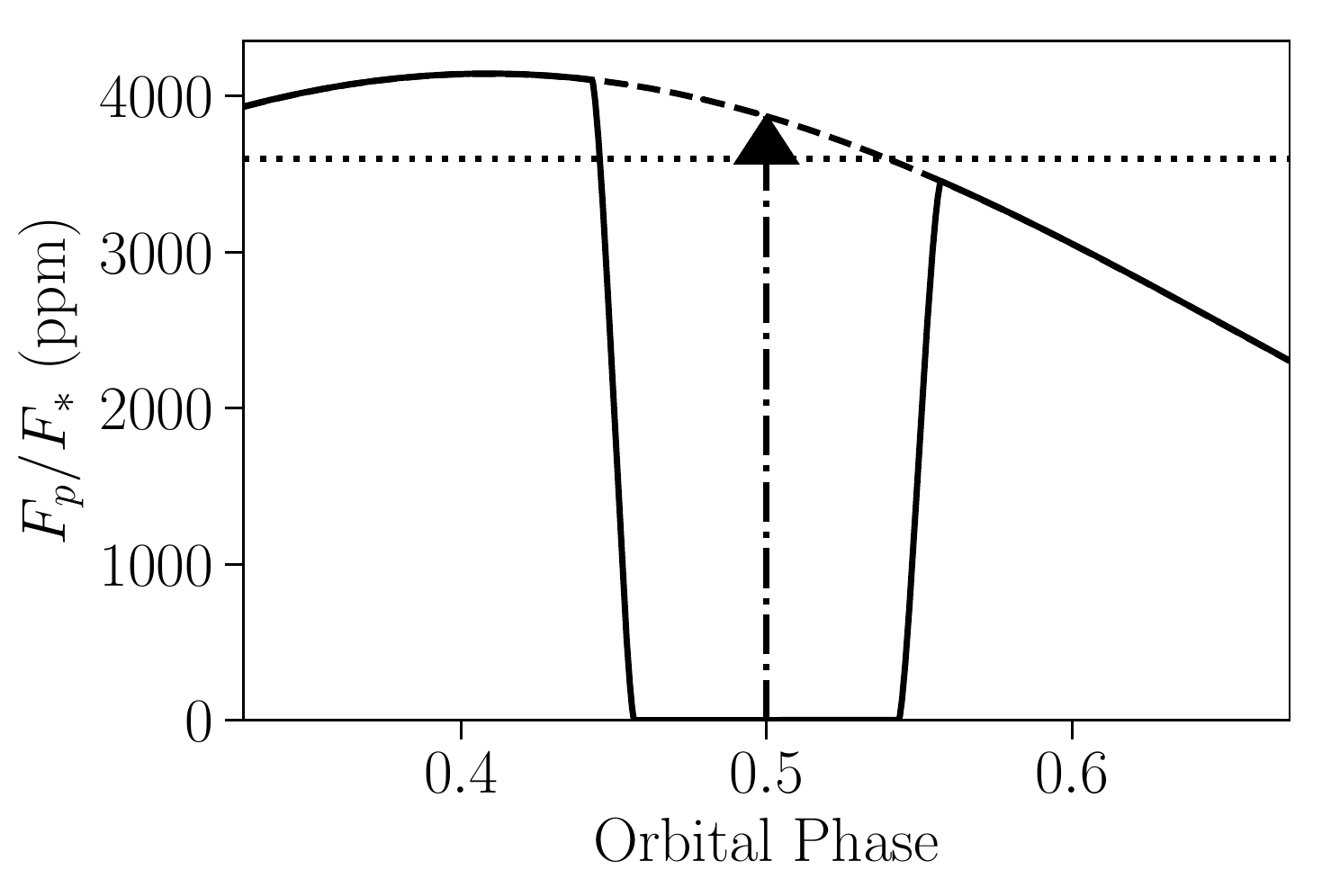}\hfill%
	\includegraphics[width=0.48\linewidth]{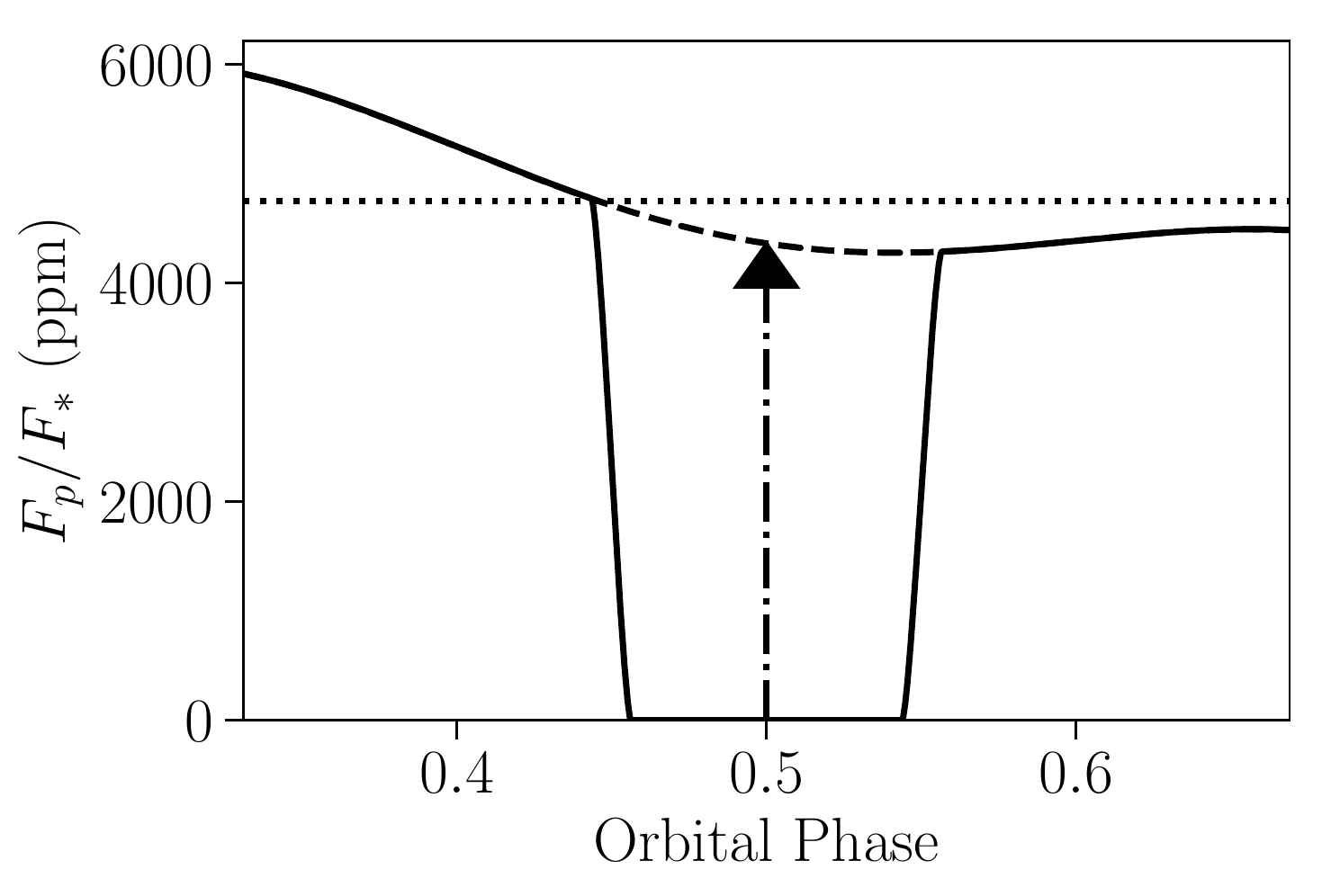}
	\caption{Bias present in eclipse-only observations of exoplanets. \textit{Left}: Our fiducial model for the 2010 phase curve at 3.6~$\mu$m is shown with a solid line, while the model neglecting the secondary eclipse is shown with a dashed line. It is with respect to this line that we calculate our eclipse depth (shown with a dash-dotted arrow), while the eclipse depth that would be measured using eclipse-only observations is shown with a dotted line. This bias occurs because there is insufficient evidence of phase variations with eclipse-only observations, so a flat or a sloped line is used instead. \textit{Right}: The same bias at 4.5~$\mu$m but in the opposite direction due to the abnormal concave-up phase variations near the 4.5~$\mu$m eclipse.}\label{fig:biases}
\end{figure*}

It is important to note that our reported ``eclipse depths'' ($F_{\rm day}/F_*$) are measured with respect to the phase curve value expected at the centre of eclipse and are not measured with respect to pre-ingress and post-egress flux measurements as would be the case for observations of only the eclipse. Given our fitted phase curve parameters for WASP-12b, the difference between our reported value and using the average of pre-ingress and post-egress baselines is ${\sim}9$\% of $F_{\rm day}/F_*$ at both \textit{Spitzer} bandpasses (assuming these baseline durations are both the same duration as the eclipse duration). This bias in eclipse observations occurs because the phase variations before ingress and after egress are flattened out by most decorrelation routines when solely observing the eclipse. For most exoplanets whose phase variations will be concave down around eclipse (like WASP-12b when seen at 3.6~$\mu$m), eclipse observations will underestimate $F_{\rm day}/F_*$. For the unusual case of WASP-12b's 4.5~$\mu$m phase variations which are concave up near eclipse, eclipse observations will overestimate $F_{\rm day}/F_*$. This effect is particularly important for short period planets which undergo significant rotation throughout the duration of eclipse observations and whose strong day--night temperature contrast cause strong phase variations over this time span. Among other things, this may explain the discrepancies between reported 3.6~$\mu$m and 4.5~$\mu$m eclipse depths from full-orbit phase curves \citep{cowan2012} and eclipse-only observations \citep{madhusudhan2011,stevenson2014a}, and the associated inference of C/O ratio. See Figure \ref{fig:biases} for a demonstration of this effect.

\section{Discussion and Conclusions}\label{sec:discussion}

By independently analyzing and then combining two sets of 3.6~$\mu$m and 4.5~$\mu$m \textit{Spitzer} phase curves of the UHJ WASP-12b, we have conclusively detected strong and persistent second order sinusoidal variations at 4.5~$\mu$m and placed stringent upper limits on these variations at 3.6~$\mu$m. These observations of WASP-12b raise several questions which will require further study to resolve.

Our two emission hypotheses could be distinguished with phase curve observations of the $\sim$1.6~$\mu$m and/or 2.29~$\mu$m CO emission bands and/or with phase curve observations at wavelengths longer than 5~$\mu$m which should exhibit strong second order sinusoidal variations if the 3.6~$\mu$m vs.~4.5~$\mu$m amplitude discrepancy is the result of blackbody emission. The high precision and wavelength coverage achievable with the James Webb Space Telescope should allow these two emission hypotheses to be tested. The $\sim$1.6~$\mu$m CO emission band also lies within the Hubble/WFC3 bandpass and may be detectable with phase curve observations.

Critically, future models must also address the fact that the fitted planetary radius is significantly smaller at 4.5~$\mu$m than at 3.6~$\mu$m; this may be the result of unocculted emitting gas. Combined hydrodynamic and radiative transfer simulations are required to fully understand this system. These simulations will allow us to determine the location and spatial extent of the emitting gas, and they may resolve the apparent tension between the constraint from these observations that the gas is well aligned with the star--planet axis, while NUV observations which probe lower density gas show that the gas flows significantly ahead of the planet. Understanding the nature of the increased emission at 4.5~$\mu$m will also require modelling the mass loss and the UV dissociation and potential recombination of CO molecules as they flow from the planet's upper atmosphere through a gas stream and potentially experience a shock. These models may also assist in understanding the observed hot spot variability seen at 3.6~$\mu$m.

Finally, while WASP-12b is one of the exoplanets closest to overflowing it's Roche lobe (see Figure \ref{fig:rocheLimits}), there are several other UHJs with similar characteristics with published Spitzer phase curves that do not show strong second order sinusoidal variations at 4.5~$\mu$m: particularly WASP-19b \citep{wong2016}, WASP-33b \citep{zhang2018}, and WASP-103b \citep{kreidberg2018b}. One potential explanation is that WASP-12b's orbit may be decaying \citep{maciejewski2016,patra2017} while the other exoplanets may be more stable; this could potentially be explained if WASP-12b was locked in a high obliquity state due to a resonance with a perturbing planet which could drive orbital decay and inflate the planet beyond it's Roche lobe \citep{millholland2018}. Alternatively, the high energy irradiation from WASP-12A may be stronger than the other UHJ host stars. Further research is required to understand why WASP-12b is the only exoplanet known to be exhibiting these exceptionally strong second order sinusoidal variations at 4.5~$\mu$m.

\section*{Acknowledgements}

T.J.B.~acknowledges support from the McGill Space Institute Graduate Fellowship, the Natural Sciences and Engineering Research Council of Canada's Postgraduate Scholarships-Doctoral Fellowship, and from the Fonds de recherche du Qu\'ebec -- Nature et technologies through the Centre de recherche en astrophysique du Qu\'ebec. The research leading to these results has received funding from the European Research Council (ERC) under the European Union's Horizon 2020 research and innovation programme (grant agreement no. 679633; Exo-Atmos). We have also made use of open-source software provided by the Python, Astropy, SciPy, and Matplotlib communities.




\bibliographystyle{mnras}
\bibliography{mega_science} 




\appendix

\renewcommand{\thefigure}{A\arabic{figure}}
\renewcommand{\thetable}{A\arabic{table}}
\setcounter{figure}{0}
\setcounter{table}{0}

\subsection*{Appendix A: Correction for Dilution by Stellar Companions}
To correct for the dilution of our lightcurves by the nearby stellar companions WASP-12BC, we apply the dilution factors from \citet{stevenson2014b}: \mbox{$\alpha_{\rm comp} = 0.1149 \pm 0.0039$} and \mbox{$0.1196 \pm 0.0042$} for 3.6~$\mu$m and 4.5~$\mu$m respectively. Since our phase curve amplitudes are normalized by the eclipse depth, no corrections need to be made to $C_1$, $D_1$, $C_2$, or $D_2$. Additionally, while the planetary radii need to be corrected for dilution from WASP-12BC, the ratio $R_{p,2}/R_p$ remains the same for models with ellipsoidal variations. Following \citet{stevenson2014a,stevenson2014b}, the multiplicative correction factor is
$$
	C_{\rm corr}(\lambda) = 1+g(\beta,\lambda)\alpha_{\rm comp}(\lambda),
$$
where $g(\beta,\lambda)$ is the fraction of WASP-12BC's flux which falls within our aperture of size $\beta$. We estimated $g(\beta,\lambda)$ using \texttt{STINYTIM}\footnote{\url{http://irsa.ipac.caltech.edu/data/SPITZER/docs/dataanalysistools/tools/contributed/general/stinytim/}}, the point response function modelling software for Spitzer. We made $10\times$ oversampled point response functions calculated at the pixel position (25,25) assuming a \mbox{$T = 3660$} blackbody source \citep[the effective temperature of WASP-12BC;][]{stevenson2014a}. We found \mbox{$g(2.5,4.5~\mu$m)} = 0.8147, \mbox{$g(3.2,4.5~\mu$m)} = 0.8608, \mbox{$g(4.3,3.6~\mu$m)} = 0.9089, and \mbox{$g(2.9,3.6~\mu$m)} = 0.8580. For the $3\times3$ pixel stamp used in M.~Zhang's PLD analyses, we find \mbox{$g(3{\times}3, 3.6~\mu$m)} = 0.6518 and \mbox{$g(3{\times}3, 4.5~\mu$m)} = 0.6291. For P.~Cubillos' analyses, we find \mbox{$g(3.0,4.5~\mu$m)} = 0.8533, \mbox{$g(2.5,4.5~\mu$m)} = 0.6957, \mbox{$g(2.5,3.6~\mu$m)} = 0.8254, and \mbox{$g(4.0,3.6~\mu$m)} = 0.9015. We also checked $g(2.25, 3.6~\mu$m$)$ and $g(2.25, 4.5~\mu$m$)$ to compare our calculation to that of \citet{stevenson2014a}; we find values of 0.8007 and 0.7586, where \citet{stevenson2014a} found 0.7116 and 0.6931. This discrepancy is likely caused by an incorrect angular separation used in the previous work's calculation.

The planet's radius was then corrected using
$$
	R_{p, {\rm corr}}(\lambda) = \sqrt{C_{\rm corr}(\lambda)} \, R_{p, {\rm meas}}(\lambda),
$$
with the elongated axis, $R_{p,2}$, in bi-axial ellipsoid models corrected similarly. The dayside flux was corrected using
$$
	F_{\rm day,corr}(\lambda) = C_{\rm corr}(\lambda) \, F_{\rm day,meas}(\lambda),
$$
with the white noise amplitude, $\sigma_{F}$, corrected similarly.

\subsection*{Appendix B: Computing Astrophysical Parameters}
Tables A2--A9 present many of the fitted astrophysical values from all models run in all three independent analyses. $T_{b,\rm day}$ and $T_{b,\rm night}$ are the apparent brightness temperatures of the planet's day and night hemispheres which we calculate using \textit{only the contribution from the first order sinusoid}. In doing so, we are assuming that the second order sinusoidal variations are attributable to something other than the planet, although the second-order sinusoidal variations end up having negligible contributions during transit and eclipse anyway. These brightness temperatures are calculated by inverting the Planck function \citep{cowan2011b}, using
\begin{equation*}
	T_b(\lambda) = \frac{hc}{\lambda k_B}\bigg[\ln\bigg(1+\frac{\exp(hc/\lambda k_B T_{*,\rm b})-1}{\psi(\lambda)}\bigg)\bigg]^{-1},
\end{equation*}
where $h$ is Planck's constant, $c$ is the speed of light, $k_B$ is the Boltzmann constant, $\lambda$ is the wavelength. For $T_{b,\rm day}$, \mbox{$\psi=(F_{\rm day}/F_*)(R_p/R_*)^{-2}$}, and for $T_{b,\rm night}$, \mbox{$\psi=(F_{\rm day}/F_*)(1-2\,C_1)(R_p/R_*)^{-2}$}. The stellar brightness temperature, $T_{*,\rm b}$ was calculated by fitting blackbodies to the relevant wavelengths from a \texttt{PHOENIX} stellar model \citep{husser2013} with previously measured \citep{hebb2009} values of $T_{\rm *, eff}=6300$~K and $\log(g)=4.5$. We find \mbox{$T_{*,\rm b}=6000$~K} for 4.5~$\mu$m and 5800~K for 3.6~$\mu$m. The tabulated first and second order offsets are measured in degrees after the secondary eclipse and are calculated using:
$$\psi_{\rm 1} = -(180/\pi)\arctan(D_1/C_1)$$
$$\psi_{\rm 2} = 180-0.5(180/\pi)\arctan(D_2/C_2).$$

\subsection*{Appendix C: Tidal Distortion Calculations}
To assess the impact of stellar and planetary tidal distortion, we model the stellar/planetary surfaces using the dimensionless Roche potential, defined by
\begin{align*}
	\Omega(r,\theta,\phi) = \frac{1}{r} &+ q\bigg(\frac{1}{\sqrt{1-2r\sin\theta\cos\phi+r^2}} - r\sin\theta\cos\phi \bigg) \\
	                                    &+ \frac{q+1}{2}r^2\sin^2\theta,
\end{align*}
where $r$ is the distance from the host star, $\theta$ is the polar angle, $\phi$ is the azimuthal angle, and $q$ is the mass ratio, $M_*/M_p$. We find that the star's radius should be 0.0085\% longer along the star--planet axis compared to the perpendicular equatorial axis, while the planet's radius should be 5.5\% longer along the star--planet axis compared to the dawn--dusk axis seen at transit.

\begin{figure*}
	\centering
	\includegraphics[width=0.6\linewidth]{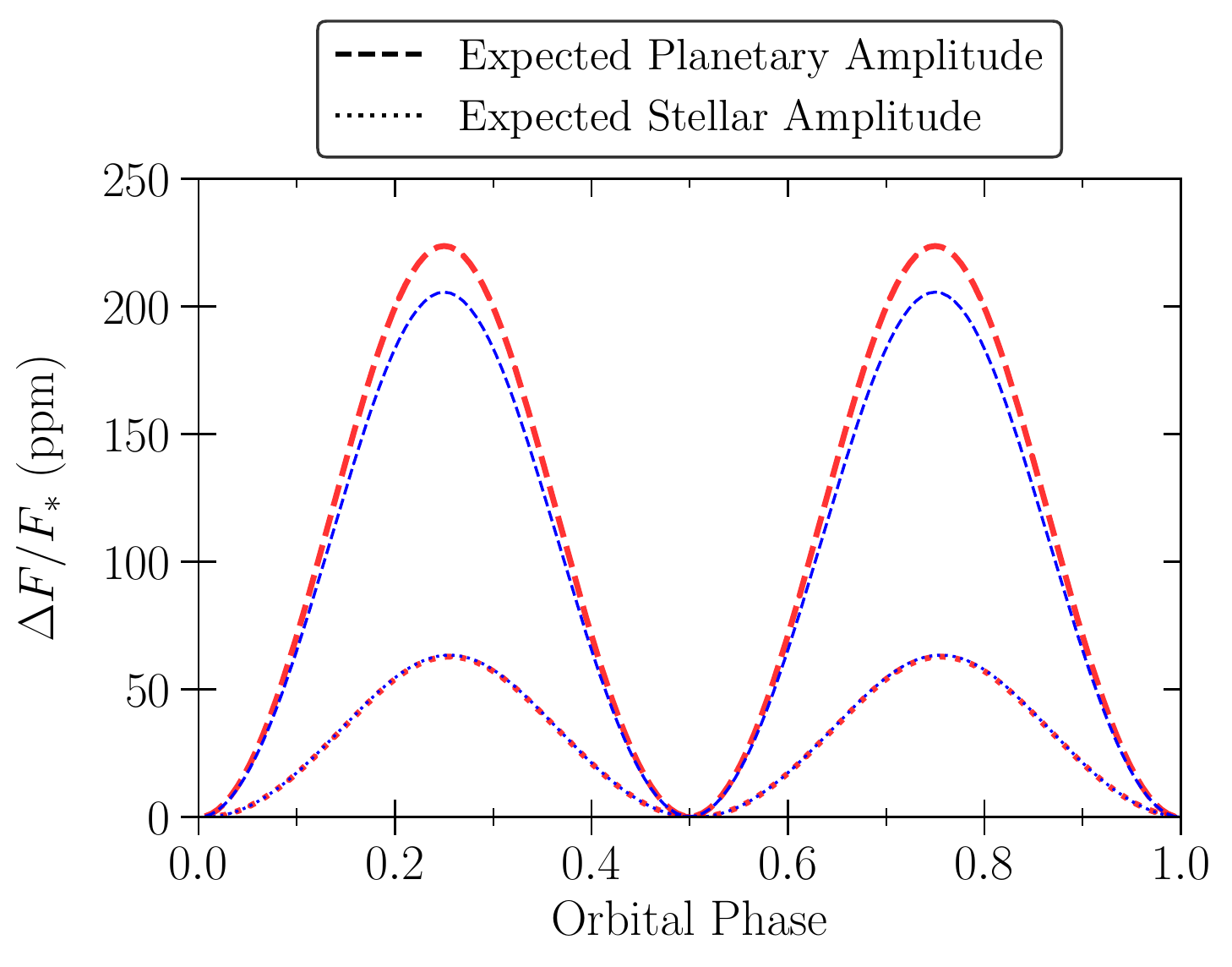}
	\caption{Expected amplitude of tidal distortion from the host star compared to that from the planet. Thin blue lines show the amplitudes at 3.6~$\mu$m, while thick red lines show the amplitudes at 4.5~$\mu$m.}\label{fig:expectDist}
\end{figure*}

We first assume that the planet and star have a constant temperature of 3000~K and 6300~K, respectively, and then perturb these temperatures to account for gravity darkening using the $T_{\rm eff} \propto g_{\rm eff}^{\beta}$ model \citep{lara2011} where $\beta$ is 0.24 for the appreciably distorted planet and 0.25 for the more spherical host star. Next, we convert these temperature maps into flux maps using the Planck blackbody function. We then compute disk-integrated phase curves \citep{cowan2013} while also accounting for the variations in apparent areas of the two objects. Our calculations show that the planet's expected variations are only ${\sim}3.5$ times stronger than that of the host star at \textit{Spitzer}/IRAC wavelengths (see Figure \ref{fig:expectDist} for a depiction).

\begin{figure*}
	\centering
	\includegraphics[width=0.6\linewidth]{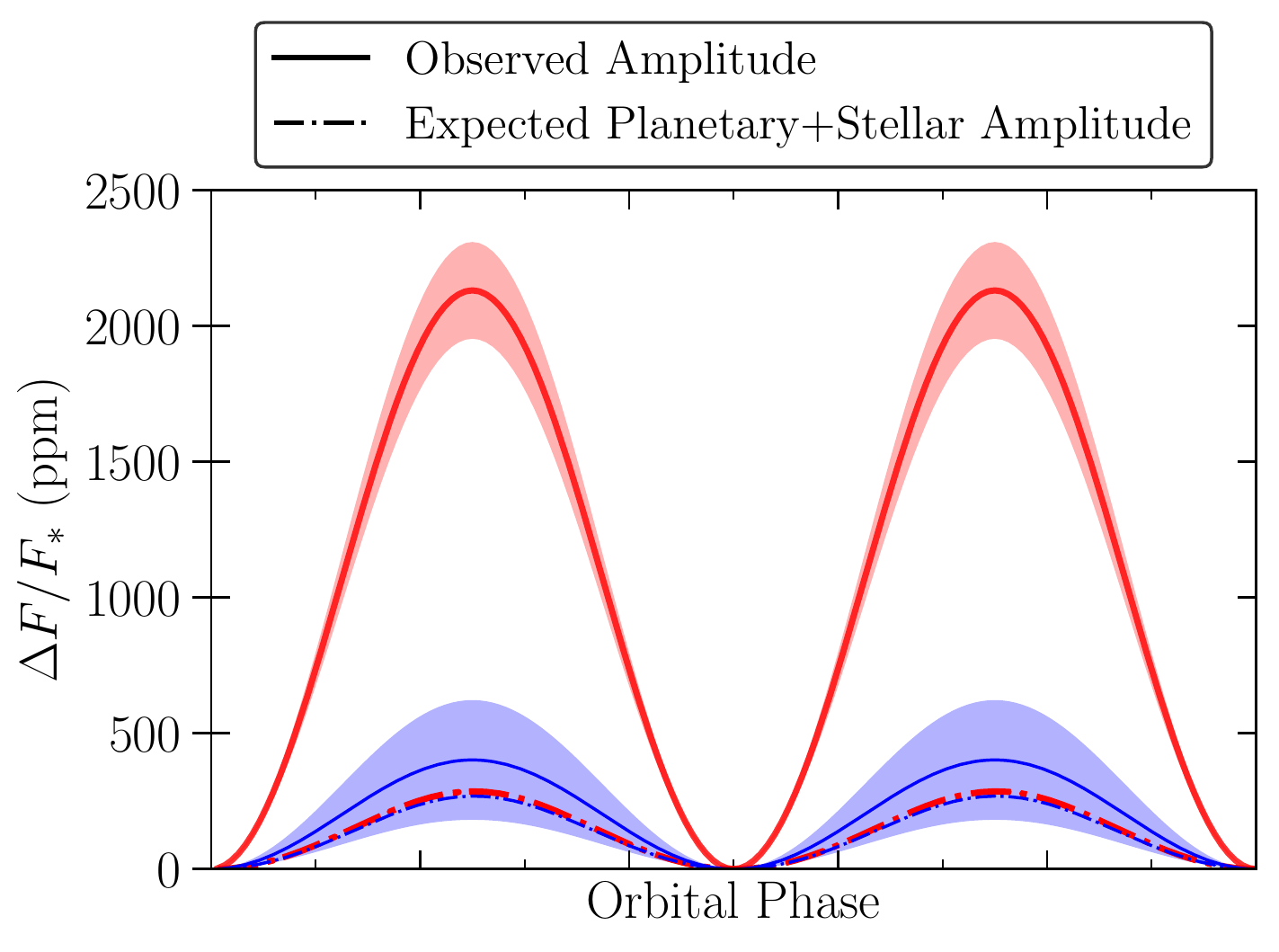}
	\caption{Observed second order sinusoidal variations at 3.6~$\mu$m and 4.5~$\mu$m and their 1$\sigma$ uncertainties compared to the expected amplitude of tidal distortion from the host star and the planet. The 3.6~$\mu$m phase curve is consistent with the expected amplitudes while the 4.5~$\mu$m phase curve is highly discrepant. Thin blue lines show the amplitudes at 3.6~$\mu$m, while thick red lines show the amplitudes at 4.5~$\mu$m.}\label{fig:expectDistVsData}
\end{figure*}

Our predicted ellipsoidal and gravity darkening variations are consistent with past predictions \citep{budaj2011} and with the amplitude of the Zhang PLD model with second order sinusoidal variations fitted to the 3.6~$\mu$m data collected in 2010 (we set the offset to zero as there is no significant detection of an offset in this phase curve). However, the expected ellipsoidal and gravity darkening variations are highly discrepant with the observed amplitude at 4.5~$\mu$m (see Figure \ref{fig:expectDistVsData}). Running simulations where the planet fills its Roche lobe ($R_{p,2}/R_p \approx 1.4$), our ellipsoidal variations and gravity darkening model would be able to explain the full amplitude of the 4.5~$\mu$m phase curve, but the variations remain mostly monochromatic and the model drastically over predicts the variations in the 3.6~$\mu$m phase curve.

\subsection*{Appendix D: Red Noise Tests}
\begin{figure*}
    \begin{flushleft}
        \LARGE \hspace{1.7in} 2010 \hspace{3.2in} 2013
    \end{flushleft}
    
    \centering
	
	\includegraphics[width=0.48\linewidth]{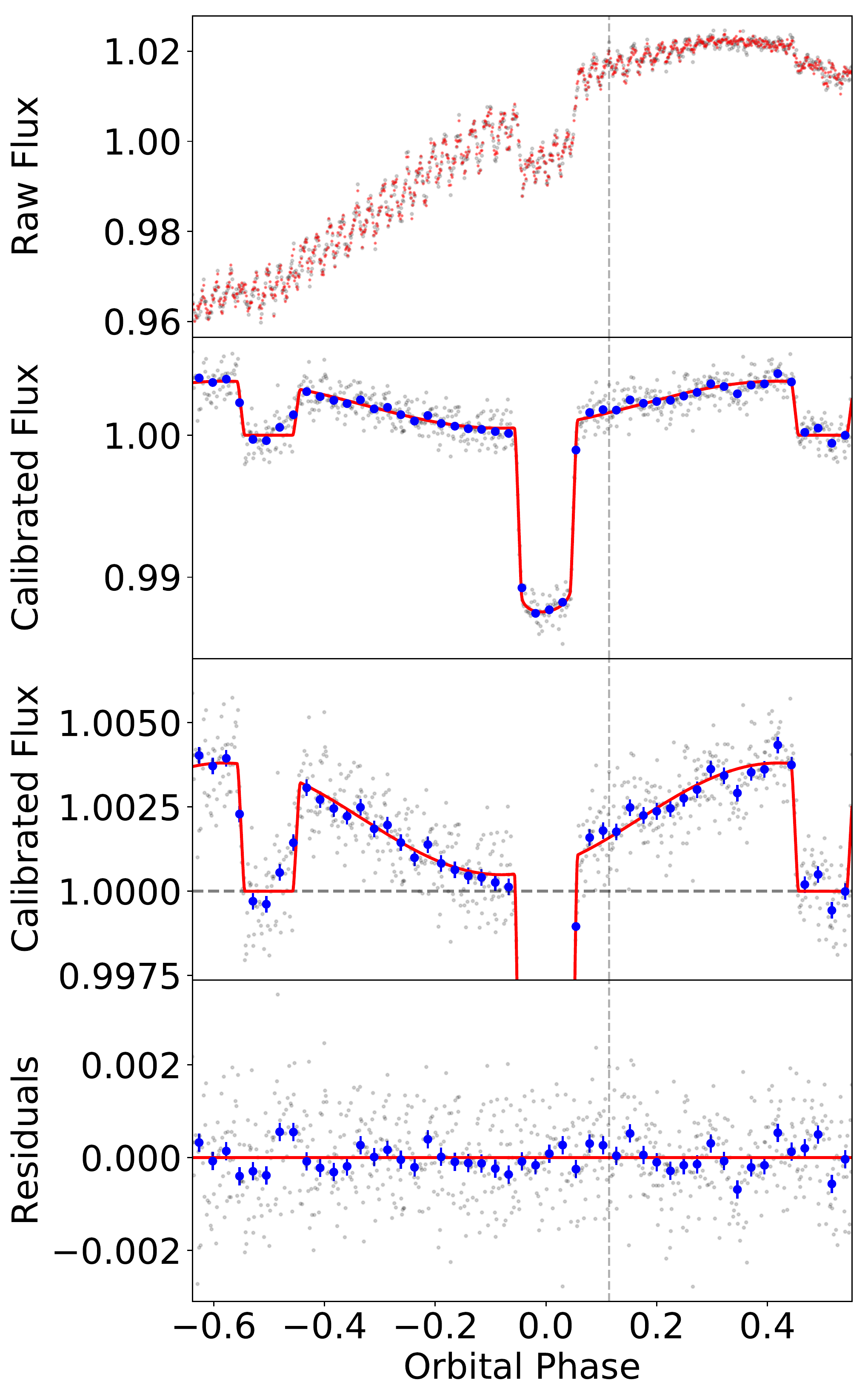}\hfill%
	\includegraphics[width=0.48\linewidth]{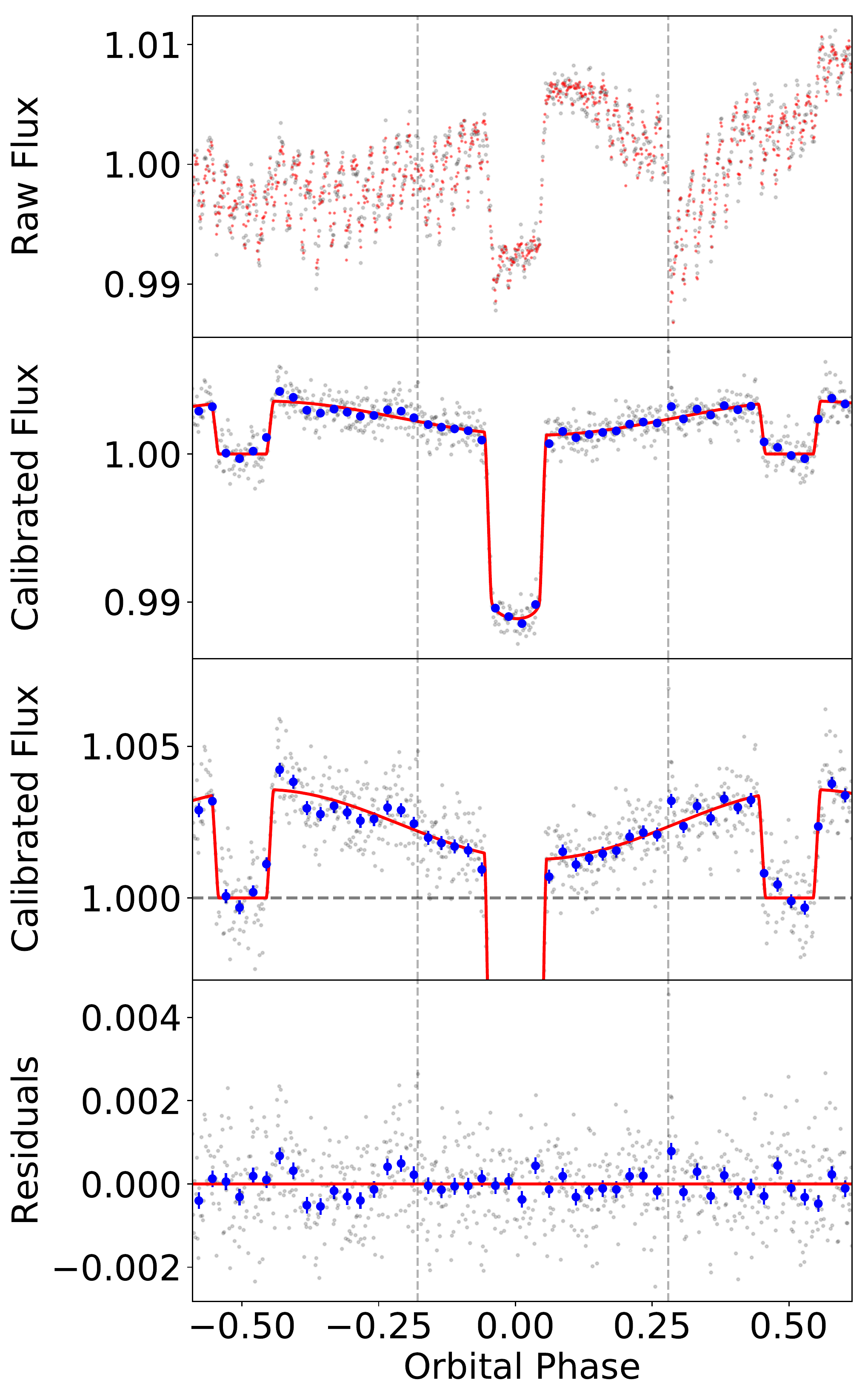}
	
	\vspace{0.5cm}
	
	\includegraphics[width=0.48\linewidth]{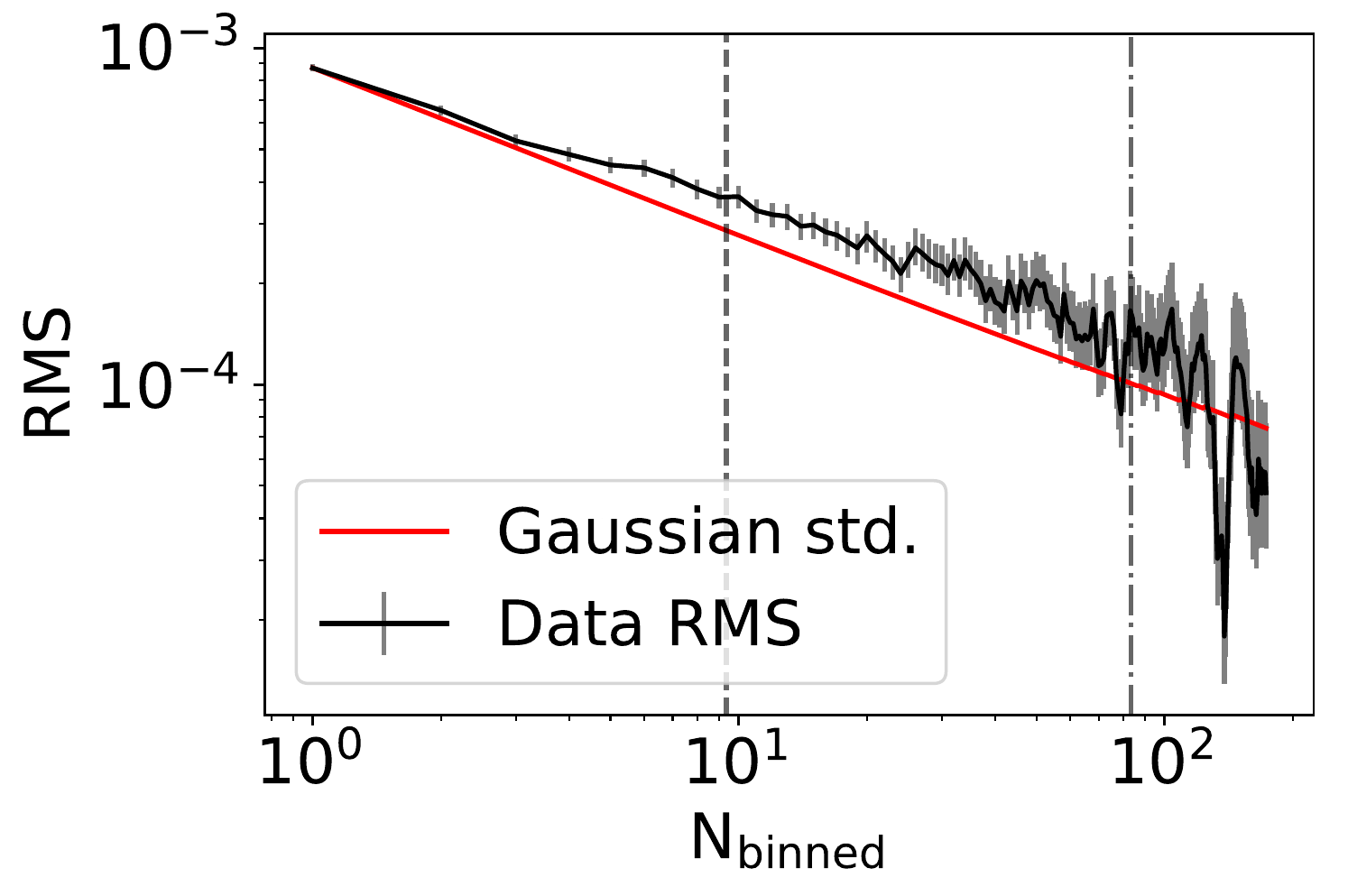}\hfill%
	\includegraphics[width=0.48\linewidth]{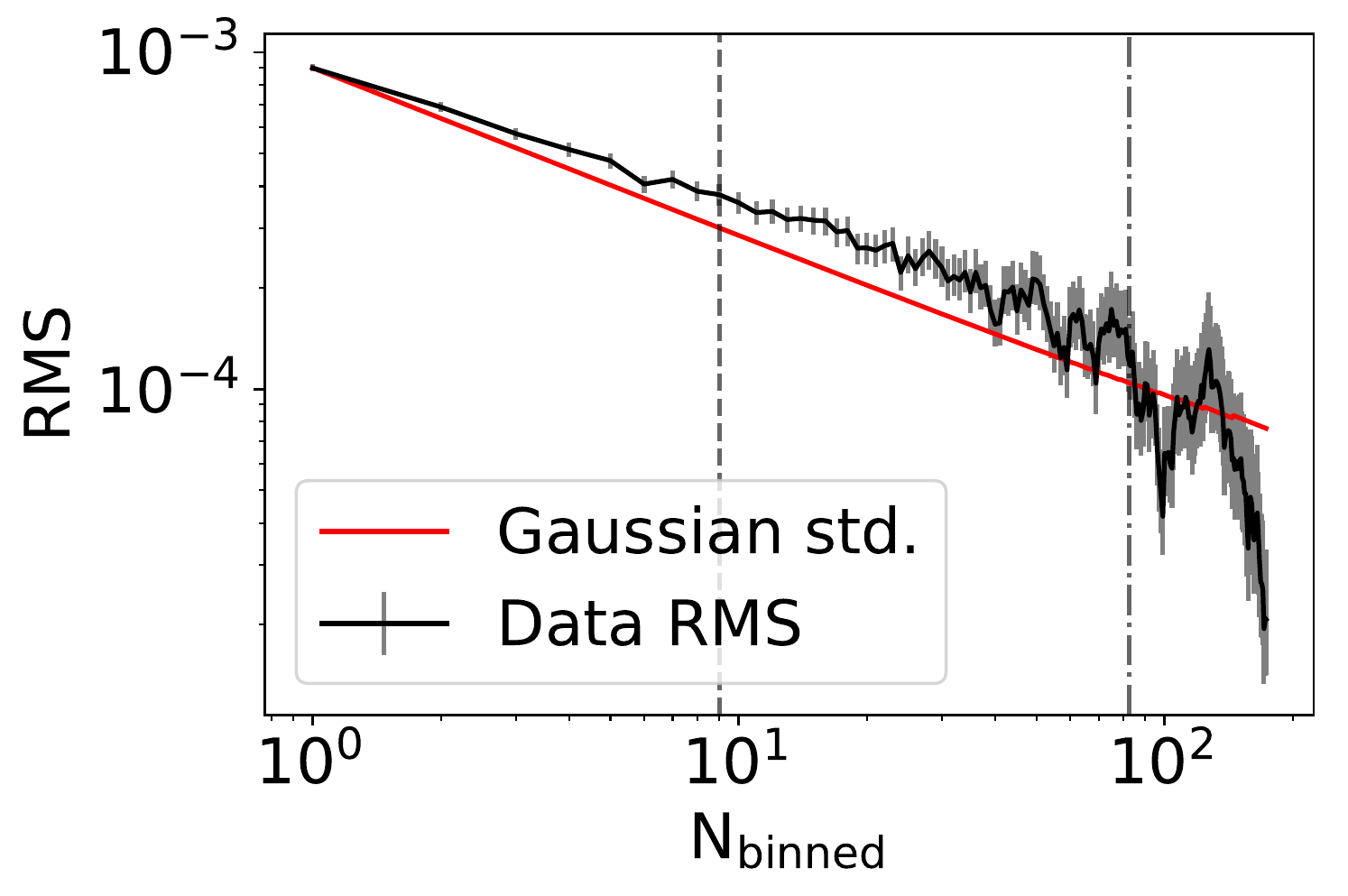}
	\caption{\textit{Top row:} WASP-12b 2010 (left) and 2013 (right) 3.6~$\mu$m observations, fit using the Zhang PLD detector model and second- (2010) and first order (2013) phase variations model. Vertical dashed lines mark the transitions between AORs. %
	\textit{Bottom row:} Red noise test for the 2010 (left) and 2013 (right) 3.6~$\mu$m observations of WASP-12b for the above fits. The black line shows the decrease in the observed standard deviation in the residuals as $N_{\rm binbed}$ (the number of datapoints binned together) increases. The red line shows the expected decrease in standard deviation, assuming the noise is entirely white. The close match between the two curves suggests that little-to-no red-noise remains in the residuals. A vertical, dashed line shows the timescale for transit/eclipse ingress and egress, while the dash-dotted line shows the \mbox{$t_1$--$t_4$} transit duration.}\label{fig:zhang1}
\end{figure*}

The bottom rows of Figures \ref{fig:zhang1}--A10 show the observed standard deviation in the residuals versus the number of data cubes binned together for each lightcurve made using the \texttt{binrms} routine from the \texttt{Multi-Core Markov-Chain Monte Carlo (MC3)}\footnote{\url{http://pcubillos.github.io/MCcubed/}} package \citep{cubillos2017}; this allows us to test for any red noise remaining in our residuals \citep{winn2007,cowan2012}. These figures show that minimal red noise remains after our fiducial models have been subtracted from the data (the photometric uncertainty decays roughly as $\sqrt{N_{\rm binned}}$). There is, however, some lower-frequency noise in the 2010 4.5~$\mu$m observations between the first eclipse and the transit that cannot be modelled by any of the three decorrelation pipelines.

\begin{table*}
\caption{A summary of all the priors used in the three independent analyses. Uniform priors were used where there are inequalities below, Gaussian priors were used where uncertainties are indicated, variables were fixed where only a value is indicated, and parameters were unconstrained where Free is written.}\label{tab:priors}
\begin{tabular}{c|c|c|c}
                             & Zhang PLD                         & Bell SPCA                       & Cubillos POET                      \\ \hline
$t_0$ (BMJD)                 & $54508.20396 < t_0 < 54508.74968$ & 56176.16825800 $\pm$ 0.00007765 & Free                               \\ \hline
$R_p/R_*$                    & $> 0$                             & $0 < R_p/R_* < 1$               & $> 0$                              \\ \hline
$a/R_*$                      & 3.039                             & 3.039 $\pm$ 0.034               & 3.039 $\pm$ 0.034                  \\ \hline
$i$ (degrees)                & 83.37                             & 83.37 $\pm$ 0.68                & 83.37 $\pm$ 0.68 (fitted $\cos i$) \\ \hline
$P$ (days)                   & 1.09142245                        & 1.0914203                       & 1.0914203                          \\ \hline
$F_p/F_*$                    & Free                              & $0 < F_p/F_* < 1$               & $> 0$                              \\ \hline
$C_1$                        & Free                              & Positive Phasecurve             & Positive Phasecurve                \\ \hline
$D_1$                        & Free                              & Positive Phasecurve             & Positive Phasecurve                \\ \hline
$C_2$                        & Free                              & Positive Phasecurve             & Positive Phasecurve                \\ \hline
$D_2$                        & Free                              & Positive Phasecurve             & Positive Phasecurve                \\ \hline
$R_{p, 2}/R_*$               & $> 0$                             & $0 < R_{p, 2} < 1$              & $> 0$                              \\ \hline
$\sigma_F/F_*$ (white noise) & $0 < \sigma_F/F_* < 1$            & $> 0$                           & Free                               \\ \hline
Limb Darkening               & Sing 2010 Model                   & \begin{tabular}[c]{@{}l@{}}$0 < q_1 < 1$;\\ $0 < q_2 < 1$\end{tabular}  %
                                                                                                   & \begin{tabular}[c]{@{}l@{}}$0 < q_1 < 1$;\\ $0 < q_2 < 1$\end{tabular} %
                                                                                                                                        \\ \hline
$e$                          & 0                                 & 0                               & \begin{tabular}[c]{@{}l@{}}$t_{\rm eclipse}$: Free;\\ $t_{14, \rm eclipse} > 0$;\\
                                                                                                                                $t_{12, \rm eclipse} > 0$;\\ $t_{34} = t_{12}$\end{tabular} %
                                                                                                                                        \\ \hline
Instrumental Variables       & \begin{tabular}[c]{@{}l@{}}Free (PLD coefficients,\\ Slope in time)\end{tabular} %
                                                                 & \begin{tabular}[c]{@{}l@{}}Free (Polynomial coefficients, \\ Slope in time for 2013 3.6~$\mu$m)\end{tabular} %
                                                                                                   & \begin{tabular}[c]{@{}l@{}}Free (Slope in time for\\ 2013 4.5~$\mu$m)\end{tabular}  %
                                                                                                                                        \\ \hline
\end{tabular}
\end{table*}

\subsection*{Appendix E: NUV Evidence for Mass Loss}
Across the NUV, WASP-12b appears to be larger than the planet's Roche radius, implying significant mass loss \citep{fossati2010b,haswell2012,nichols2015}. The first Hubble Space Telescope, Cosmic Origins Spectrograph transit observation of WASP-12b \citep{fossati2010b} also detected an early ingress in the NUV; this suggests the presence of a stream of gas stripped from the planet flowing in toward the star \citep{lai2010,bisikalo2013a,matsakos2015} which forms a bow shock ahead of the planet \citep{vidotto2010,llama2011,bisikalo2013a,cherenkov2014,matsakos2015,turner2016}, although the position of this shock can vary \citep{vidotto2011, llama2013}. There is also evidence for variable NUV ingress times \citep{haswell2012,nichols2015} which suggests variable mass-loss rates and/or a variations in the planet--shock distance \citep{vidotto2011}. The non-detection of stellar activity indicators from WASP-12A \citep{knutson2010,fossati2013} may also suggest that WASP-12b is undergoing mass loss. The final resting place of the gas stripped from WASP-12b is debated, with some suggesting an accretion disk interior to the planet's orbit \citep{lai2010,li2010} and others suggesting an extended circumstellar torus of gas with the planet embedded inside \citep{debrecht2018}.

\subsection*{Appendix F: Discussion of Variability}
To date, no \textit{Spitzer} phase curve observation has shown variability in the phase curve offset of an exoplanet, although significant near-infrared variability has been seen for brown dwarfs and isolated planetary mass objects \citep{artigau2009,radigan2012}, and \textit{Kepler} phase curves of the hot Jupiter HAT-P-7b have been reported to vary \citep{armstrong2016}. Variability is expected for WASP-12b due to coupling between the planet's partially ionized atmosphere and the planet's magnetic field \citep{rogers2017}. The timescale of this variability is set by the Alfv\'en timescale \citep[${\sim}115$ days for WASP-12b assuming magnetic effects occur on the dayside where the atmosphere is dominated by atomic hydrogen;][]{rogers2017,dang2018}. Variability may also arise in the presence of time-variable cloud coverage, although optically reflective clouds on the planet's dayside were stringently rejected using \textit{Hubble}/STIS optical eclipse spectroscopy of WASP-12b \citep{bell2017}. Any time-variability in the gas streaming from the planet could also obscure different portions of the planet over time and lead to an apparent variation in the 3.6~$\mu$m phase curve.

\subsection*{Appendix G: Model Selection}
The preferred model for each phase curve was chosen to be the model with the lowest Bayesian Information Criterion (BIC), defined as
$$
	{\rm BIC} = -2\ln(L) + N_{\rm par}\ln(N_{\rm dat}),
$$
where $N_{\rm par}$ is the number of model parameters and $N_{\rm dat}$ is the number of data. The log-likelihood is
$$
	\ln(L) = -\frac{\chi^2}{2} - N_{\rm dat}\ln(\sigma_F) - \frac{N_{\rm dat}}{2}\ln(2\pi)
$$
where $\sigma_F$ is the fitted photometric uncertainty (assumed to be constant throughout the observation) and
$$
	\chi^2 = \frac{\sum_i\big(F_{{\rm obs}, i} - F_{{\rm model}, i}\big)^2}{\sigma_F^2}
$$
is a measure of the badness-of-fit, where $F_{{\rm obs}, i}$ are the observed flux measurements. We adopt the threshold that models with a $\Delta {\rm BIC} \leq 5$ with respect to the favoured model cannot be strongly ruled out.

\begin{table*}
\begin{center}
	\caption{3.6 and 4.5~$\mu$m phase curve parameters from the preferred models for 2010 and 2013. Fiducial models are indicated with bolding.}\label{tab:s1}
	\textbf{3.6~$\bm\mu$m}\\
	\includegraphics[width=\textwidth, trim=0 35 0 0, clip]{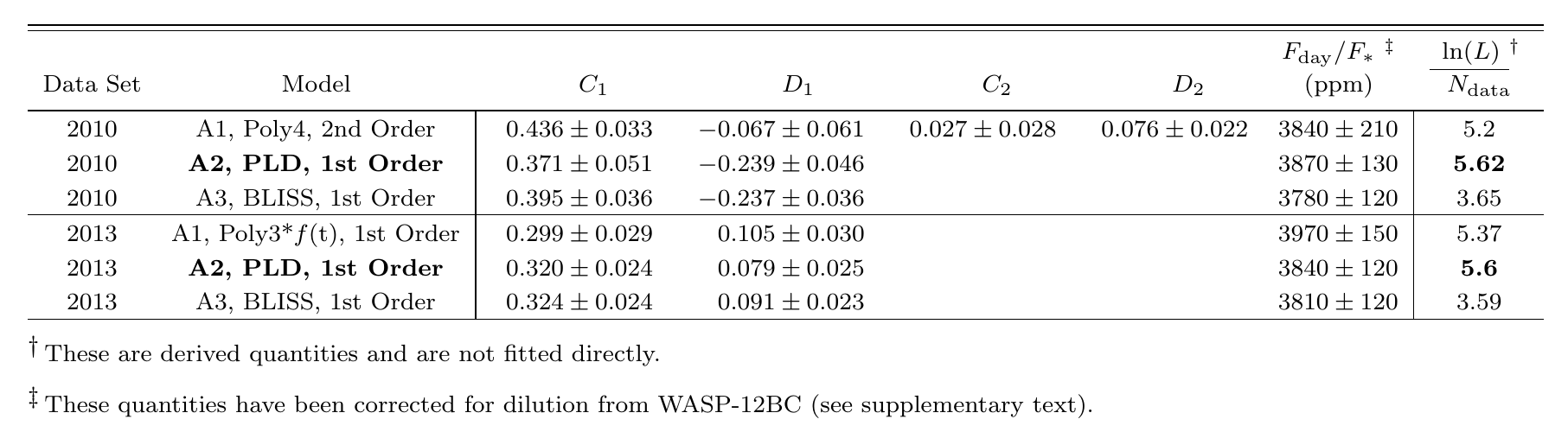}
	
	\textbf{~\\ 3.6~$\bm\mu$m, cont.}\\
	\includegraphics[width=\textwidth, trim=17 35 17 0, clip]{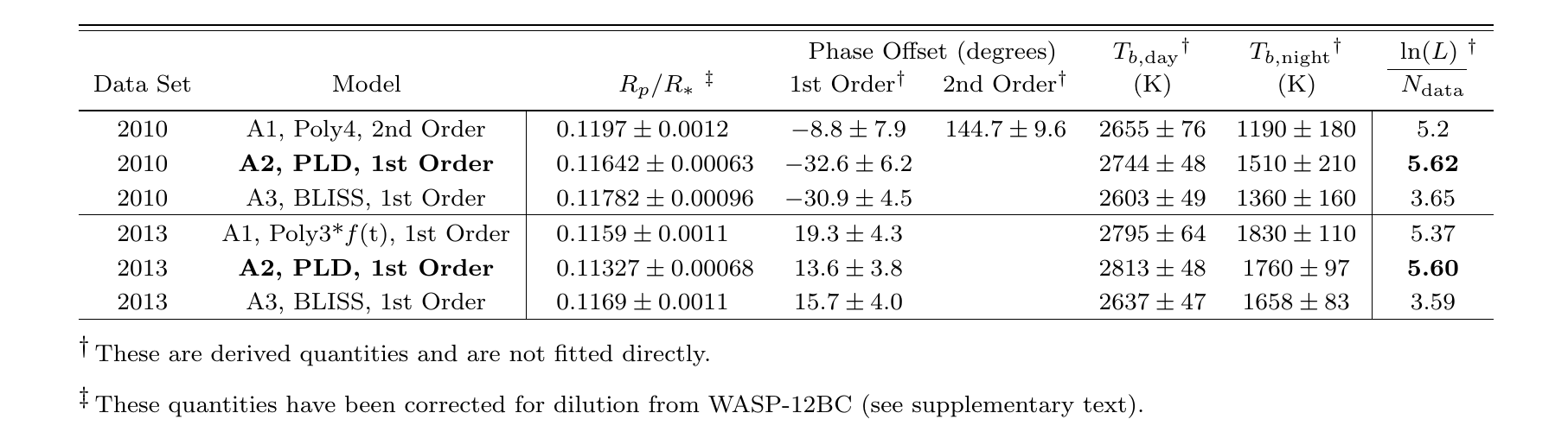}
	
	\textbf{~\\ 4.5~$\bm\mu$m}\\
	\includegraphics[width=\textwidth, trim=0 35 0 0, clip]{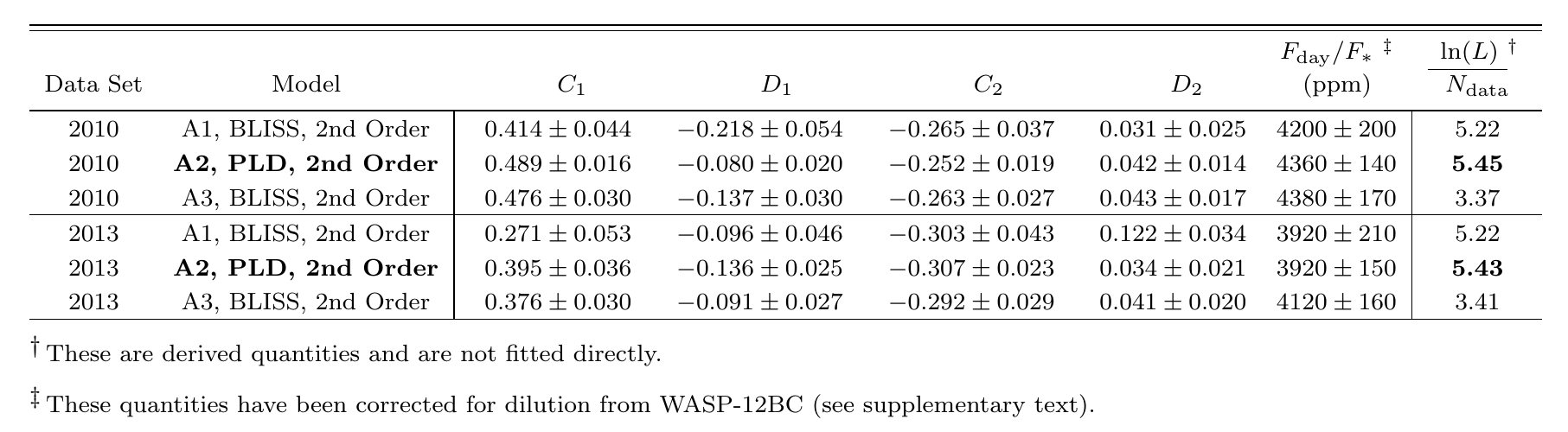}
	
	\textbf{~\\ 4.5~$\bm\mu$m, cont.}\\
	\includegraphics[width=\textwidth, trim=21 0 21 0, clip]{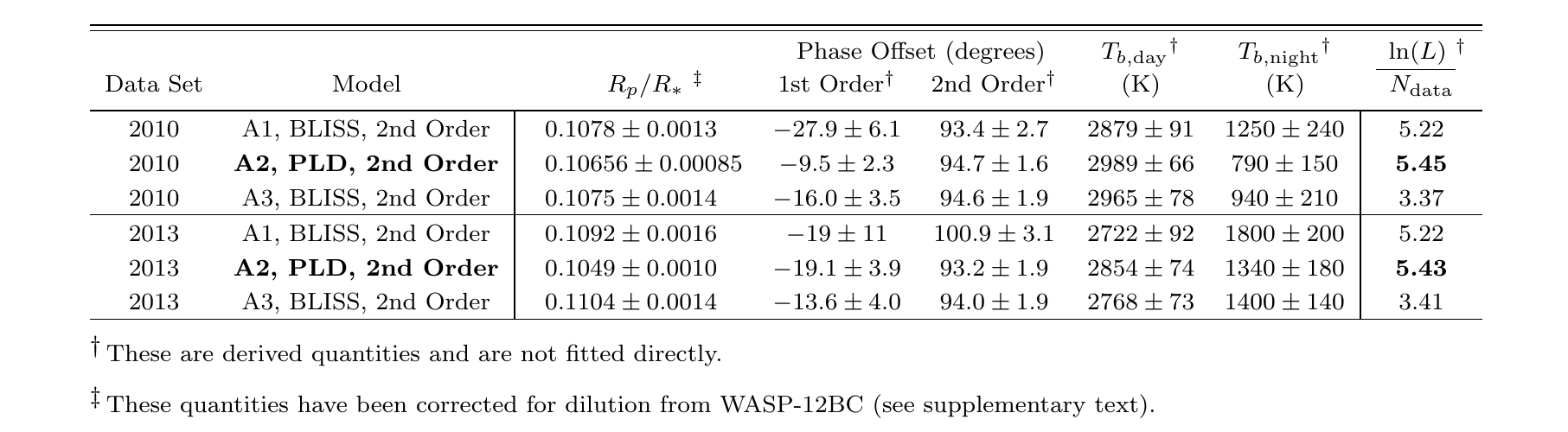}
\end{center}
\end{table*}

\clearpage
\begin{figure*}
	\centering
	\includegraphics[width=0.7\linewidth]{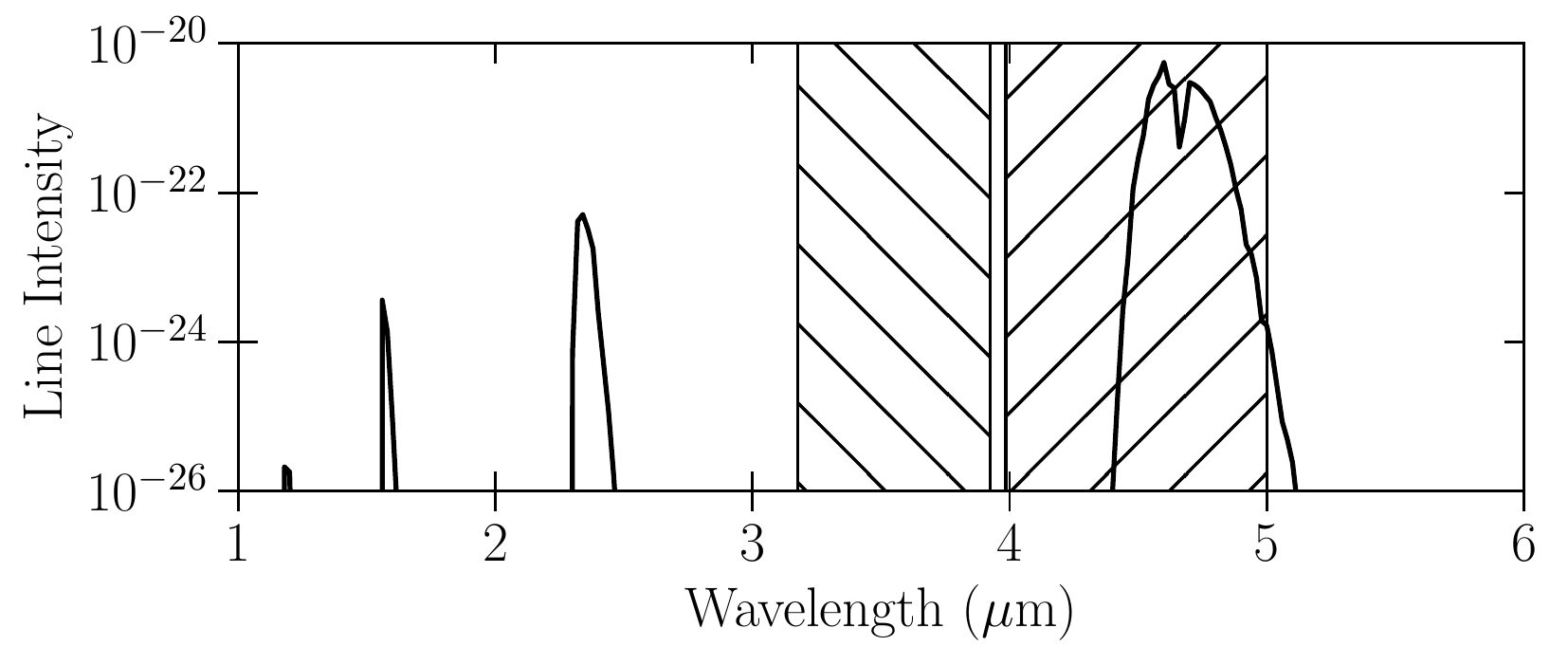}
	\caption{CO line intensities at 296~K from HITEMP \citep{rothman2010} in units of cm$^{-1}$/(molecule $\times$ cm$^{-2}$) which has been binned to a spectral resolution of 10~cm. The bandwidths of 
\textit{Spitzer}/IRAC channels 1 and 2 are respectively shown with downward sloping and upward sloping hatched regions.}\label{fig:coIntensities}
\end{figure*}

\begin{figure*}
	\centering
	\includegraphics[width=\linewidth]{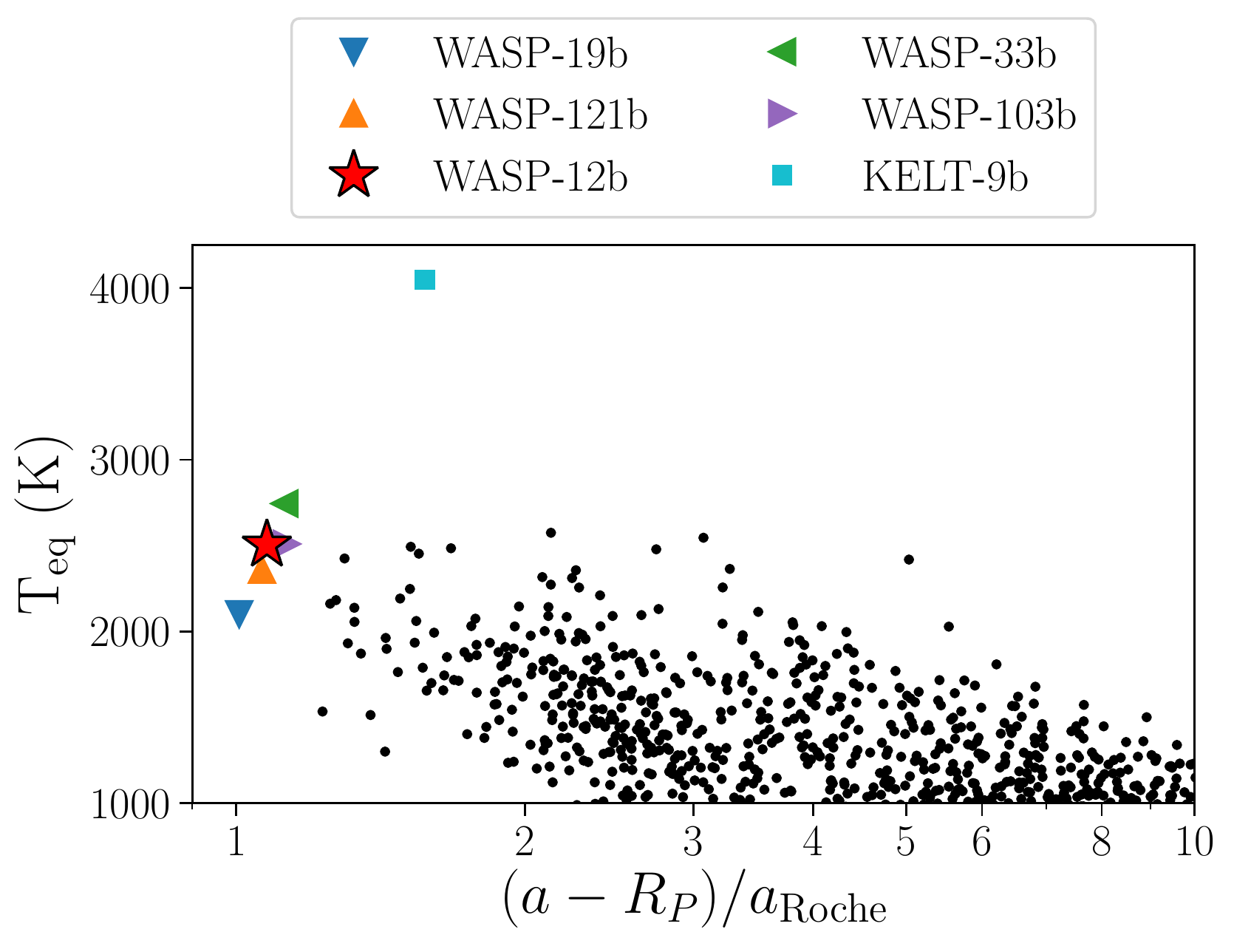}
	\caption{A comparison of WASP-12b to other exoplanets. The $y$-axis is the planets' equilibrium temperature ($\rm T_{eq}=0.25^{0.25}T_*\sqrt{R_*/a}$), and this $x$-axis is the distance of the substellar point on the planets from their L1 Lagrange point \citep{roche1847}, where $a_{\rm Roche}=2.44(R_p)(M_*/M_p)^{1/3}$. While WASP-12b is one of the exoplanets closest to overflowing it's Roche lobe, there are several others with similar characteristics for which Spitzer phase curves do not show strong second order sinusoidal variations at 4.5~$\mu$m \citep{wong2016,zhang2018,kreidberg2018b}. One potential explanation is that WASP-12b's orbit may be decaying \citep{maciejewski2016,patra2017} while the other exoplanets may be more stable.}\label{fig:rocheLimits}
\end{figure*}

\clearpage

\begin{figure*}
    \begin{flushleft}
        \LARGE \hspace{1.7in} 2010 \hspace{3.2in} 2013
    \end{flushleft}
    
	\centering
	\includegraphics[width=0.48\linewidth]{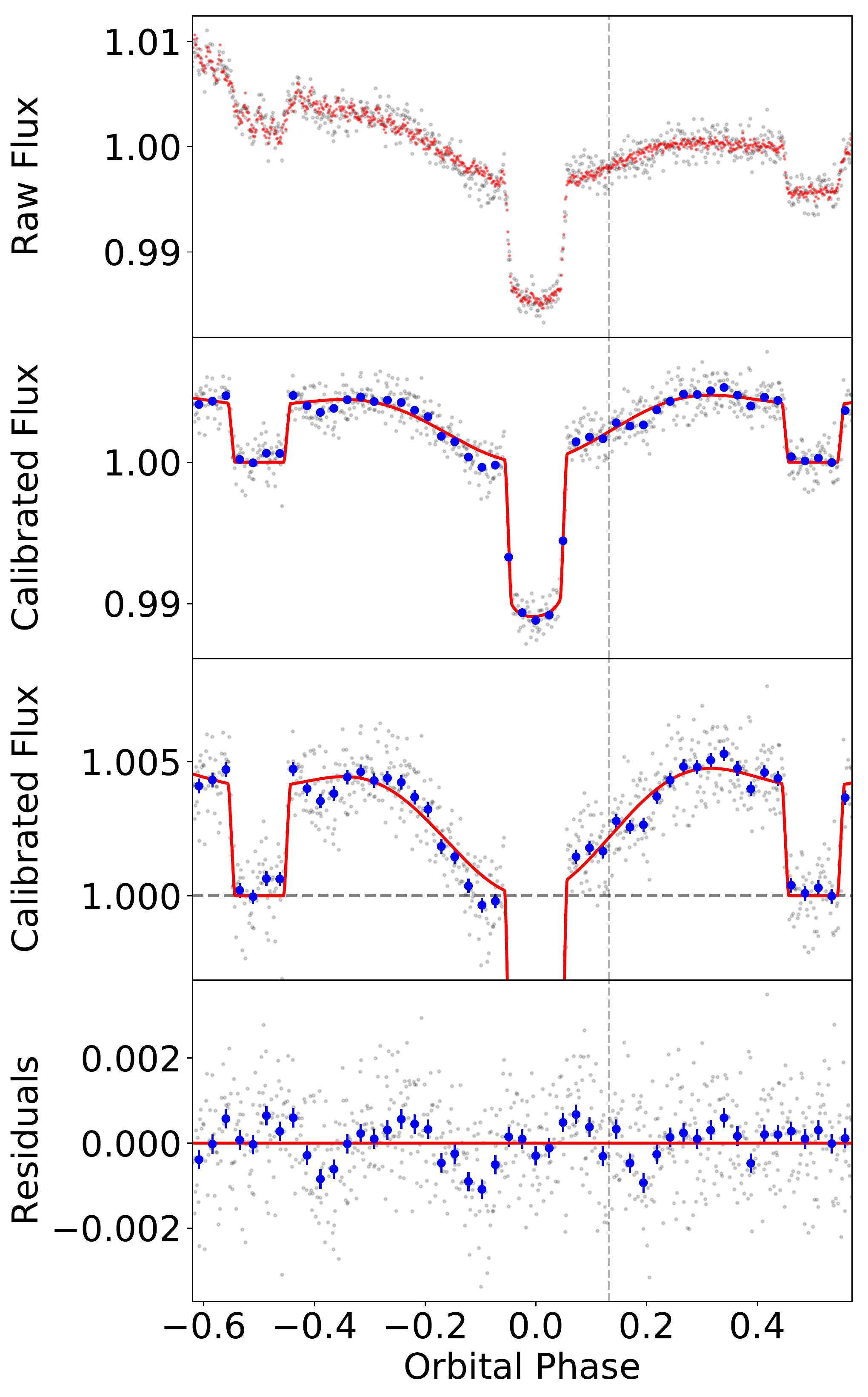}\hfill%
	\includegraphics[width=0.48\linewidth]{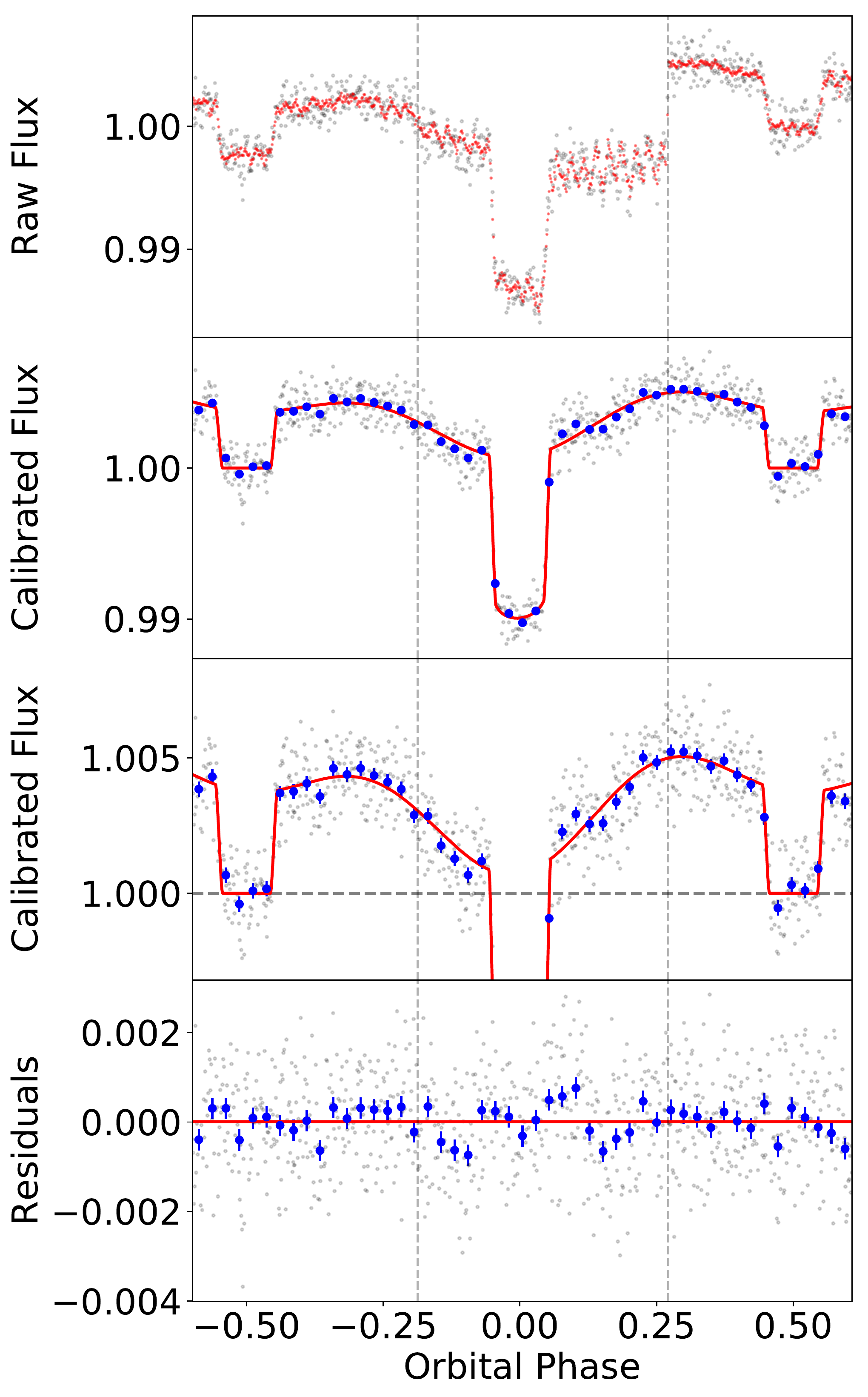}
	
	\vspace{0.5cm}
	
	\includegraphics[width=0.48\linewidth]{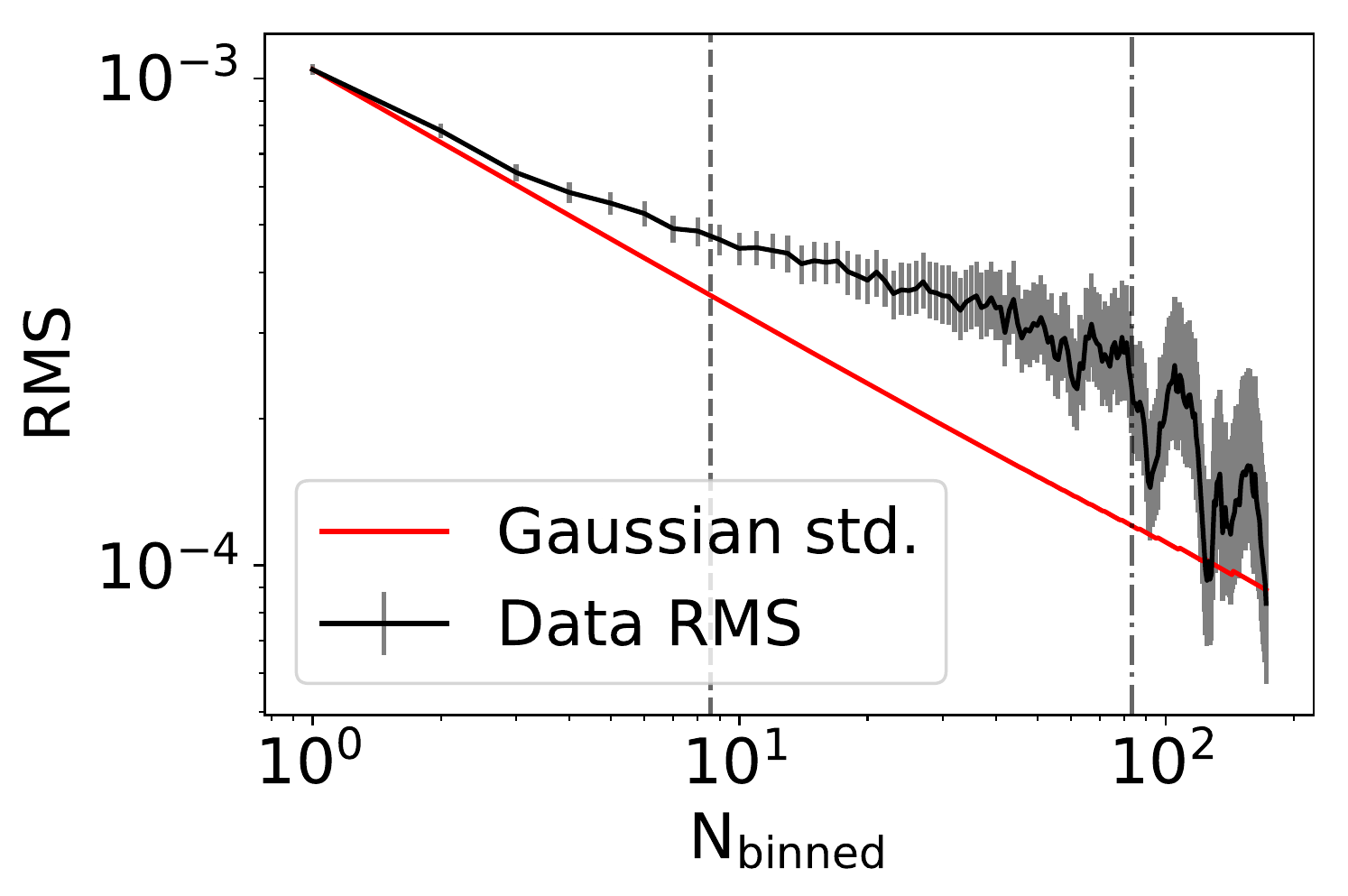}\hfill%
	\includegraphics[width=0.48\linewidth]{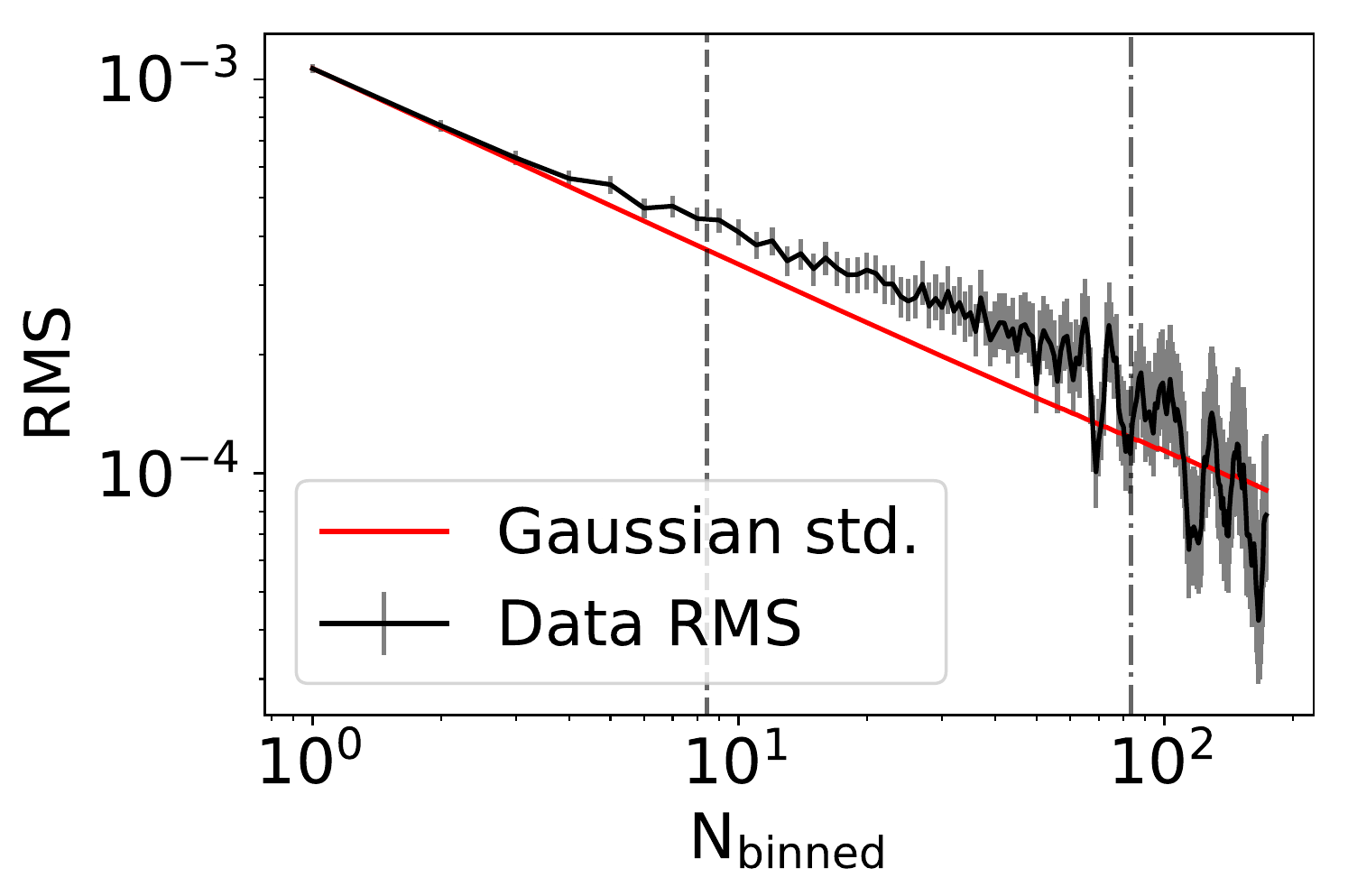}
	\caption{Same figure as the left panel of Figure \ref{fig:zhang1} but for the 2010 (left column) and 2013 (right column) 4.5~$\mu$m observations of WASP-12b, both fit using the Zhang PLD detector model and the second order phase variations model.}\label{fig:zhang2}
\end{figure*}

\clearpage


\bsp	
\label{lastpage}

\begin{figure*}
    \begin{center}
        \Huge
        \textbf{Supplementary Information}
        
        \vspace{1cm}
    \end{center}

    \begin{flushleft}
        \LARGE \hspace{1.7in} 2010 \hspace{3.2in} 2013
    \end{flushleft}
    
	\centering
	\includegraphics[width=0.48\linewidth]{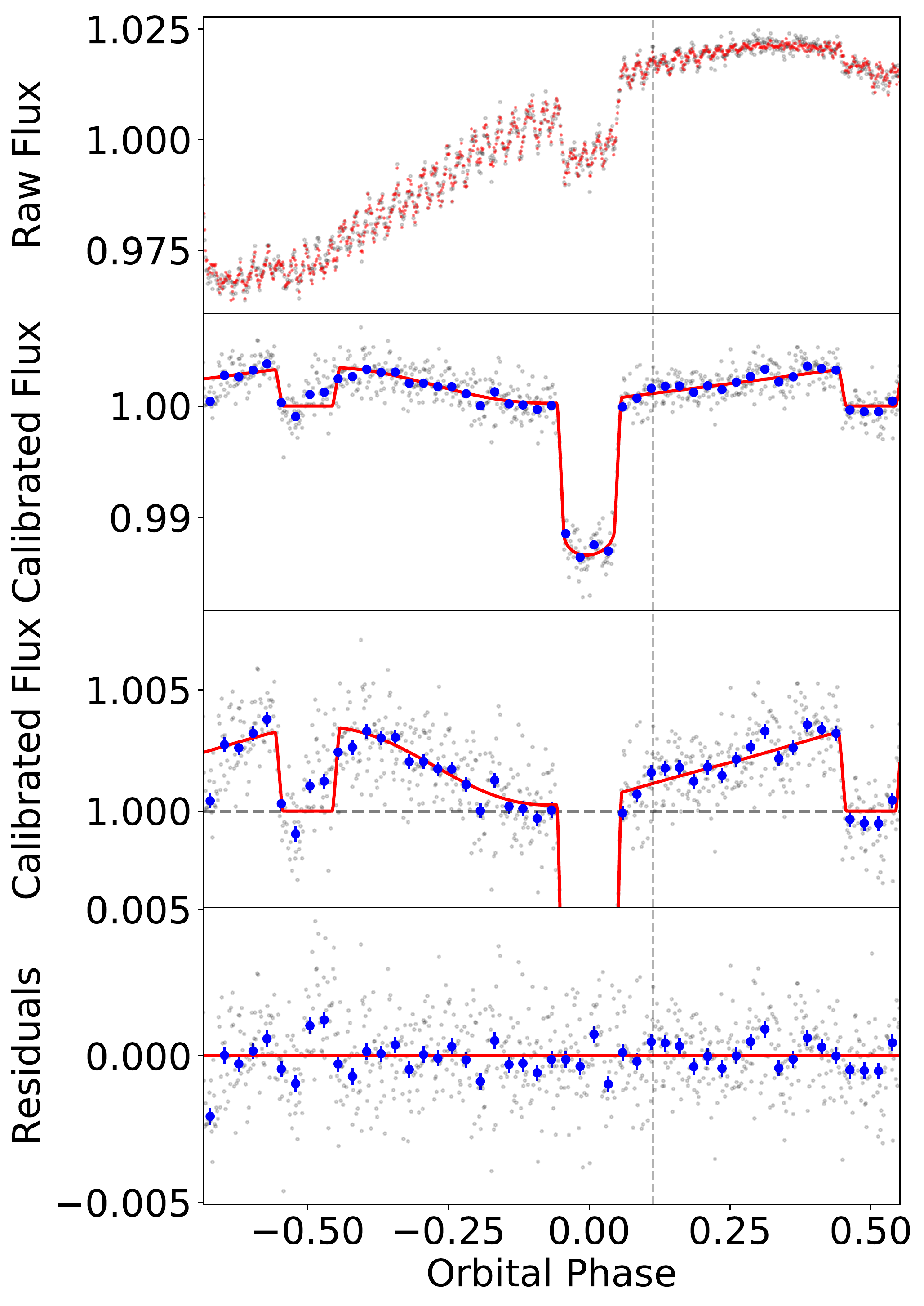}\hfill%
	\includegraphics[width=0.48\linewidth]{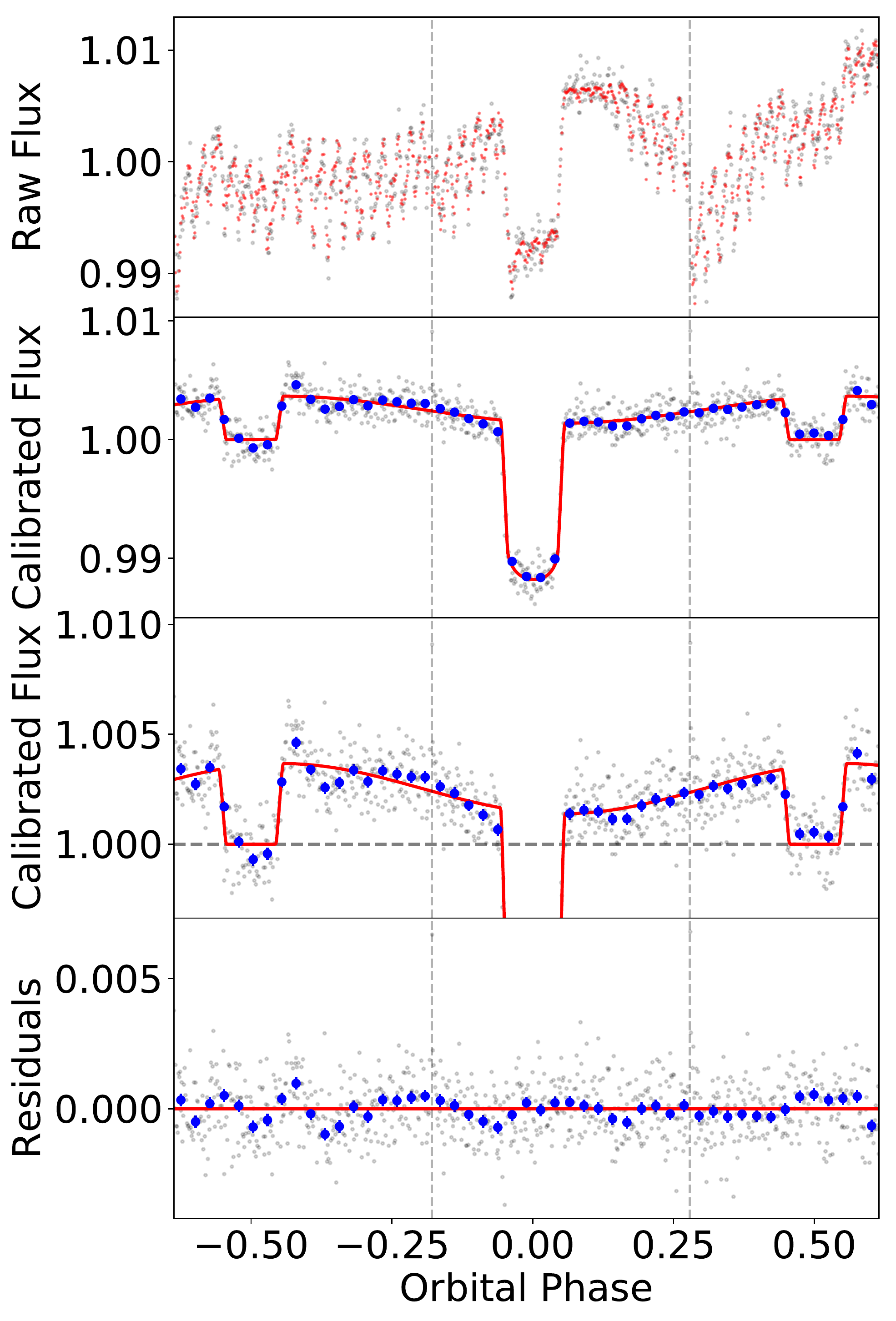}
		
	\vspace{0.5cm}
	
	\includegraphics[width=0.48\linewidth]{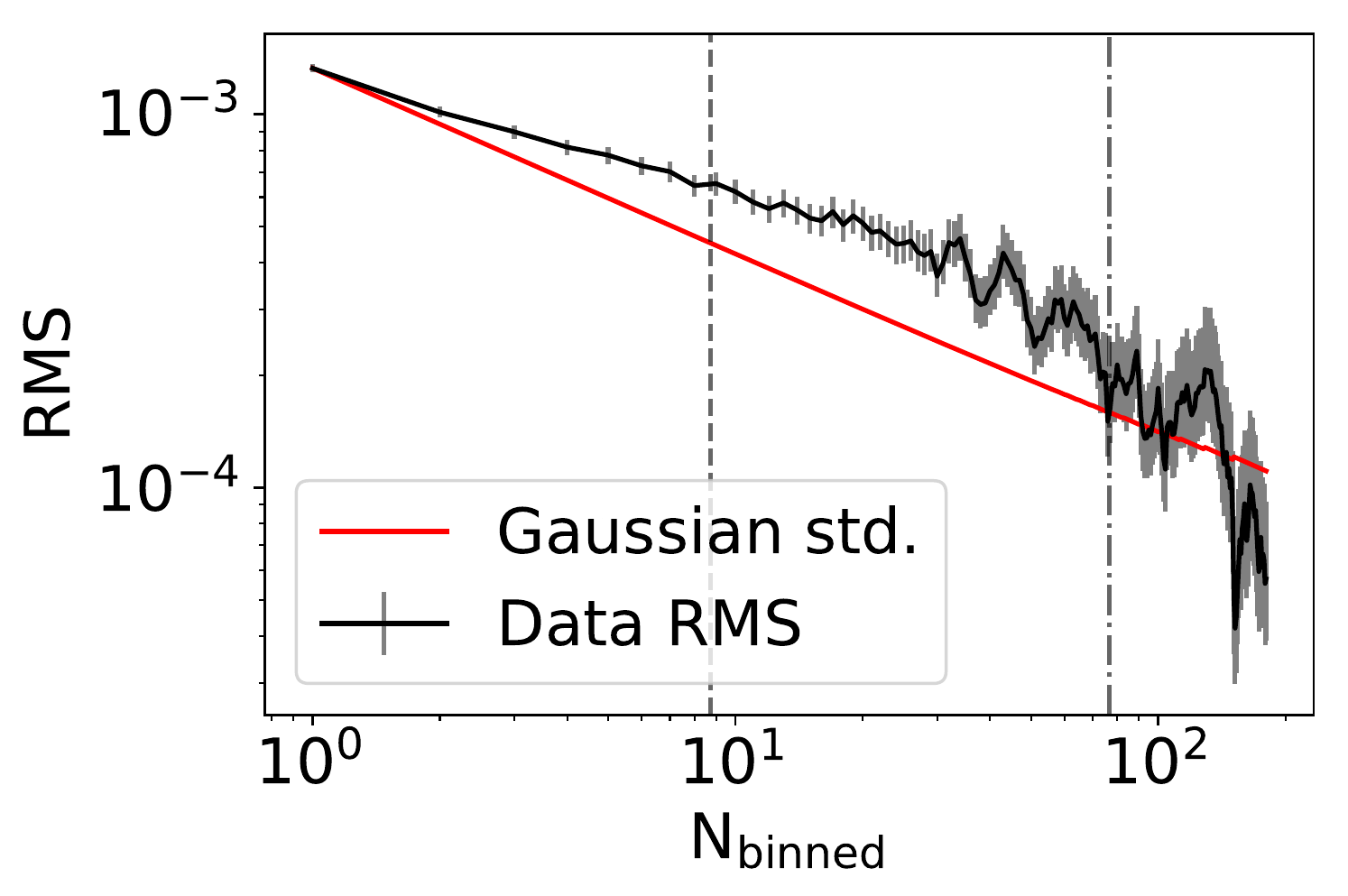}\hfill%
	\includegraphics[width=0.48\linewidth]{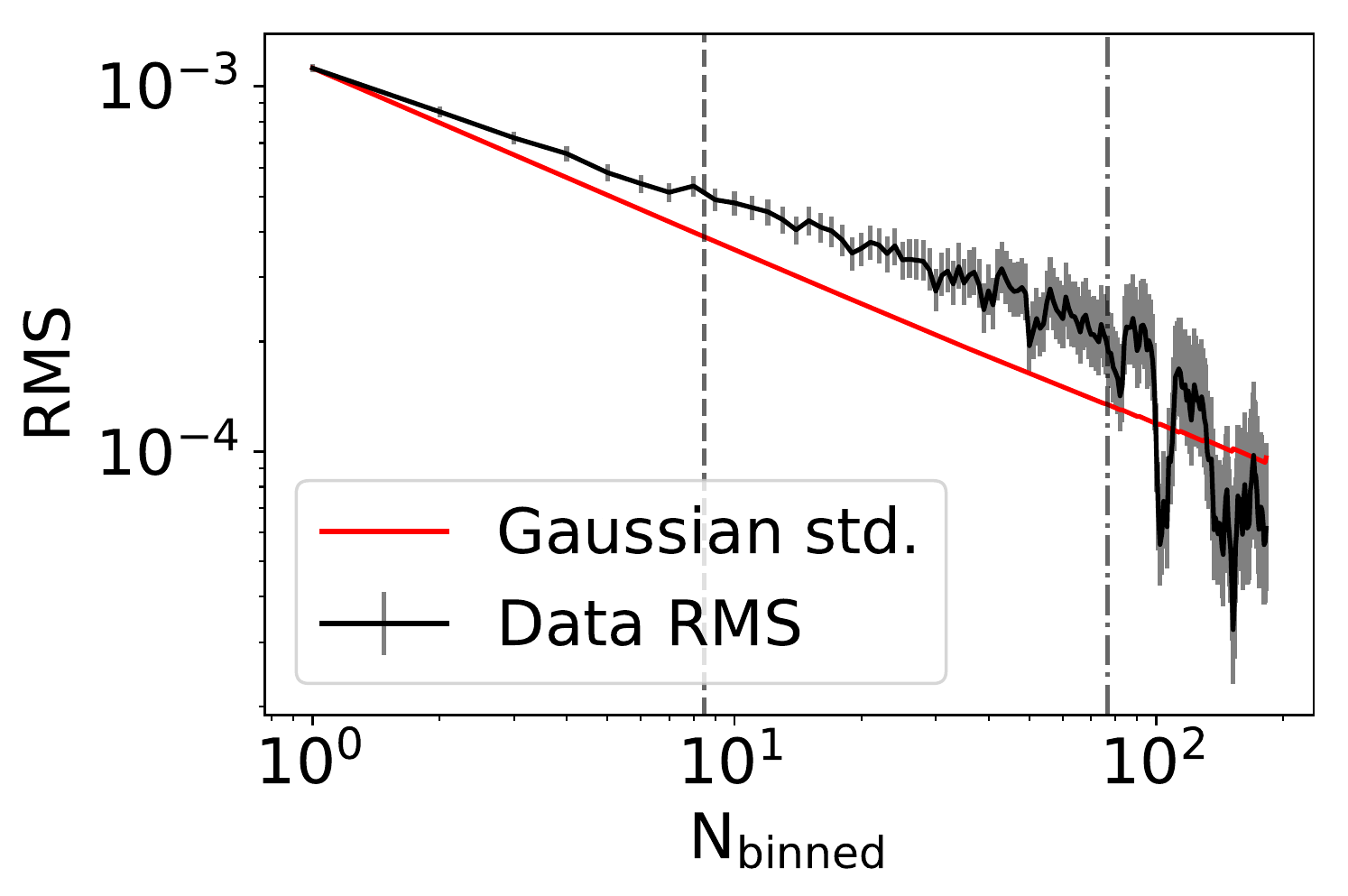}
	\caption{Same as Figure A5 but for the 2010 (left) and 2013 (right) 3.6~$\mu$m observations of WASP-12b, fit using the SPCA Poly4 (2010) and SPCA Poly3*$f(t)$ (2013) detector models and the second- (2010) and first order (2013) phase variations model.}\label{fig:bell1}
\end{figure*}

\begin{figure*}
	\begin{flushleft}
        \LARGE \hspace{1.7in} 2010 \hspace{3.2in} 2013
    \end{flushleft}
    
	\centering
	\includegraphics[width=0.48\linewidth]{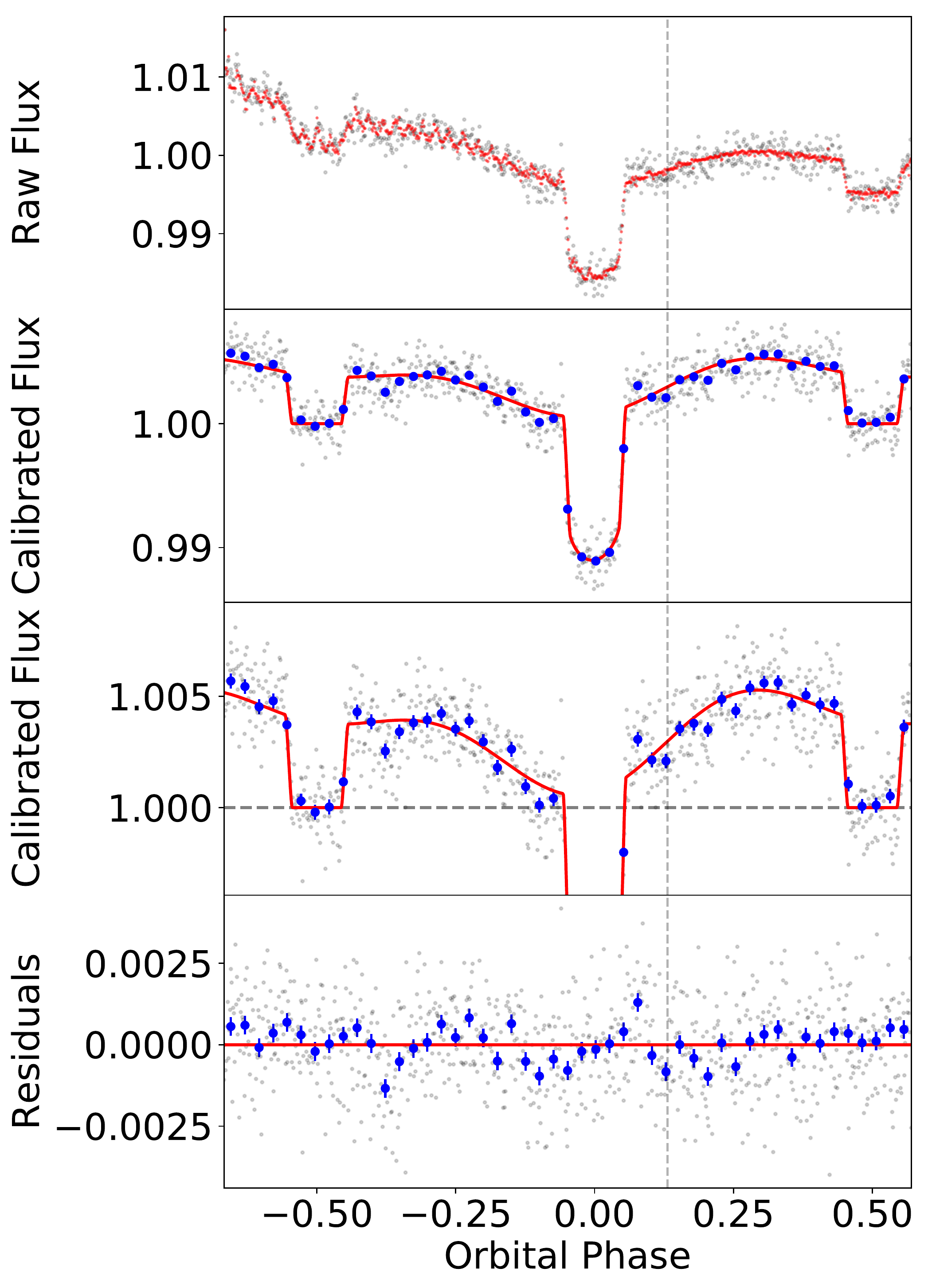}\hfill%
	\includegraphics[width=0.48\linewidth]{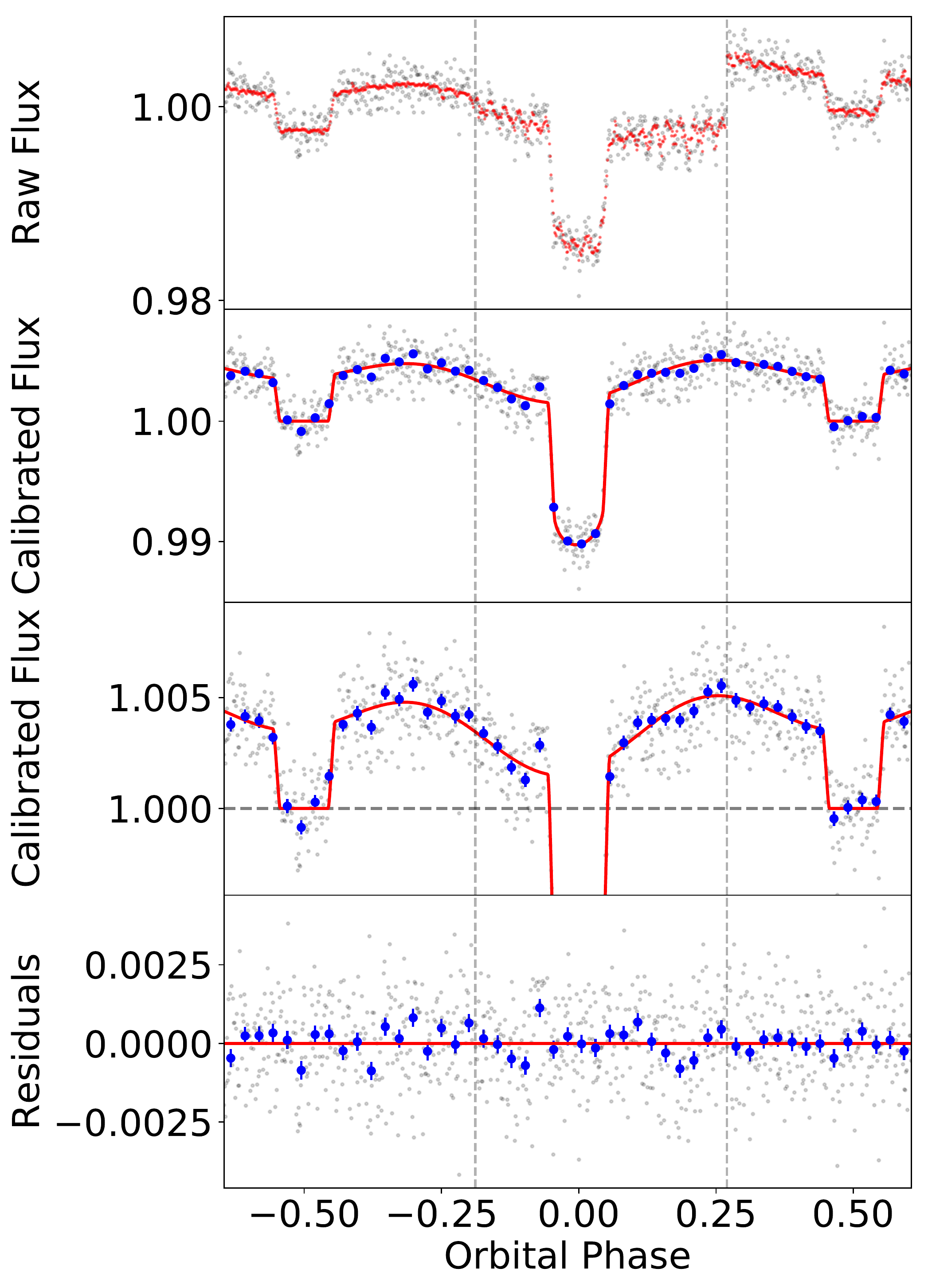}
		
	\vspace{0.5cm}
	
	\includegraphics[width=0.48\linewidth]{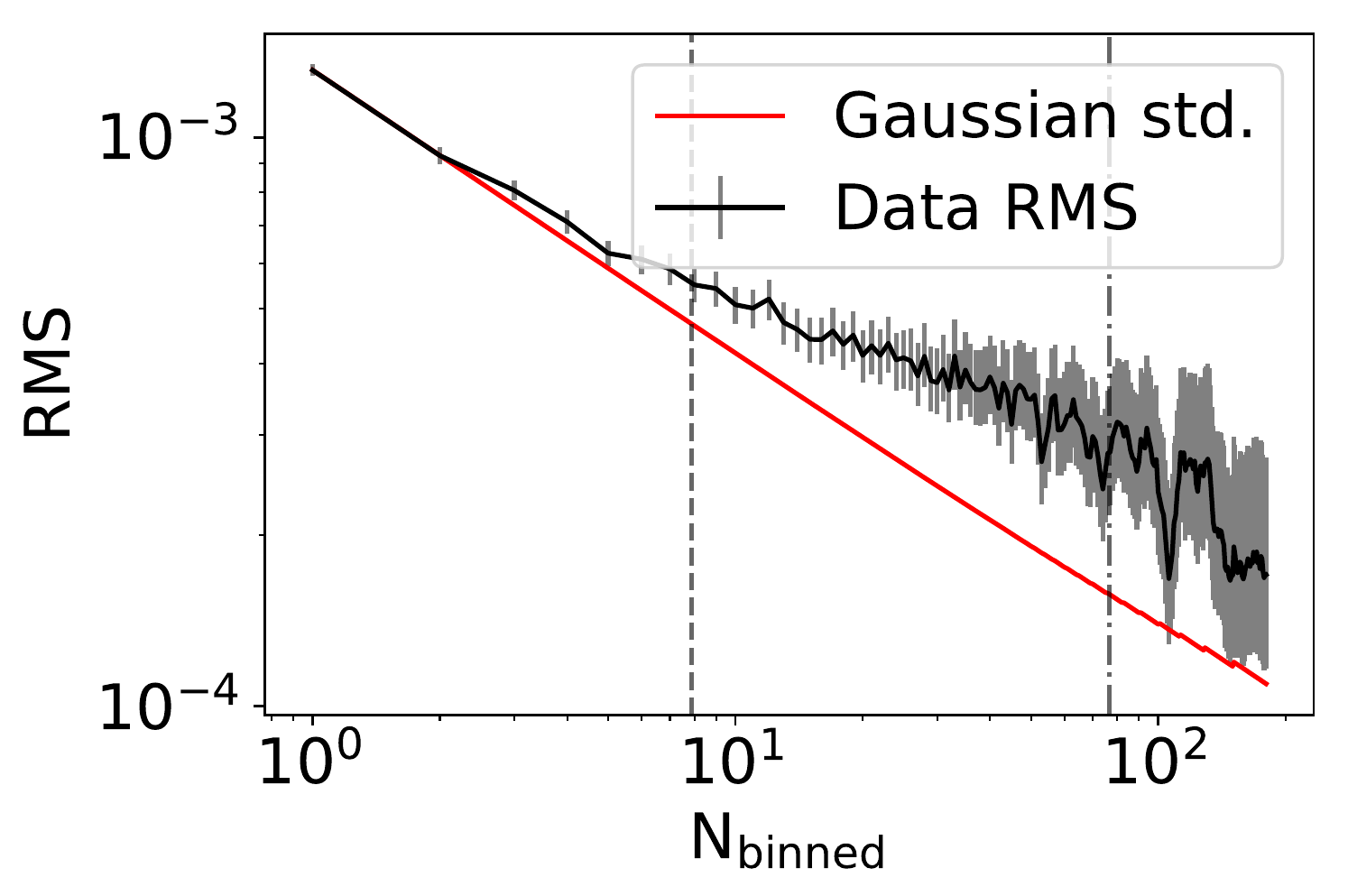}\hfill%
	\includegraphics[width=0.48\linewidth]{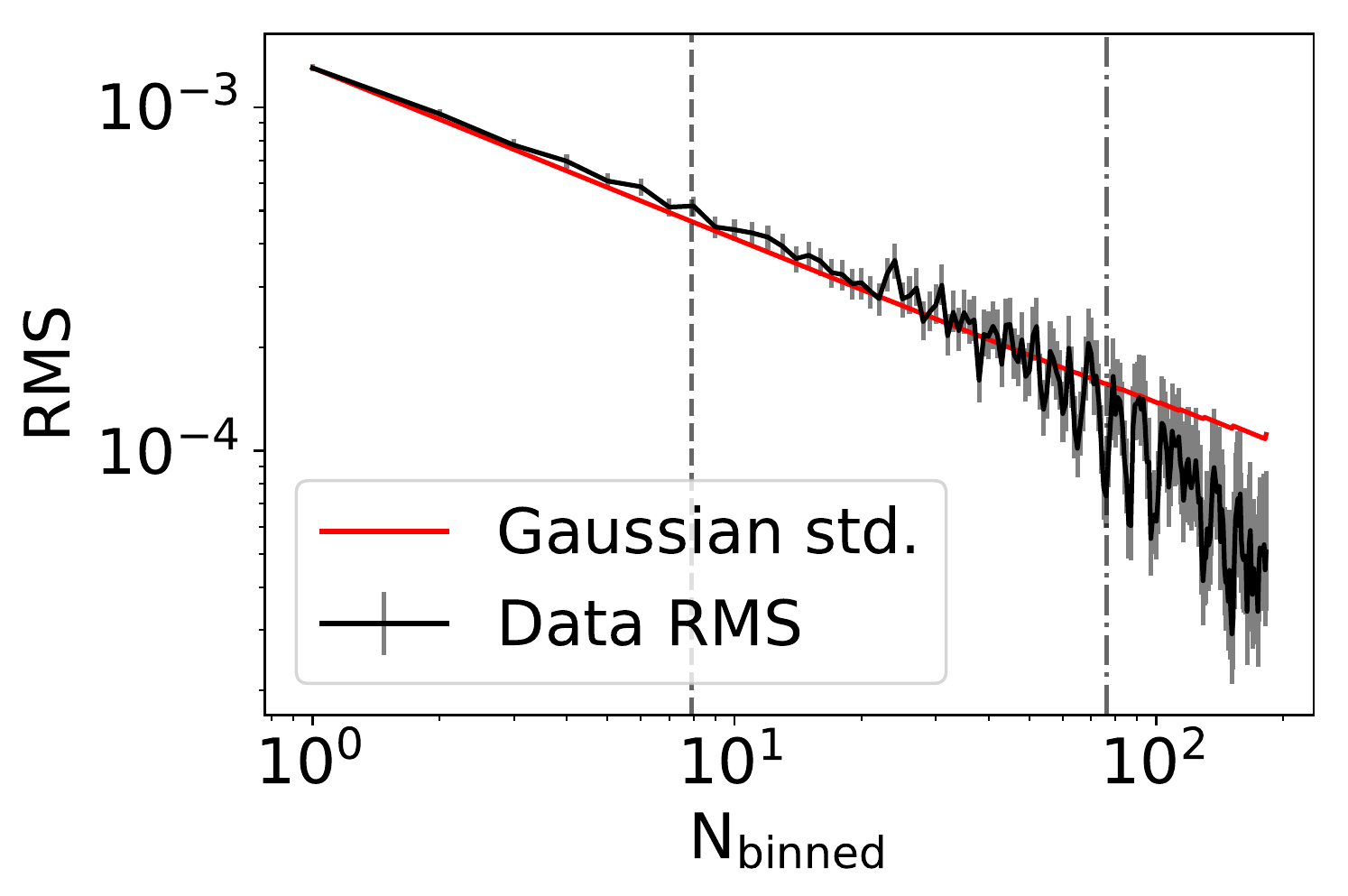}
	\caption{Same figure as Figure A5 but for the 2010 (left) and 2013 (right) 4.5~$\mu$m observations of WASP-12b, both fit using the SPCA BLISS detector model and the second order phase variations model.}\label{fig:bell2}
\end{figure*}

\begin{figure*}
	\begin{flushleft}
        \LARGE \hspace{1.7in} 2010 \hspace{3.2in} 2013
    \end{flushleft}
    
	\centering
	\includegraphics[width=0.48\linewidth]{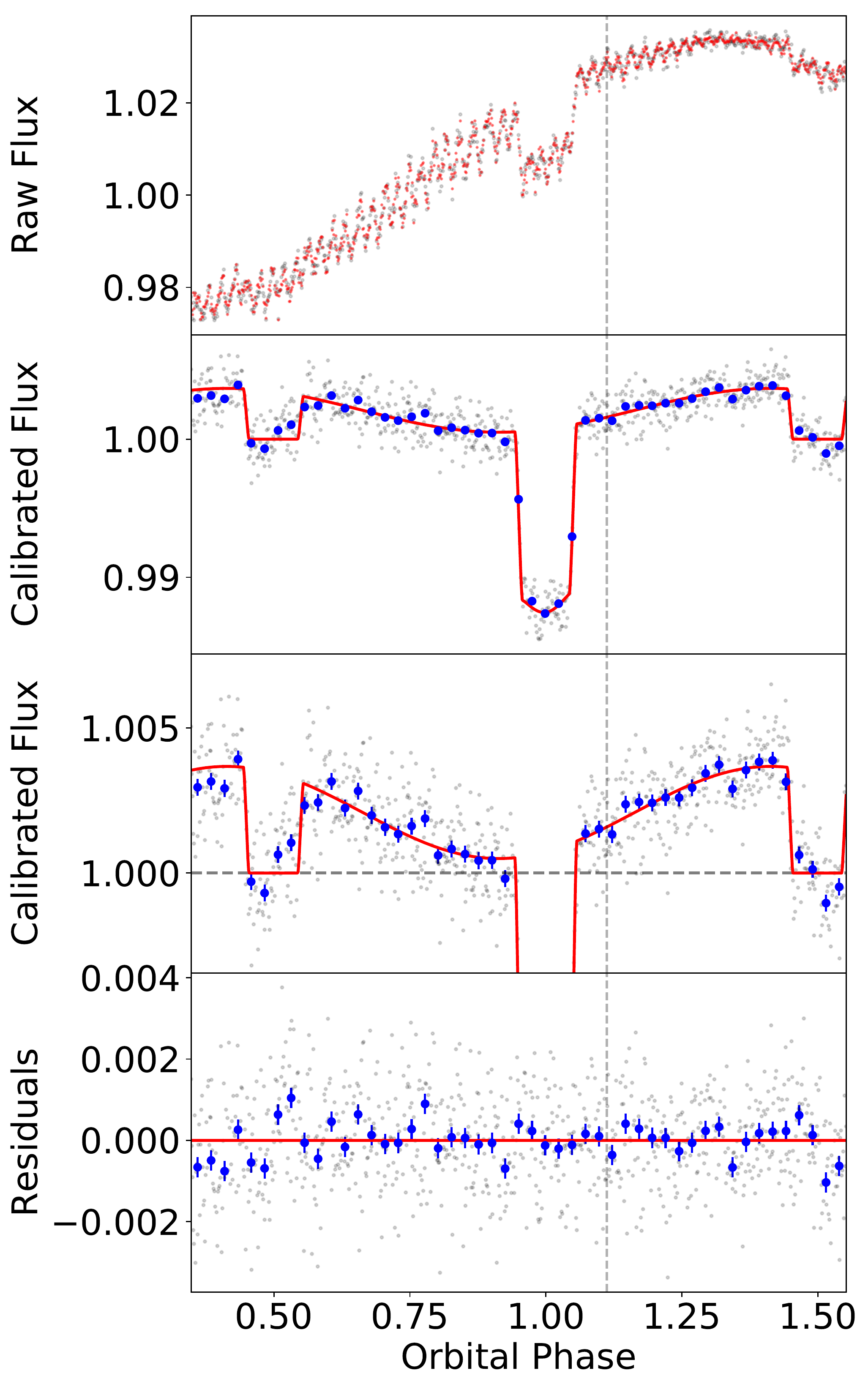}\hfill%
	\includegraphics[width=0.48\linewidth]{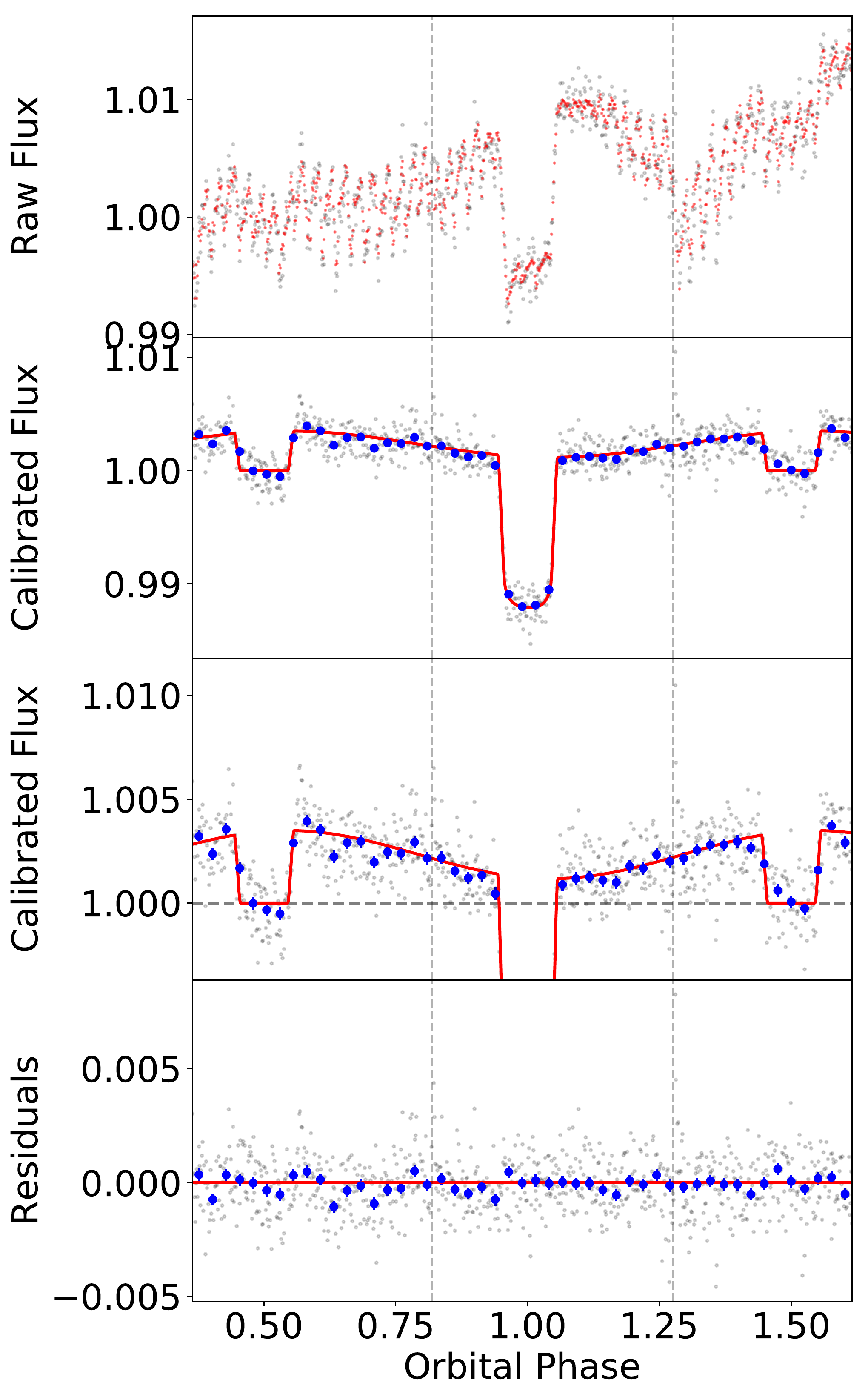}
		
	\vspace{0.5cm}
	
	\includegraphics[width=0.48\linewidth]{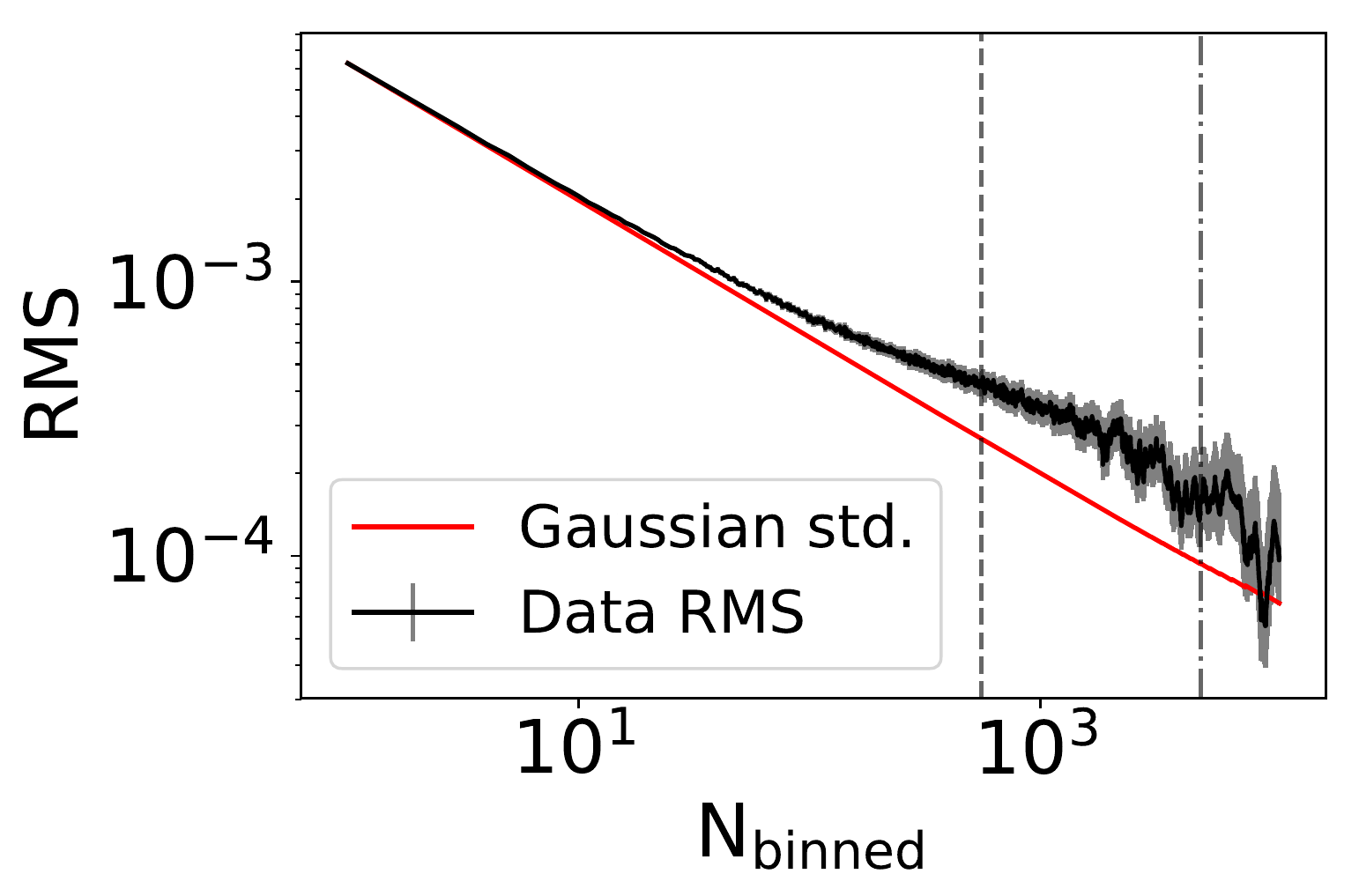}\hfill%
	\includegraphics[width=0.48\linewidth]{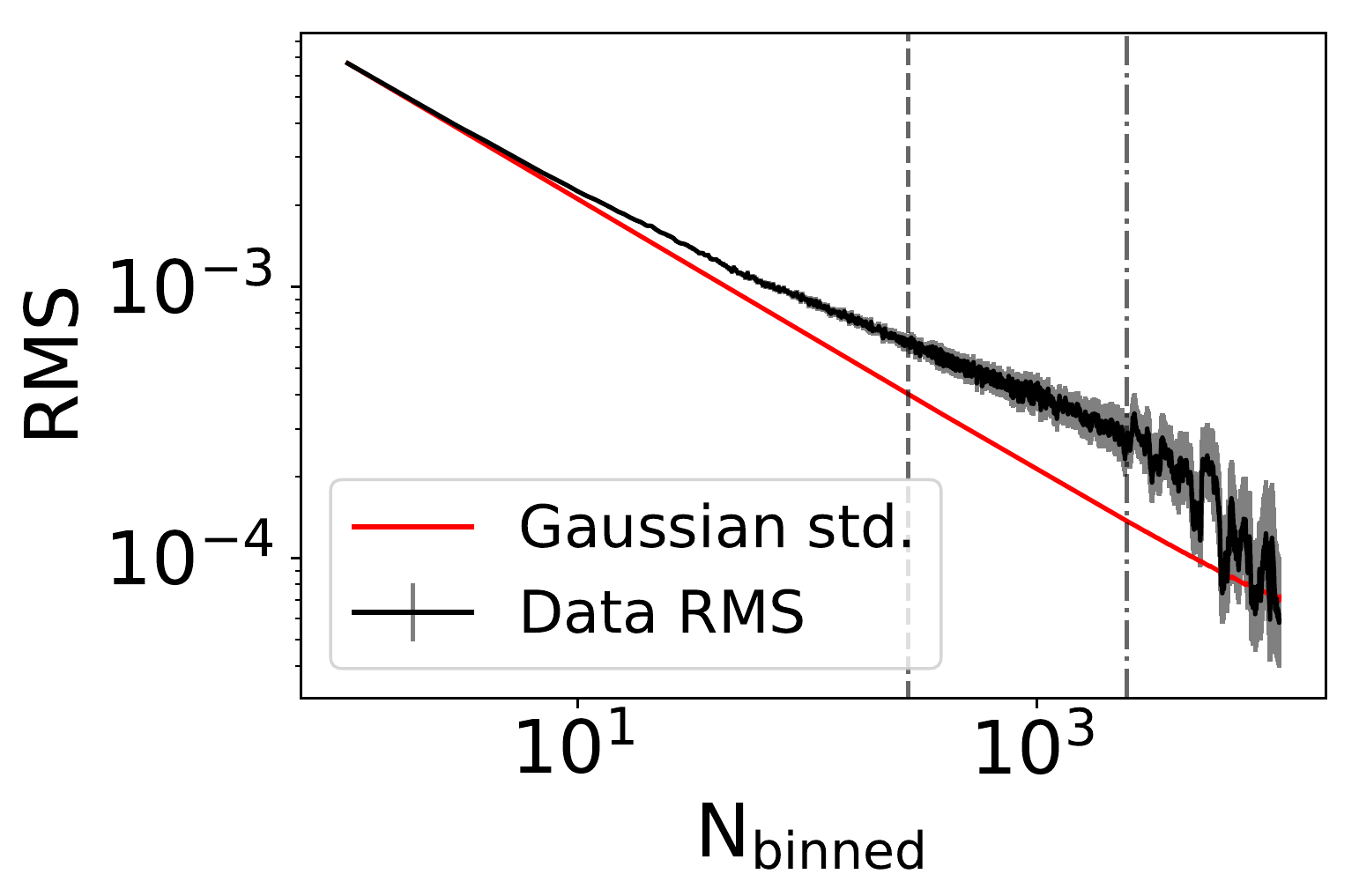}
	\caption{Same figure as Figure A5 but for the 2010 (left) and 2013 (right) 3.6~$\mu$m observations of WASP-12b, both fit using the \texttt{POET} pipeline.}\label{fig:cubillos1}
\end{figure*}

\begin{figure*}
	\begin{flushleft}
        \LARGE \hspace{1.7in} 2010 \hspace{3.2in} 2013
    \end{flushleft}
    
	\centering
	\includegraphics[width=0.48\linewidth]{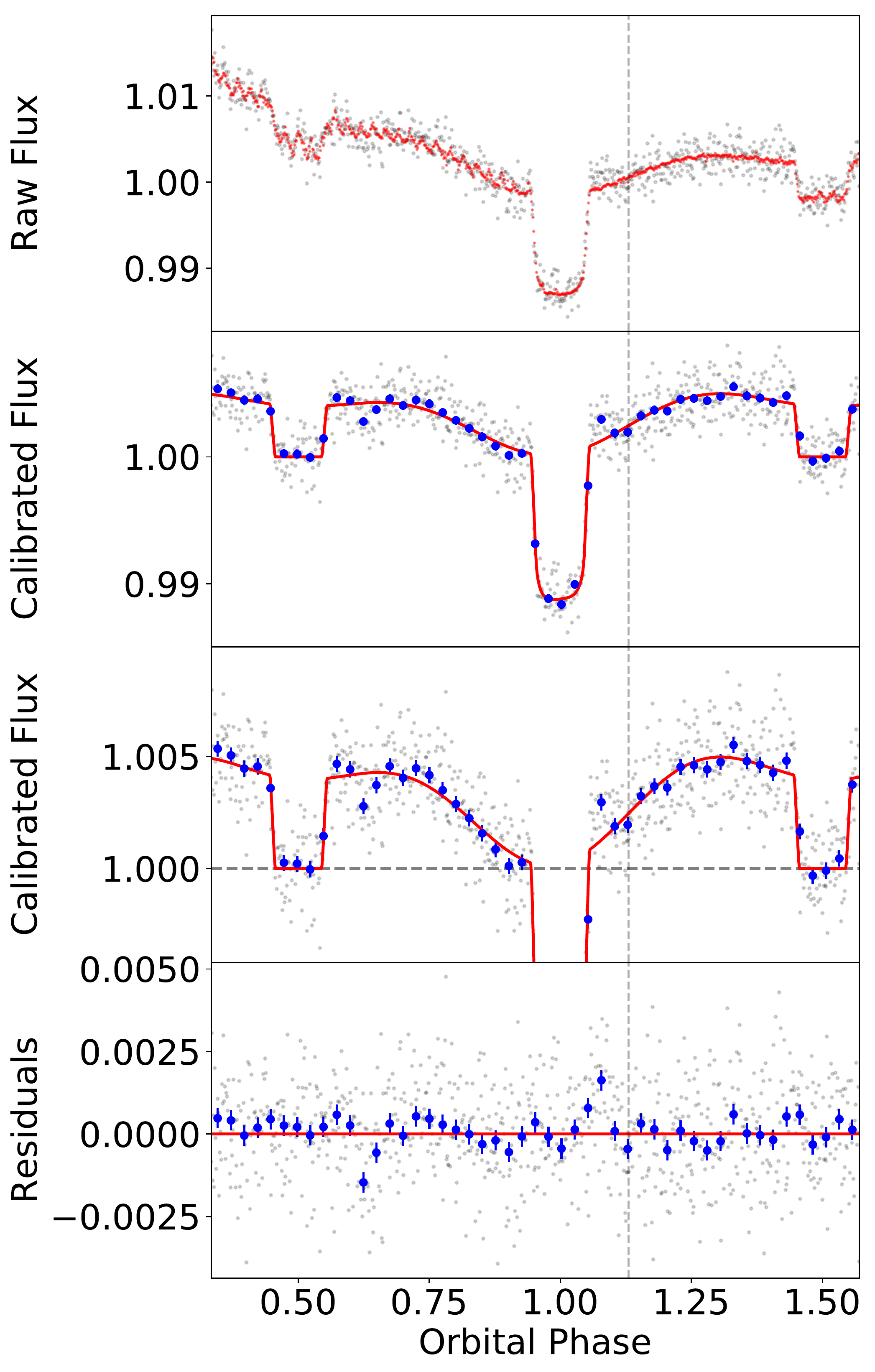}\hfill%
	\includegraphics[width=0.48\linewidth]{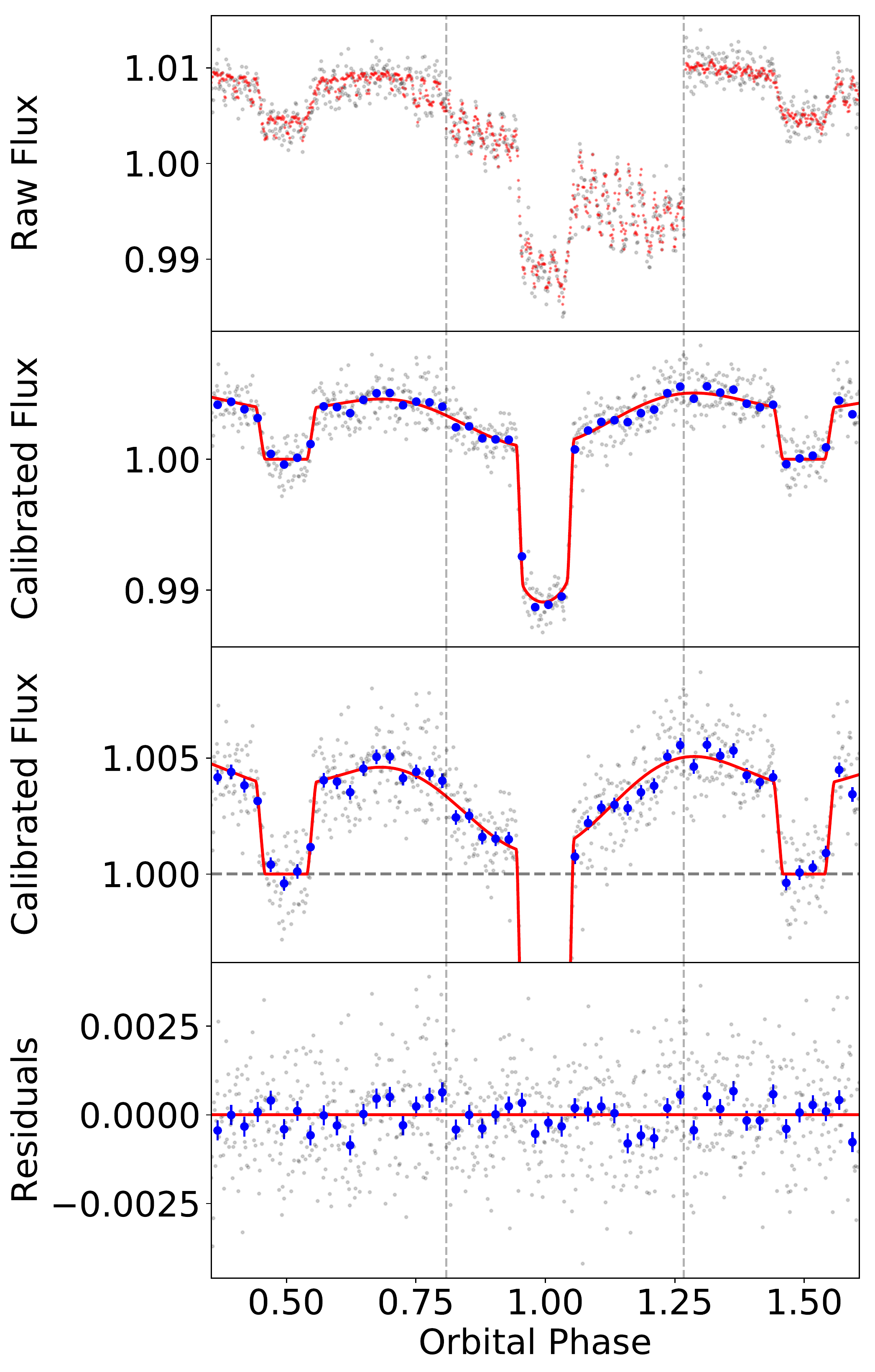}
		
	\vspace{0.5cm}
	
	\includegraphics[width=0.48\linewidth]{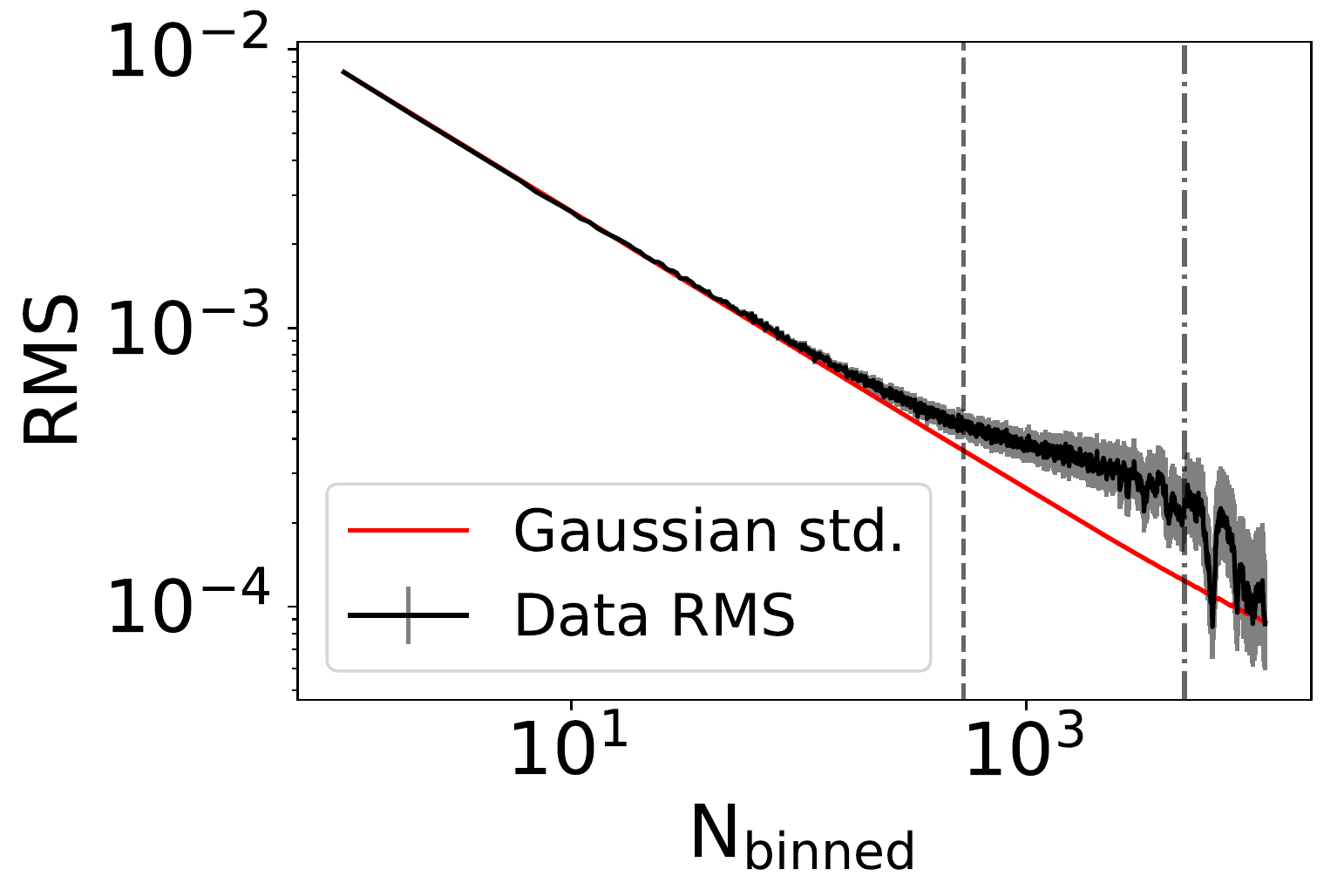}\hfill%
	\includegraphics[width=0.48\linewidth]{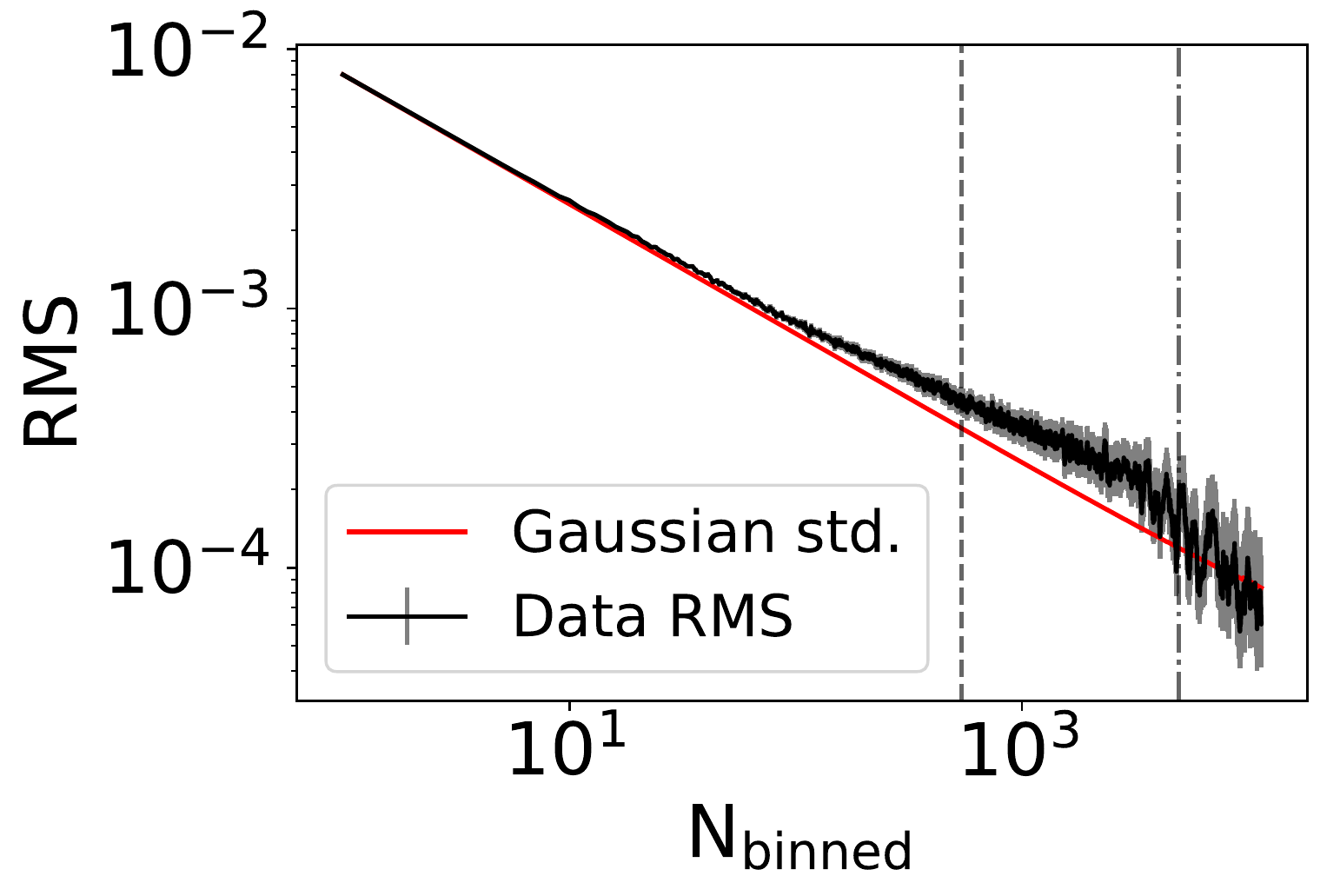}
	\caption{Same figure as Figure A5 but for the 2010 (left) and 2013 (right) 4.5~$\mu$m observations of WASP-12b, both fit using the \texttt{POET} pipeline.}\label{fig:cubillos2}
\end{figure*}

\clearpage

\begin{table*}
\begin{center}
	\includegraphics[height=0.975\textheight]{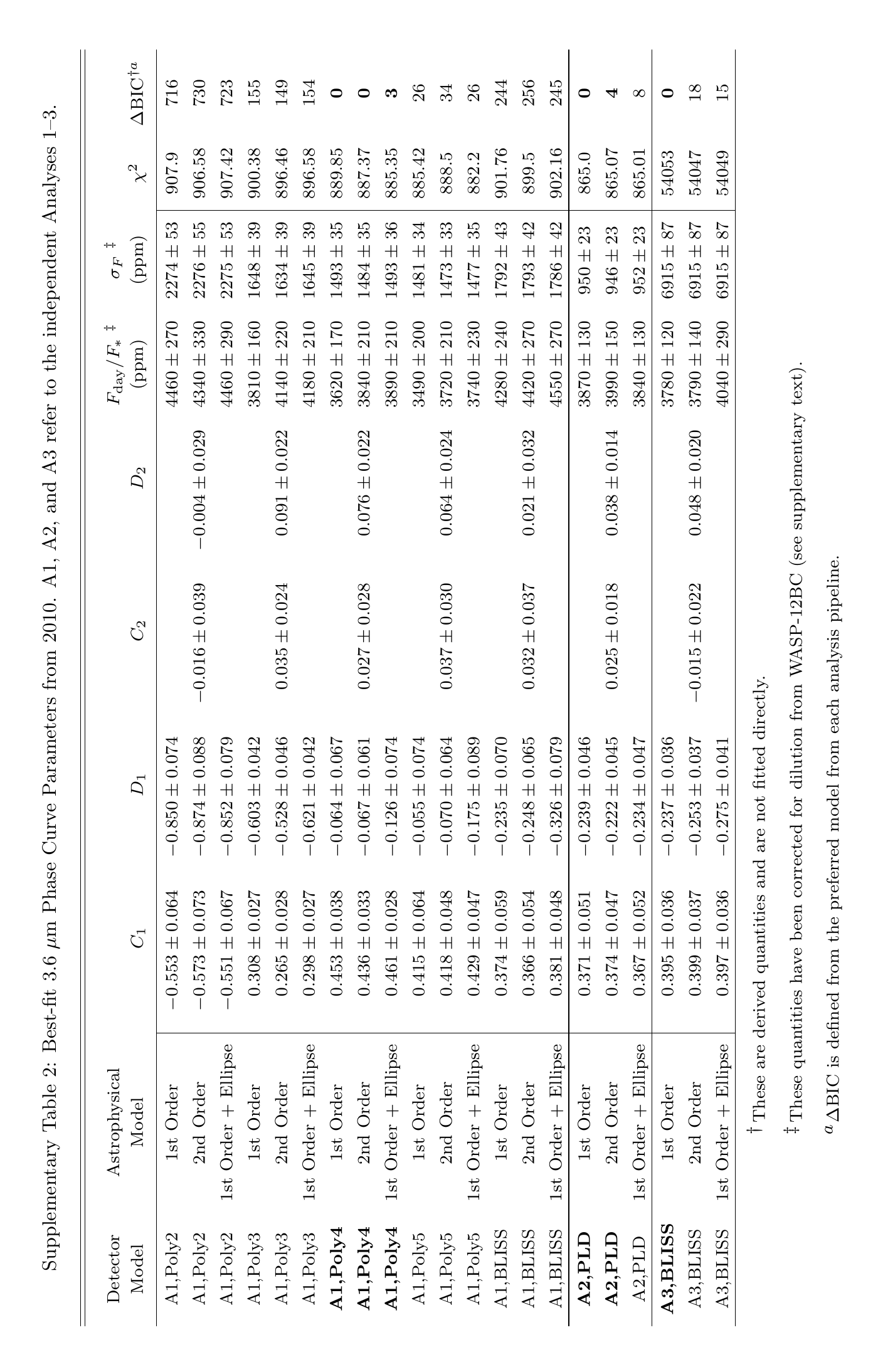}
\end{center}
\end{table*}

\begin{table*}
\begin{center}
	\includegraphics[height=0.975\textheight]{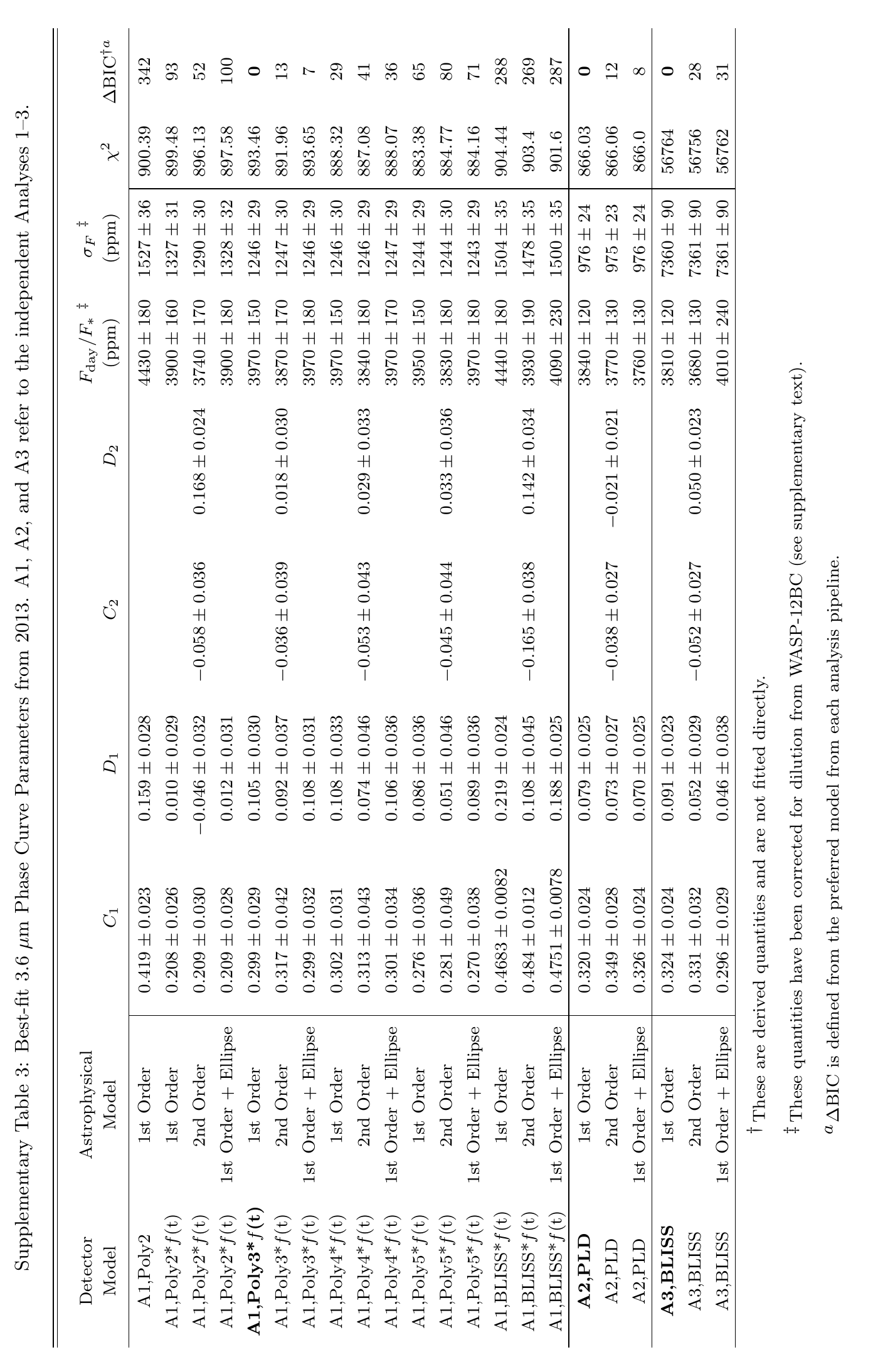}
\end{center}
\end{table*}

\begin{table*}
\begin{center}
	\includegraphics[height=0.975\textheight]{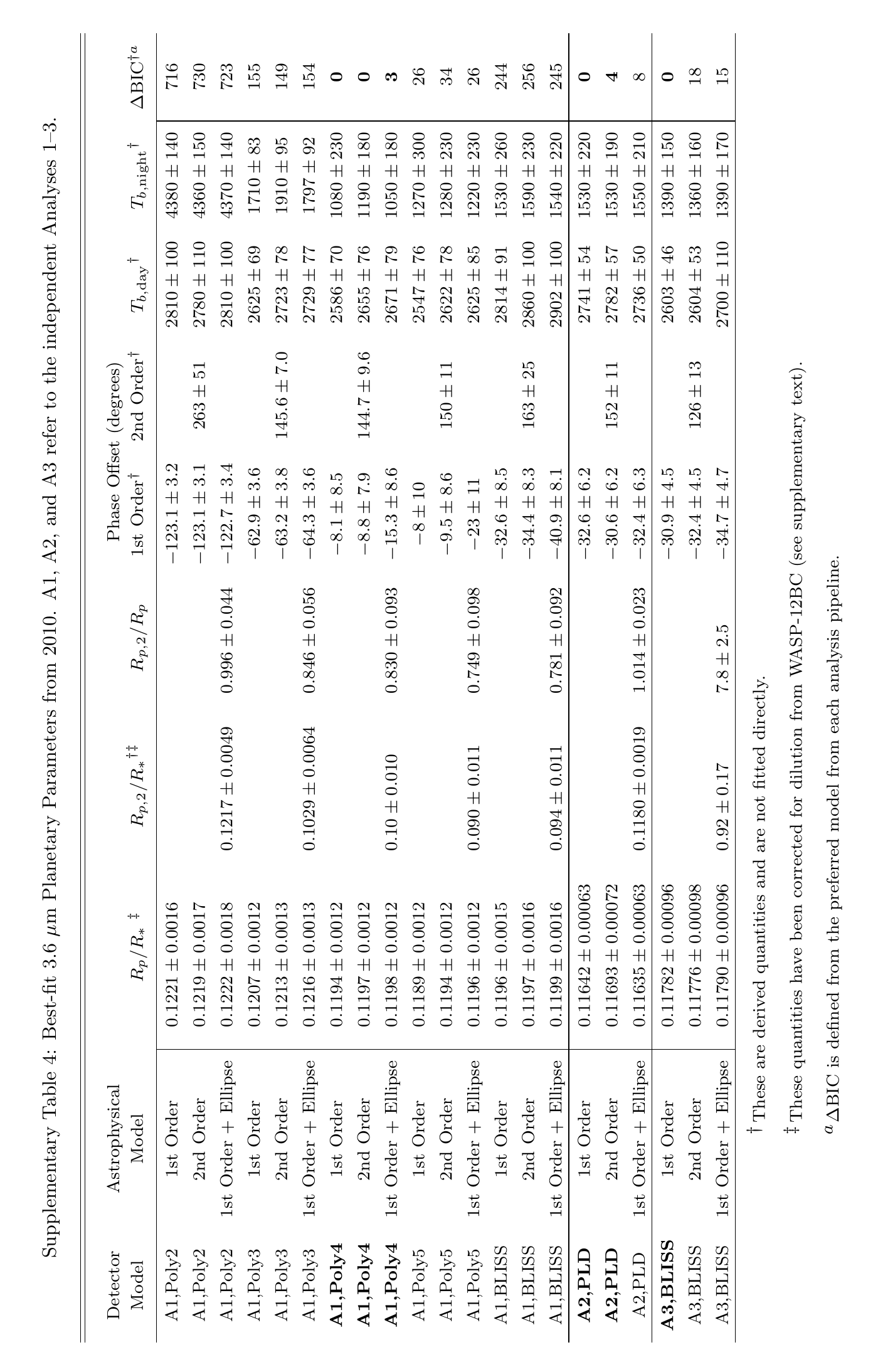}
\end{center}
\end{table*}

\begin{table*}
\begin{center}
	\includegraphics[height=0.975\textheight]{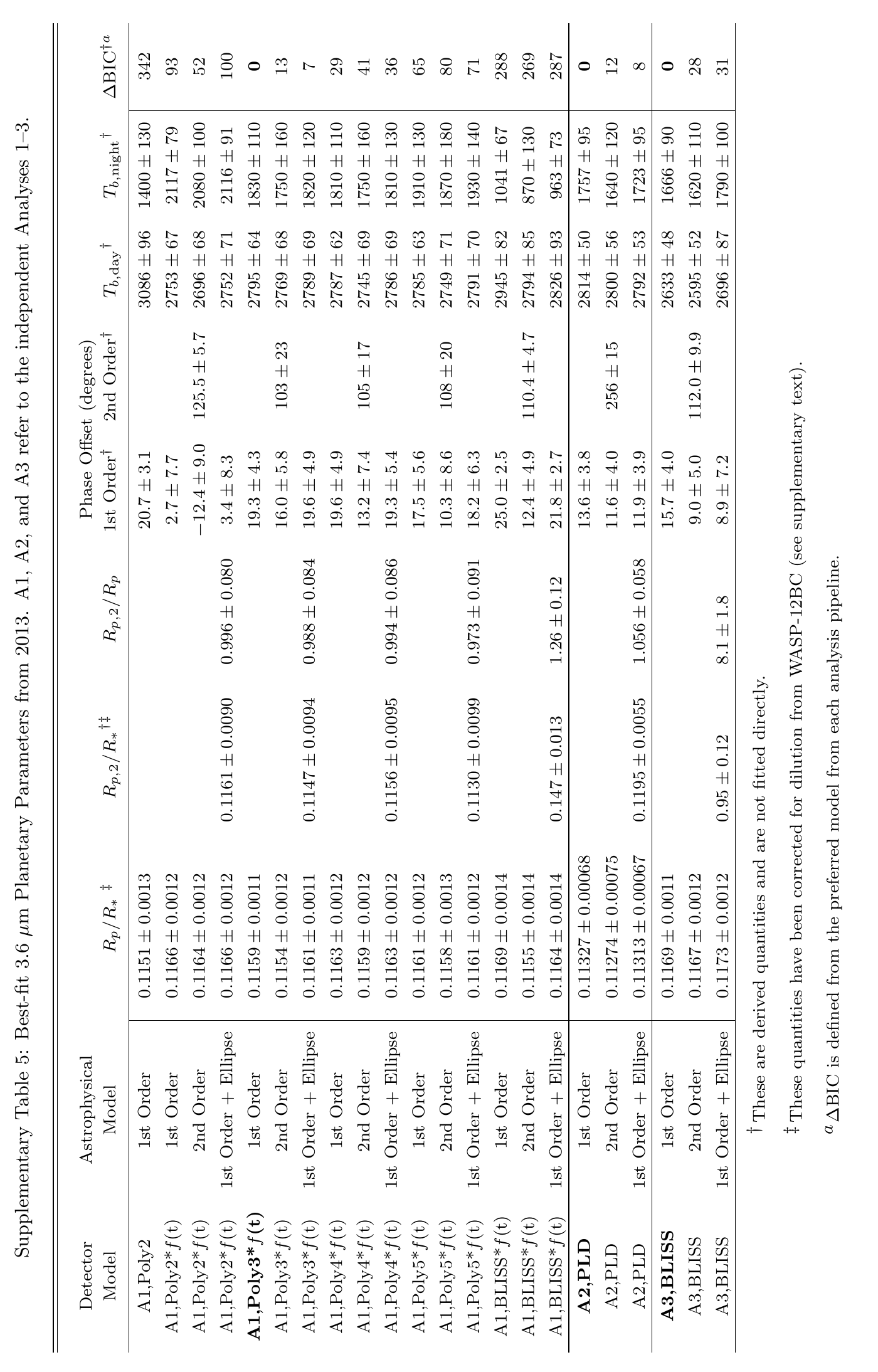}
\end{center}
\end{table*}

\begin{table*}
\begin{center}
	\includegraphics[height=0.975\textheight]{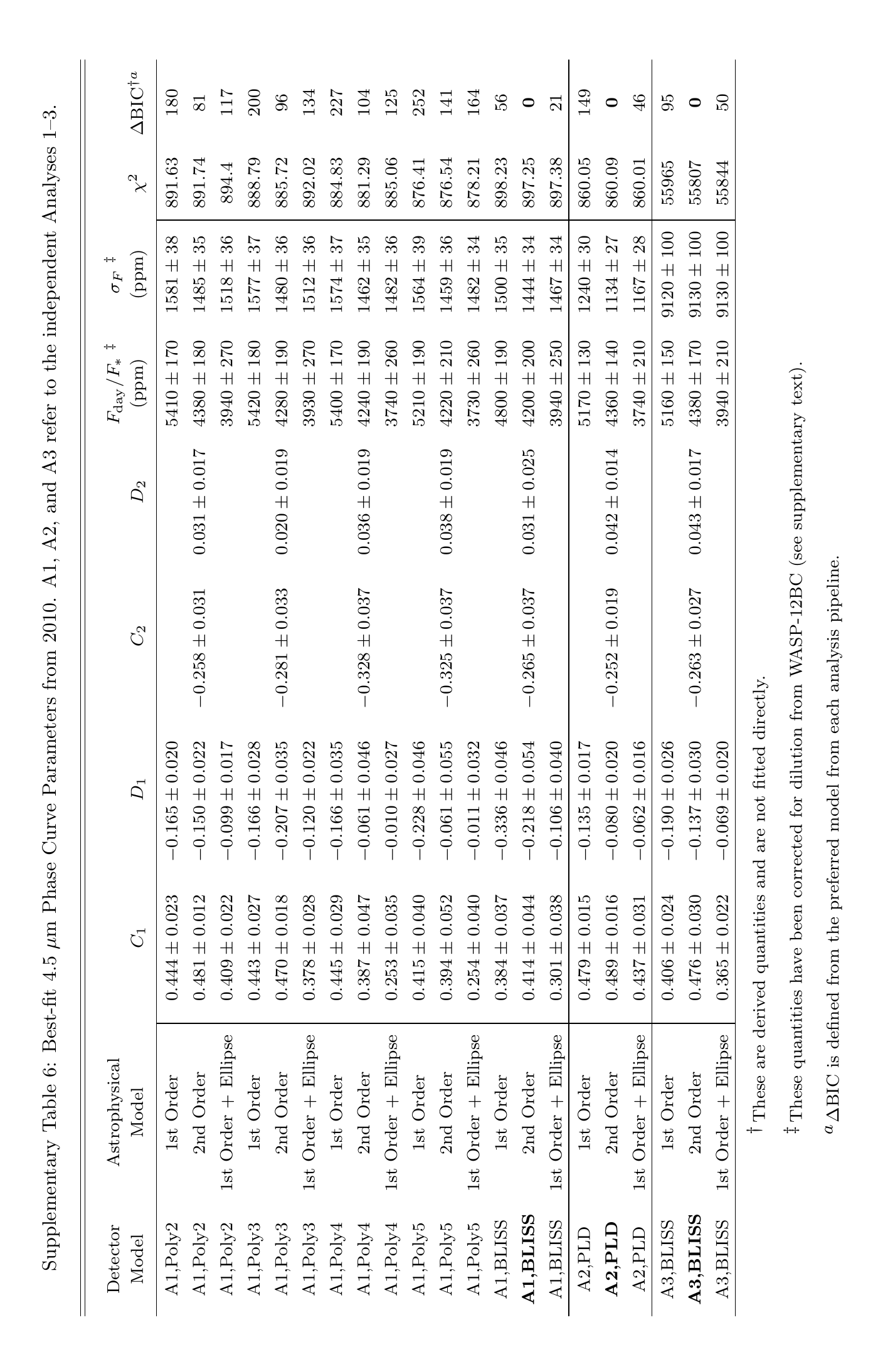}
\end{center}
\end{table*}

\begin{table*}
\begin{center}
	\includegraphics[height=0.975\textheight]{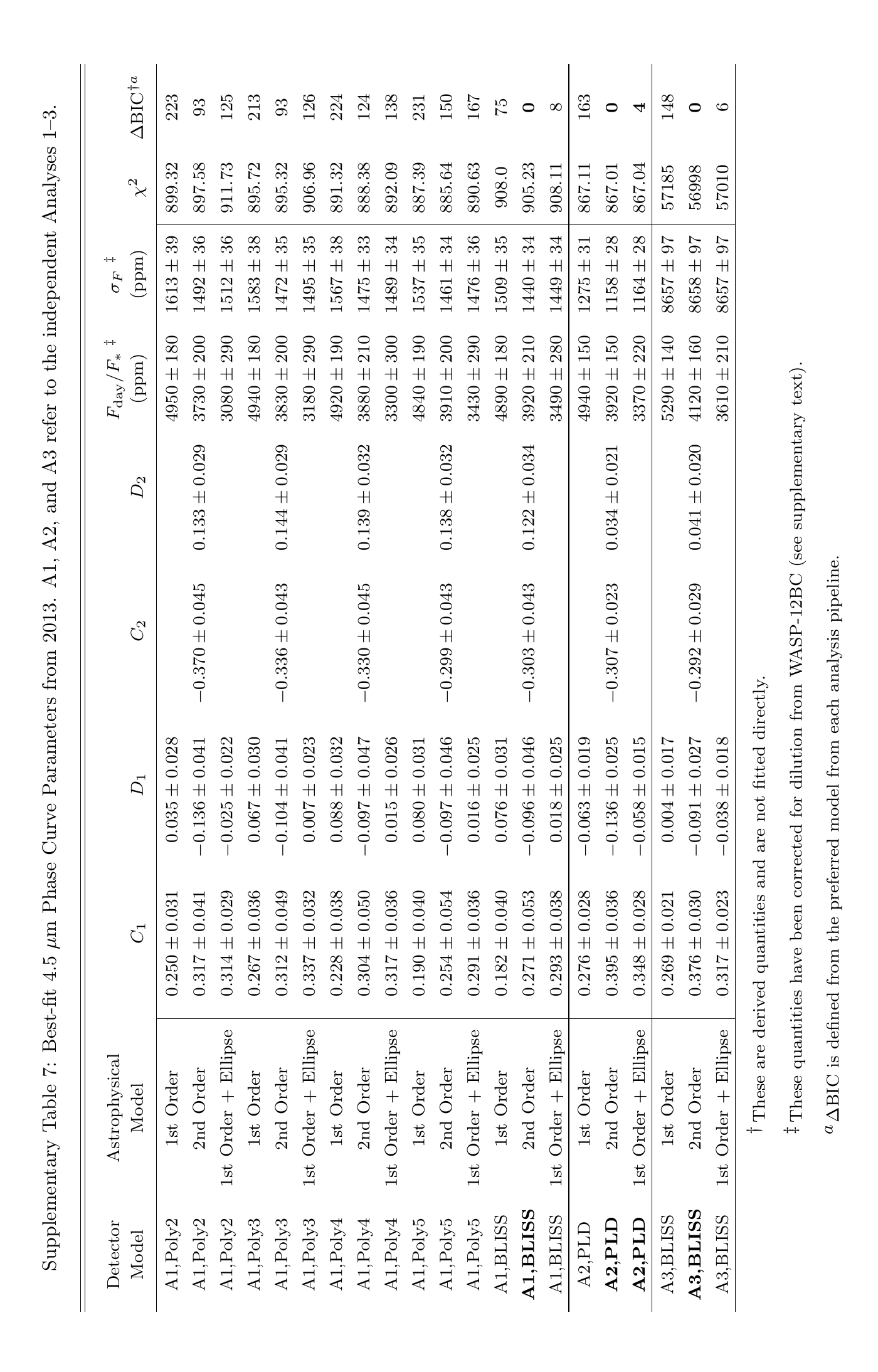}
\end{center}
\end{table*}

\begin{table*}
\begin{center}
	\includegraphics[height=0.975\textheight]{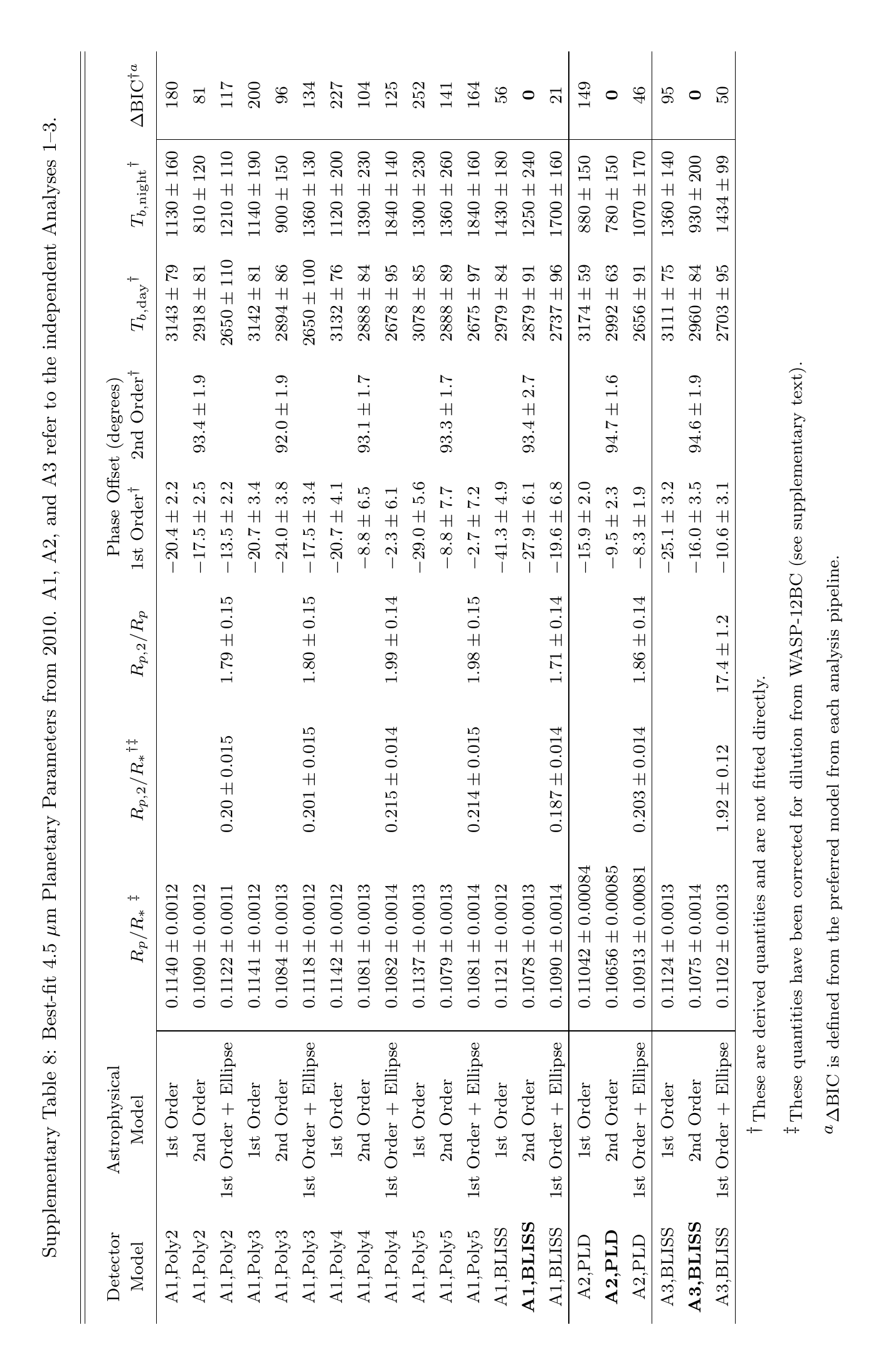}
\end{center}
\end{table*}

\begin{table*}
\begin{center}
	\includegraphics[height=0.975\textheight]{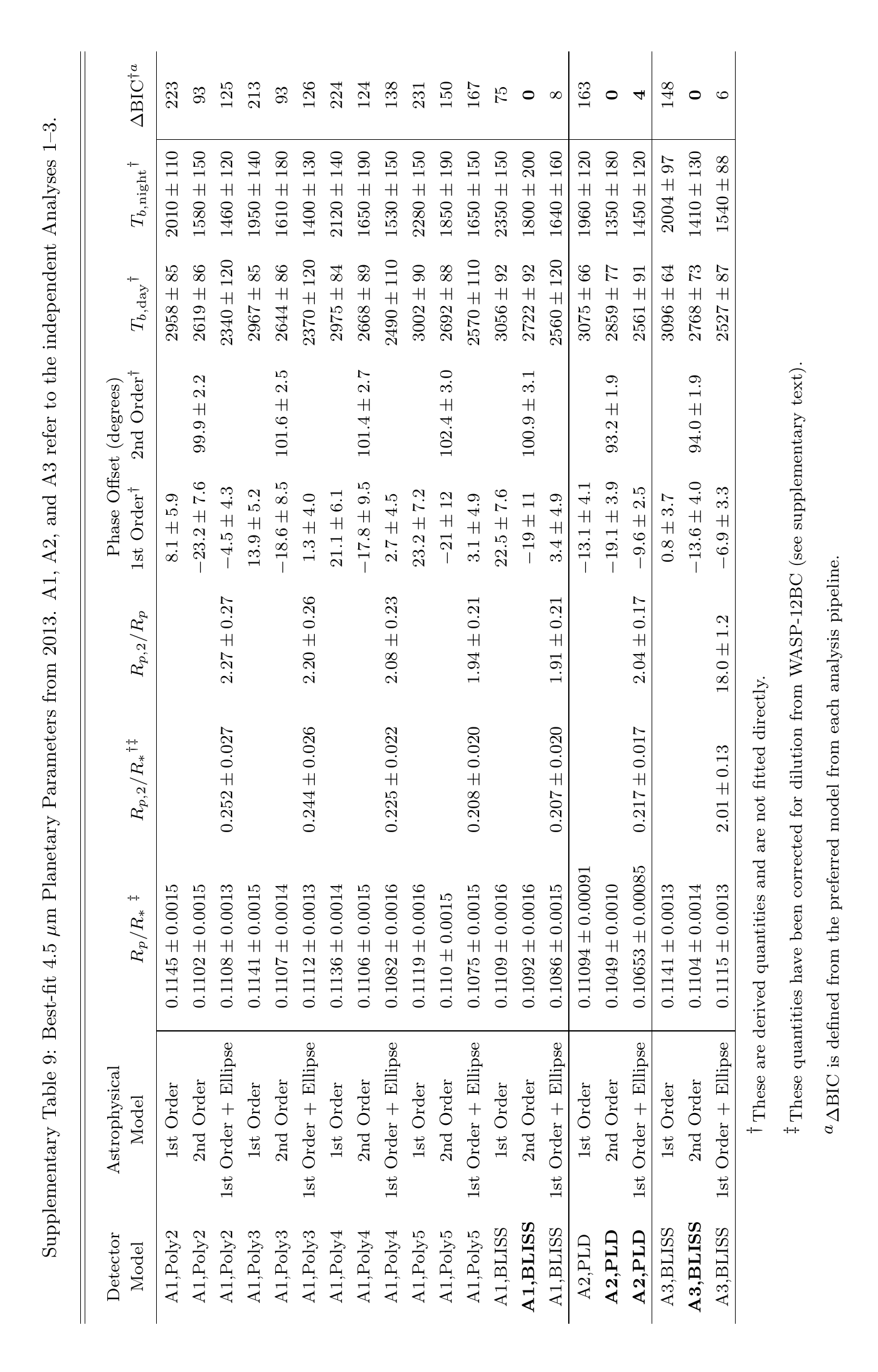}
\end{center}
\end{table*}

\end{document}